\documentclass[final,3p,authoryear]{elsarticle}

\usepackage{url}
\usepackage{hyperref}
\hypersetup{
  colorlinks   = true, 
  urlcolor     = blue, 
  linkcolor    = blue,  
  citecolor    = blue   
}
\usepackage{xspace}
\newcommand*{\eg}{e.g.,\@\xspace}
\newcommand*{\ie}{i.e.,\@\xspace}

\usepackage[T1]{fontenc}
\usepackage[utf8]{inputenc}

\usepackage[export]{adjustbox}
\usepackage[english]{babel}
\usepackage{subcaption}
\usepackage{caption,setspace}
\usepackage[algoruled]{algorithm2e}
\usepackage{csquotes}

\usepackage{flafter}
\usepackage{placeins}

\usepackage{amssymb}
\usepackage{mathtools}
\usepackage[nameinlink,capitalize]{cleveref}

\newcommand{\bval}[1]{b~=~#1~s/mm$^2$}
\DeclarePairedDelimiter{\abs}{\lvert}{\rvert}

\DeclareGraphicsExtensions{.png,.jpg}
\graphicspath{{images/}}

\makeatletter
\g@addto@macro\@floatboxreset\centering
\makeatother

\journal{Medical Image Analysis}

\newcommand{\review}[1]{{#1}}

\begin{document}
\begin{frontmatter}

    \title{Automated characterization of noise distributions in diffusion MRI data}

    \author[umc]{Samuel St-Jean}\corref{corr}\ead{samuel@isi.uu.nl}
    \author[umc]{Alberto De Luca}\ead{alberto@isi.uu.nl}
    \author[cubric]{Chantal M. W. Tax}\ead{taxc@cardiff.ac.uk}
    \author[umc]{Max A. Viergever}\ead{max@isi.uu.nl}
    \author[umc]{Alexander Leemans}\ead{a.leemans@umcutrecht.nl}

    \address[umc]{Image Sciences Institute, Department of Radiology, University Medical Center Utrecht, Heidelberglaan 100, 3584 CX Utrecht, the Netherlands}
    \address[cubric]{Cardiff University Brain Research Imaging Centre (CUBRIC), School of Psychology, Cardiff University, Maindy Road, Cardiff, CF24 4HQ, United Kingdom}

    \cortext[corr]{Corresponding author}


\begin{abstract}
Knowledge of the noise distribution in magnitude diffusion MRI images is the centerpiece to quantify uncertainties arising from the acquisition process.
The use of parallel imaging methods, the number of receiver coils and imaging filters applied by the scanner, amongst other factors, dictate the resulting signal distribution.
Accurate estimation beyond textbook Rician or noncentral chi distributions often requires information about the acquisition process (\eg coils sensitivity maps or reconstruction coefficients),
which is usually not available.
We introduce two new automated methods using the moments and maximum likelihood equations of the Gamma distribution
to estimate noise distributions as they explicitly depend on the number of coils,
making it possible to estimate all unknown parameters using only the magnitude data.
A rejection step is used to make the framework automatic and robust to artifacts.
Simulations using stationary and spatially varying noncentral chi noise distributions were created
for two diffusion weightings with SENSE or GRAPPA reconstruction and 8, 12 or 32 receiver coils.
Furthermore, MRI data of a water phantom with different combinations of parallel imaging were acquired on a 3T Philips scanner along with noise-only measurements.
Finally, experiments on freely available datasets from a single subject acquired on a 3T GE scanner are used
to assess reproducibility when limited information about the acquisition protocol is available.
Additionally, we demonstrated the applicability of the proposed methods for a bias correction and denoising task
on an in vivo dataset acquired on a 3T Siemens scanner.
A generalized version of the bias correction framework for non integer \review{degrees of freedom} is also introduced.
The proposed framework is compared with three other algorithms with datasets from three vendors,
employing different reconstruction methods.
Simulations showed that assuming a Rician distribution can lead to misestimation of the noise distribution in parallel imaging.
Results on the acquired datasets showed that signal leakage in multiband can also lead to a misestimation of the noise distribution.
Repeated acquisitions of in vivo datasets show that the estimated parameters are stable and have lower variability than compared methods.
Results for the bias correction and denoising task show that the proposed methods reduce the appearance of noise at high b-value.
The proposed algorithms herein can estimate both parameters of the noise distribution automatically,
are robust to signal leakage artifacts and perform best when used on acquired noise maps.
\end{abstract}


\end{frontmatter}


\section{Introduction}
\label{sec:intro}

Diffusion magnetic resonance imaging (dMRI) is a non invasive imaging technique which allows
probing microstructural properties of living tissues. Advances in parallel imaging techniques \citep{Pruessmann1999,Griswold2002},
such as accelerated acquisitions (\eg partial k-space \citep{Storey2007a}, multiband imaging \citep{Nunes2006,Moeller2010} and compressed sensing \citep{Lustig2007,Paquette2014a}),
have greatly reduced the inherently long scan time in dMRI.
New acquisition methods and pulse sequences in dMRI are also pushing the limits of spatial resolution while reducing scan time \citep{Holdsworth2019},
which also affects the signal distribution in ways that are challenging to model.
Estimation of signal distributions deviating from theoretical cases is challenging and oftentimes requires
information such as coil sensitivities or reconstruction matrices.
This information may not be recorded at acquisition time or is even not available from the scanner,
making techniques relying on these parameters difficult to apply in practice.
Even though the magnitude signal model is still valid nowadays, the use of image filters \citep{Dietrich2008},
acceleration methods subsampling k-space
(\eg the SENSE (SENsitivity ENcoding) \citep{Pruessmann1999}, GRAPPA (GeneRalized Autocalibrating Partial Parallel Acquisition) \citep{Griswold2002,Heidemann2012}
or the homodyne detection methods \citep{Noll1991})
and spatial correlation between coil elements \citep{Dietrich2008,Aja-Fernandez2014} influence, amongst other factors, the parameters of the resulting signal distribution.
\review{See \eg \citep{Aja-Fernandez2015a,Aja-Fernandez2009} for a review on estimating noise distributions in MRI and common statistical distributions encountered therein.}

With the recent trend towards open data sharing and large multicenter studies using standardized protocols \citep{Duchesne2019,Emaus2015},
differences in hardware, acquisition or reconstruction algorithms may inevitably lead to different signal distributions.
This may affect large scale longitudinal studies investigating neurological changes due to these \enquote{scanner effects} \citep{Sakaie2018}
as the acquired data may be fundamentally different across sites in terms of statistical properties of the signal.
Algorithms have been developed to mitigate these potential differences \citep{Tax2019,Mirzaalian2018}, but characterization of the signal distribution
from various scanners is challenging due to the black box nature of the acquisition process, especially in routine clinical settings.
While some recent algorithms for dMRI are developed to include information about the
noise distribution \citep{Collier2018,Sakaie2017}, there is no method, to the best of our knowledge,
providing a fully automated way to characterize the noise distribution
using information from the magnitude data itself only.
Due to this gap between the physical acquisition process and noise estimation theory,
noise distributions are either assumed as Rician (with parameter $\sigma_g$ related to the standard deviation) or noncentral chi (with fixed degrees of freedom $N$)
and concentrate in estimating the noise standard deviation $\sigma_g$ \citep{Veraart2015a,Koay2009b,Tabelow2014}.
This assumption inevitably leads to misestimation of the true signal distribution as $N$ and $\sigma_g$ are interdependent for some reconstruction algorithms \citep{Aja-Fernandez2013}.
Reconstruction filters preserving only the real part of the signal also cause $N$ to deviate from the Rician noise distribution,
producing instead a half-Gaussian signal distribution \citep{Dietrich2008}.
Misestimation of the appropriate signal distribution could impact subsequent processing steps such as bias correction \citep{Koay2009a},
denoising \citep{St-Jean2016a} or diffusion model estimation \citep{Zhang2012d,Landman2007a,Sakaie2017},
therefore negating potential gains in statistical power from analyzing datasets acquired in different centers or from different vendors.

In this work, we propose to estimate the parameters $\sigma_g$ and $N$ from either the magnitude data or the acquired noise maps
by using a change of variable to a Gamma distribution $\textit{Gamma}(N, 1)$ \citep{Koay2009b},
whose first moments and maximum likelihood equations directly depend on $N$.
This makes the proposed method fast and easy to apply to existing data without additional information,
while being robust to artifacts by rejecting outliers of the distribution.
Preliminary results of this work have been presented at the annual meeting of the MICCAI \citep{St-jean2018a}.
This manuscript now contains additional theory, simulations including signal correlations and parallel acceleration, and
experiments on phantoms and in vivo datasets acquired with parallel and multiband acceleration.
As example applications, we perform bias correction and denoising on an in vivo dataset using the estimated distribution derived with each algorithm.


\section{Theory}
\label{sec:theory}

In this section, we introduce the necessary background on the Gamma distribution, its moments and maximum likelihood equations.
Expressing the signal with a Gamma distribution highlights equations which can be solved to estimate parameters $\sigma_g$ and $N$.

\subsection{Probability distribution functions of MRI data}
\label{sec:special_pdf}

To account for uncertainty in the acquisition process, the complex signal measured in k-space by the receiver coil array
can be modeled with a separate additive zero mean Gaussian noise for each channel with identical variance $\sigma^2_g$ \citep{Gudbjartsson1995}.
The signal acquired from the real and imaginary part of each coil in a reconstructed magnitude image can be expressed as \citep{Constantinides1997}

\begin{equation}
    m_N = \sqrt{\sum_{n=1}^N m_{Rn}^2 + m_{In}^2},
\end{equation}
where $m_{Rn}$ and $m_{In}$ are the real and imaginary parts of the signal, respectively, as measured by coil number $n$,
$N$ is the number of degrees of freedom (which can be up to the number of coils in the absence of accelerated parallel imaging)
and $m_N$ is the resulting reconstructed signal value for a given voxel.
The magnitude signal can therefore be approximated by a noncentral chi distribution and has a probability density function (pdf) given by \citep{Koay2009a,Dietrich2008}
\begin{equation}
    \textit{pdf}(m | \eta, \sigma_g, N) = \frac{m^N}{\sigma^2_g \eta^{N-1}} \exp{\left(\frac{-(m^2 + \eta^2)}{2\sigma_g^2}\right)}\, I_{N-1}\left(\frac{m\eta}{\sigma^2_g}\right),
    \label{eq:pdf_nchi}
\end{equation}
where $m$ is the noisy signal value for a given voxel, $\eta$ is the (unknown) noiseless signal value, $\sigma_g$ is the Gaussian noise standard deviation,
$N$ is the number of degrees of freedom and $I_\nu(z)$ is the modified Bessel function of the first kind.

With the introduction of multiband imaging and other modern acquisition methods,
parameters estimation of the magnitude data is not straightforward anymore.
The number of degrees of freedom $N$, which is related to the number of receiver coils,
likely deviates from heuristic estimation based on the actual number of coils as $N$ also depends on the reconstruction technique employed \citep{Sotiropoulos2013b}.
The pdf of the magnitude data can be modeled by considering spatially varying degrees of freedom $\textit{N}_{eff}$ and standard deviation $\sigma_{eff}$
(also called the \emph{effective} values) and we generally have $\textit{N}_{eff} \le N$, \citep{Dietrich2008,Aja-Fernandez2014}.

The noncentral chi distribution includes the Rician $(N=1)$, the Rayleigh $(N=1, \eta=0)$ and the central chi distribution $(\eta=0)$ as special cases \citep{Dietrich2008}.
The pdf of the central chi distribution is given by 
\begin{equation}
    \textit{pdf}(m | \eta = 0, \sigma_g, N) = \frac{m^{2N-1}}{2^{N-1}\sigma^{2N}_g \Gamma(N)} \exp{\left(\frac{-m^2}{2\sigma_g^2}\right)},
    \label{eq:pdf_chi}
\end{equation}
where $\Gamma(x)$ is the Gamma function.
With a change of variable introduced by \citep{Koay2009b}, \cref{eq:pdf_chi} can be rewritten as a
Gamma distribution $\textit{Gamma}(N, 1)$ with $t = m^2 / 2\sigma^2_g, dt = m / \sigma^2_g dm$ which has a pdf given by
\begin{equation}
    \textit{pdf}(t|N, 1) = \frac{1}{\Gamma(N)} t^{N-1} \exp{(-t)}.
    \label{eq:pdf_gamma_sec_theory}
\end{equation}
\cref{eq:pdf_gamma_sec_theory} only depends on $N$, which can be estimated from the sample values.

\subsection{Parameter estimation using the method of moments and maximum likelihood}
\label{sec:estimating_params}

\paragraph{The method of moments}

The pdf of $\textit{Gamma}(\alpha, \beta)$ is defined as
\begin{equation}
    \textit{pdf}(x|\alpha,\beta) = \frac{x^{\alpha-1}}{\Gamma(\alpha) \beta^\alpha} \exp{(-x / \beta)}
    \label{eq:pdf_gamma_pure}
\end{equation}
and has mean $\mu_{gamma}$ and variance $\sigma^2_{gamma}$ given by
\begin{gather}
    \mu_{gamma} = \alpha\beta,\,
    \sigma^2_{gamma} = \alpha\beta^2.
    \label{eq:moments_gamma_pure}
\end{gather}
Another useful identity comes from the sum of Gamma distributions, which is also a Gamma distribution \citep{weisstein_gamma}
such that if $t_i \thicksim Gamma(\alpha_i, \beta)$, then
\begin{equation}
    \sum_{i=1}^K t_i \thicksim Gamma\left(\sum_{i=1}^K\alpha_i, \beta\right).
    \label{eq:sum_gammas}
\end{equation}
From \cref{eq:moments_gamma_pure}, we obtain that the mean and the variance of the distribution $\textit{Gamma}(N, 1)$ are equal \review{with a theoretical value of} $N$.
That is, we can estimate the Gaussian noise standard deviation $\sigma_g$ and the number of coils $N$ from the sample moments of the magnitude images themselves,
provided we can select voxels without any signal contribution \review{\ie} where $\eta=0$.
Firstly, $\sigma_g$ can be estimated from \cref{eq:moments_gamma_pure} as
\begin{equation}
    \sigma_g = \frac{1}{\sqrt{2}} \sqrt{\frac{\sum_{v=1}^V m^4_v}{\sum_{v=1}^V m^2_v} - \frac{1}{V}\sum_{v=1}^V m^2_v},
    \label{eq:find_sigma}
\end{equation}
where $V$ is the number of identified noise only voxels and $m_v$ the value of such a voxel, see \cref{sec:appendix_gamma} for the derivations.
Once $\sigma_g$ is known, $N$ can be estimated from the sample mean of those previously identified voxels as
\begin{equation}
    N = \frac{1}{V}\sum_{v=1}^V t_v = \frac{1}{2V\sigma^2_g}\sum_{v=1}^V m^2_v.
    \label{eq:find_N}
\end{equation}
Derivations of \cref{eq:find_sigma,eq:find_N} are detailed in \cref{sec:appendix_gamma}.

\paragraph{Maximum likelihood equations for the Gamma distribution}

Estimation based on the method of maximum likelihood yields two equations
for estimating $\alpha$ and $\beta$. Rearranging the equations for a Gamma distribution
will give \cref{eq:find_N} and a second implicit equation for $N$ that is given by \citep{Thom1958}
\begin{equation}
    \log(\beta) + \psi(\alpha) = \frac{1}{V}\sum_{v=1}^V \log t_v,
    \label{eq:find_N_ml}
\end{equation}
where $\psi(x) = \frac{d}{dx}\log(\Gamma(x))$ is the digamma function.
For the special case $\textit{Gamma}(N, 1)$, we can rewrite \cref{eq:find_N_ml} as
\begin{equation}
    \psi(N) = \frac{1}{V}\sum_{v=1}^V \log (m^2_v / 2\sigma_g^2).
    \label{eq:gamma_simplified_ml_N}
\end{equation}
Combining \cref{eq:find_N} and \cref{eq:gamma_simplified_ml_N}, we also have an implicit equation to find $\sigma_g$
\begin{equation}
    \psi\left(\frac{1}{2V\sigma^2_g}\sum_{v=1}^V m^2_v\right) = \frac{1}{V}\sum_{v=1}^V \log (m^2_v / 2\sigma_g^2)= \frac{1}{V}\sum_{v=1}^V \log (m^2_v) -  \log(2\sigma_g^2).
    \label{eq:gamma_simplified_ml_sigma}
\end{equation}
As \cref{eq:gamma_simplified_ml_N,eq:gamma_simplified_ml_sigma} have no closed form solution, they can be solved numerically \eg using Newton's method.
See \cref{sec:appendix_gamma} for practical implementation details.


\section{Material and Methods}
\label{sec:method}

\subsection{Automated and robust background separation}

The equations we presented in \cref{sec:special_pdf} are only valid when $\eta = 0$ by construction and assume that each
selected voxel $m_v$ belongs to the same Gamma distribution.
Following a methodology similar to \citep{Koay2009b}, we assume that each 2D slice with the same spatial location belongs to the same statistical distribution
throughout each 3D volume.
This practical assumption allows selecting a large number of noise only voxels for computing statistics
as well as identifying (and subsequently discarding) potential slice acquisition artifacts which may affect one volume, but not the rest of the acquisition.
Using \cref{eq:sum_gammas}, the sum of all diffusion weighted images (DWIs) can be used to separate the voxels belonging to the Gamma distribution $\textit{Gamma}(KN, 1)$,
where $K$ is the number of acquired DWIs, from the voxels not in that specific distribution with a rejection step using
the inverse cumulative distribution function\footnotemark (cdf)\footnotetext{The inverse cdf is also known as the quantile function.}.
In the particular case $\textit{Gamma}(KN, 1)$ at a probability level $p$, the inverse cdf is $\textit{icdf}(\alpha, p) = P^{-1}(\alpha, p)$,
where $P^{-1}$ is the inverse lower incomplete regularized gamma function\footnotemark.
This relationship can be used to identify potential outliers, such as voxels which contain non background signal,
by excluding any voxel $m_v$ whose value does not fall between $\lambda_{-} = \textit{icdf}(\alpha, p/2)$ and $\lambda_{+} = \textit{icdf}(\alpha, 1-p/2)$,
\ie $m_v$ is an outlier if $m_v < \lambda_{-}$ or $m_v > \lambda_{+}$.
\footnotetext{As there is no analytical solution to the inverse cdf of a Gamma distribution, one can use the function $\text{gaminv}(p, \alpha, \beta=1)$ in Matlab
or $\text{InverseGammaRegularized}(\alpha, 1-p)$ in Mathematica to numerically estimate it.}

\begin{figure}
    \raggedright{\textbf{A)}}~\includegraphics[width=\linewidth,valign=t]{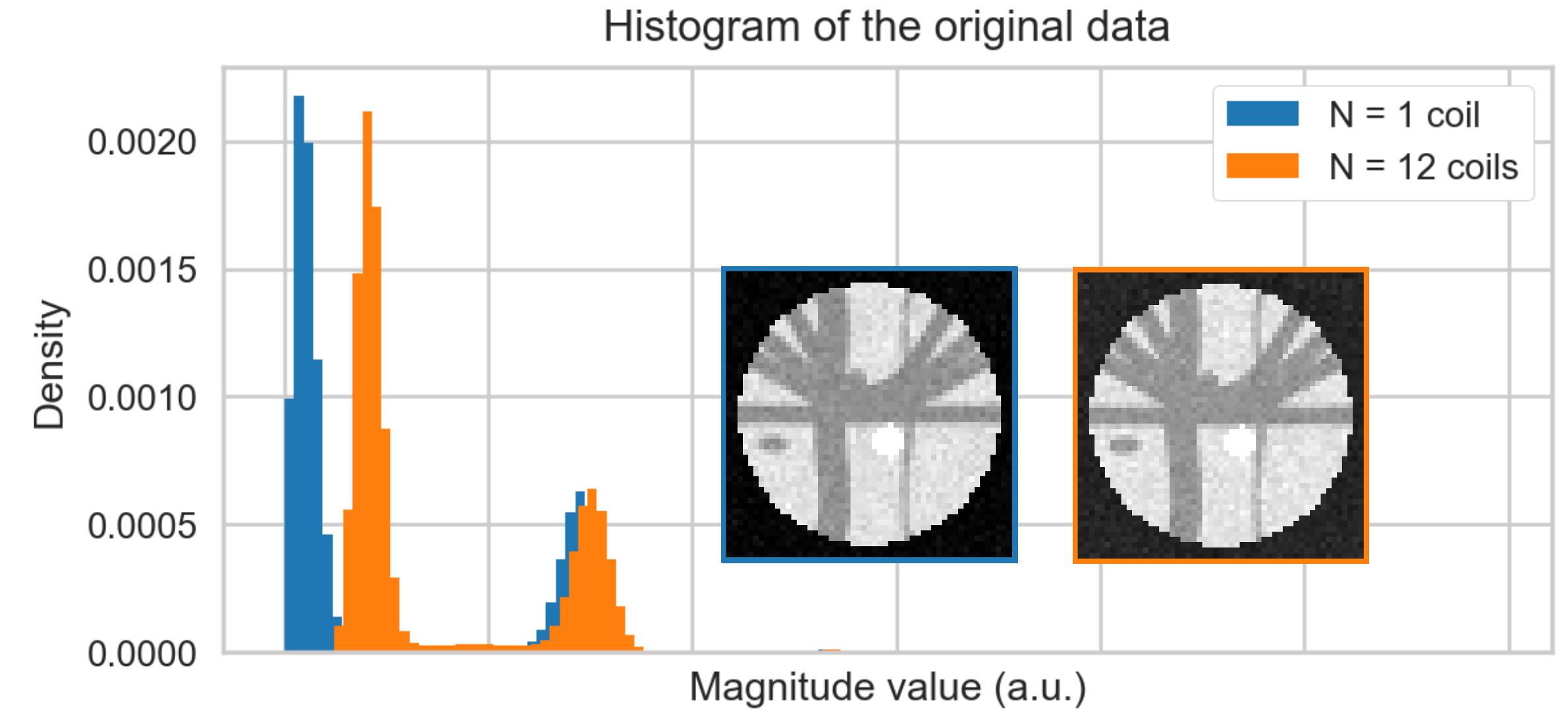}
    \raggedright{\textbf{B)}}~\includegraphics[width=\linewidth,valign=t]{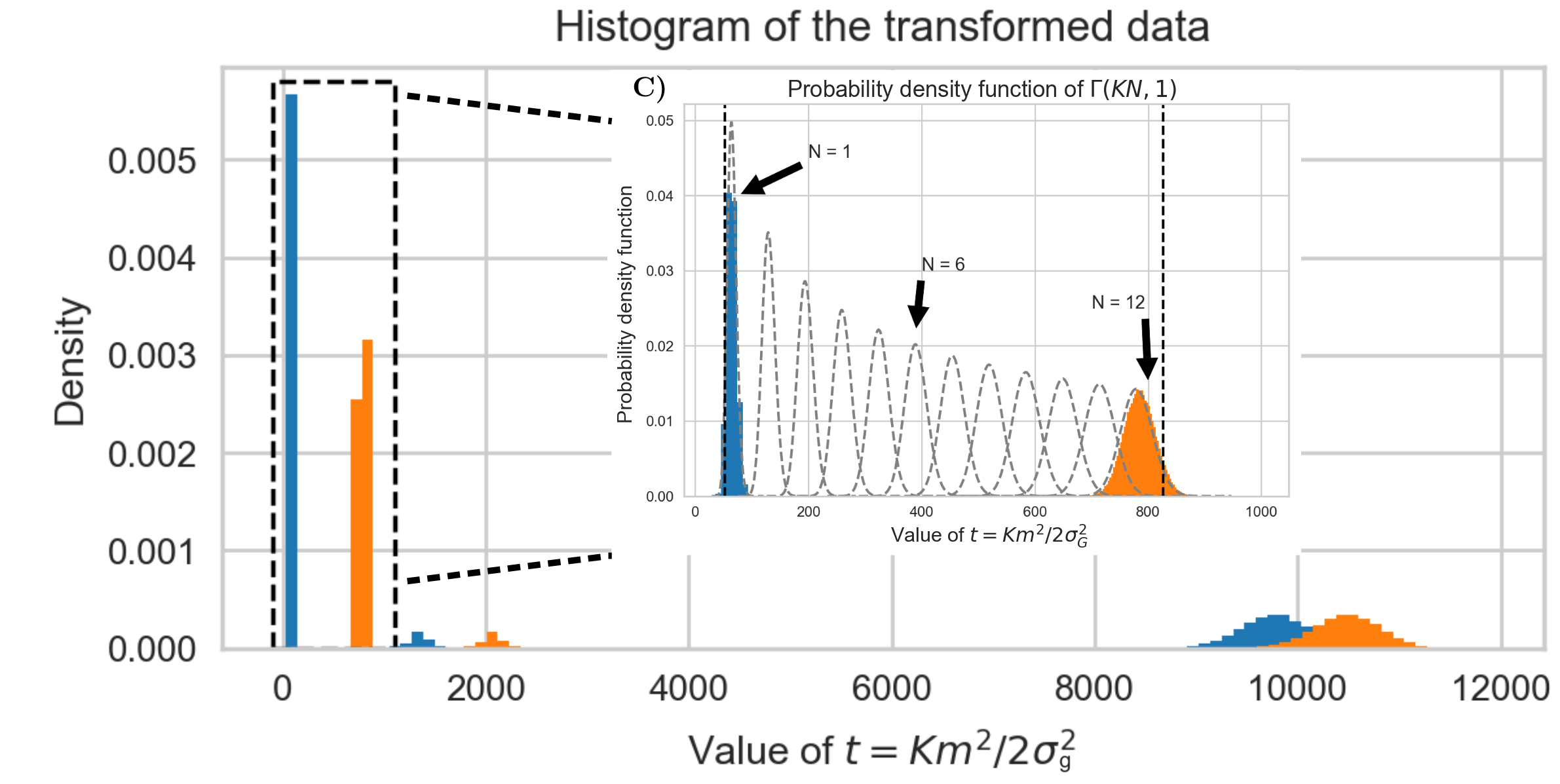}
    \caption{Histogram of the synthetic data at \bval{3000} \textbf{A)} before the change of variable to a Gamma distribution
    and \textbf{B)} after the change of variable to a Gamma distribution for $N = 1$ and $N = 12$ \review{with the true value of $\sigma_g$}.
    Summing all $K$ DWIs together separates the background voxels from the rest of the data,
    which follows a Gamma distribution $\textit{Gamma}(KN, 1)$ by construction.
    In \textbf{C)}, a view of the left part from \textbf{B)}
    with the theoretical histograms of Gamma distributions from $N=1$ up to $N=12$.
    The black dotted lines represent the lower bound $\lambda_{-}$ to the upper bound $\lambda_{+}$,
    with $p = 0.05, N_{min} = 1 \text{ and } N_{max} = 12$.
    This broad search covers the background voxels in both cases while excluding remaining voxels which do not belong to the distribution $\textit{Gamma}(KN, 1)$.}
    \label{fig:gamma_histogram}
\end{figure}

To provide a better understanding of the change of variable $t = m^2 / 2\sigma^2_g$,
\cref{fig:gamma_histogram} shows the histogram for a synthetic dataset at \bval{3000}, which will be detailed later in \cref{sec:experiments}.
Voxels belonging to the background are easily separated in terms of the Gamma distribution after transformation,
thus allowing estimation of parameters from voxels truly belonging to the noise distribution, see \cref{sec:appendix_algo} and \citep{St-jean2018a} for technical details.
Our implementation of the proposed algorithm is freely available\footnote{\url{https://github.com/samuelstjean/autodmri}} \citep{St-Jean2019d}.

\subsection{Datasets and experiments}
\label{sec:experiments}

\paragraph{Synthetic phantom datasets}

Two synthetic phantom configurations from previous dMRI challenges were used.
The first simulations were based on the ISBI 2013 HARDI challenge using phantomas \citep{Caruyer2014}.
We used the given 64 gradient directions to generate two separate noiseless single-shell phantoms
with either \bval{1000} or \bval{3000} and an additional \bval{0} volume.
The datasets were then corrupted with Rician $(N = 1)$ and noncentral chi noise profiles $(N = 4, 8 \text{ and } 12)$,
both stationary and spatially varying, at a signal-to-noise ratio (SNR) of 30 according to
\begin{equation}
    \label{eq:noising}
    \hat{I} = \sqrt{\sum_{i=0, j=0}^{N} \left(\frac{I}{\sqrt{N}} + \tau\epsilon_i \right)^2 + (\tau\epsilon_j)^2},
    \enspace \mbox{where } \epsilon_i, \epsilon_j \sim \mathbb{N}(0,  \sigma_g^2),
\end{equation}
where $I$ is the noiseless volume, $\hat{I}$ is the resulting noisy volume, $\tau$ is a mask for the spatial noise pattern, $\mathbb{N}(0,  \sigma_g^2)$
is a Gaussian distribution of mean 0 and variance $\sigma_g^2 = (\bar{m} / \textit{SNR})^2$ and
\review{$\bar{m}$ is the average signal value} of the \bval{0} image inside the white matter.
In the stationary noise case, $\tau$ is set to 1 so that the noise is uniform.
For the spatially varying noise case, $\tau$ is a sphere with a value of 1 in the center up to a value of 1.75
at the edges of the phantom, thus generating a stronger noise profile outside the phantom than for the stationary noise case.
Since all datasets are generated at SNR 30, the noise standard deviation $\sigma_g$ is the same even though the b-value or number of coils $N$ is different,
but the magnitude standard deviation \review{of the noise only voxels} $\sigma_m$ is lower than $\sigma_g$.

The second set of synthetic experiments is based on the ISMRM 2015 tractography challenge \citep{Maier-Hein2017} which consists of 25 manually delineated white matter bundles.
Ground truth data consisting of 30 gradient directions at either \bval{1000} or \bval{3000} and 3 \bval{0} images at a resolution of 2 mm isotropic
was generated using Fiberfox \citep{Neher2013e} without artifacts or subject motion.
Subsequent noisy datasets were created at SNR 20 by simulating an acquisition with 8, 12 and 32 coils
using the parallel MRI simulation toolbox\footnote{\url{https://mathworks.com/matlabcentral/fileexchange/36893-parallel-mri-noisy-phantom-simulator}}
with SENSE \citep{Pruessmann1999} or GRAPPA \citep{Griswold2002} reconstructions with an acceleration factor of $R = 2$.
The SENSE simulated datasets also included spatial correlations between coils of $\rho = 0.1$,
increasing the spatially varying effective noise standard deviation $\sigma_g$ and keeping the signal Rician distributed $(N = 1)$.
For the GRAPPA reconstructed datasets, 32 calibrating lines were sampled in the k-space center, neglecting spatial correlations ($\rho = 0$) as it is a k-space method \citep{Aja-Fernandez2015a}.
The resulting effective values of $N$ and $\sigma_g$ will be both spatially varying.
We additionally generated 33 synthetic noise maps per dataset by setting the underlying signal value to $\eta=0$ and performing
the reconstruction using the same parameters as the DWIs.
All generated datasets are available online \citep{St-Jean2018d}.

\paragraph{Acquired phantom datasets}

We acquired phantom images of a bottle of liquid
on a 3T Philips Ingenia scanner using a 32 channels head coil with a gradient strength of 45 mT/m.
We varied the SENSE factor from $R = 1, 2 \text{ or } 3$ and multiband acceleration factors from no multiband (\textit{MB}), \textit{MB} = 2 or \textit{MB} = 3
while fixing remaining acquisition parameters to investigate their influence on the resulting signal distributions, resulting in 9 different acquisitions.
The datasets consist of 5 \bval{0} volumes and 4 shells with 10 DWIs each at \bval{500}, \bval{1000}, \bval{2000} and \bval{3000} with a voxel size of 2 mm isotropic
and TE / TR = 135 ms / 5000 ms, $\Delta / \delta$ = 66.5 ms / 28.9 ms.
Six noise maps were also acquired during each of the experiments by disabling the RF pulse and gradients of the sequence.
The acquired phantom datasets are also available \citep{St-Jean2018d}.

\paragraph{In vivo datasets}

A dataset consisting of four repetitions of a single subject\footnotemark\, was also used to assess the reproducibility
of noise estimation without \textit{a priori} knowledge \citep{Poldrack2015a}.
This is the dataset we previously used in our MICCAI manuscript \citep{St-jean2018a}.
The acquisition was performed on a GE MR750 3T scanner at Stanford university,
where a 3x slice acceleration with blipped-CAIPI shift of FOV/3 was used, partial Fourier 5/8 with a homodyne reconstruction and a minimum TE of 81 ms.
Two acquisitions were made in the anterior-posterior phase encode direction and the two others
in the posterior-anterior direction.
The voxelsize was 1.7 mm isotropic with 7 \bval{0} images, 38 volumes at \bval{1500} and 38 volumes at \bval{3000}.
As the acquisition used a homodyne filter to fill the missing k-space,
this should lead in practice to a half Gaussian noise profile, a special case of the noncentral chi distribution with $N = 0.5$,
due to using only the real part of the signal for the final reconstruction \citep[Chap.~13][]{Bernstein2004,Noll1991,Dietrich2008}.
\footnotetext{{\url{https://openfmri.org/dataset/ds000031}}}

In addition, one dataset acquired on a 3T Siemens Connectom scanner from the 2017 MICCAI harmonization challenge\footnotemark\,
consisting of 16 \bval{0} volumes and 3 shells with 60 DWIs each at \bval{1200}, \bval{3000} and \bval{5000} was used \citep{Tax2019}.
The voxel size was 1.2 mm isotropic with a pulsed-gradient spin-echo echo-planar imaging (PGSE-EPI) sequence and a gradient strength of 300 mT/m.
Multiband acceleration \textit{MB} = 2 was used with GRAPPA parallel imaging with $R = 2$ and an adaptive combine reconstruction employing a 32 channels head coil.
Other imaging parameters were TE / TR = 68 ms / 5400 ms, $\Delta / \delta$ = 31.1 ms / 8.5 ms, bandwidth of 1544 Hz/pixel and partial Fourier 6/8.
\footnotetext{\url{https://www.cardiff.ac.uk/cardiff-university-brain-research-imaging-centre/research/projects/cross-scanner-and-cross-protocol-diffusion-MRI-data-harmonisation}}

\paragraph{Noise estimation algorithms for comparison}

To assess the performance of the proposed methods, we used three other noise estimation algorithms
previously used in the context of dMRI.
Default parameters were used for all of the algorithms as done in \citet{St-jean2018a}.
The local adaptive noise estimation (LANE) algorithm \citep{Tabelow2014} is designed for noncentral chi
signal estimation, but requires \textit{a priori} knowledge of $N$.
\review{Default parameters were used with $k^* = 20$ as recommended.}
Since the method works on a single 3D volume, we only use the \bval{0} image for all of the experiments
to limit computations as the authors concluded that the estimates from a single DWI are close to the mean estimate.
We also use the Marchenko-Pastur (MP) distribution fitting on the principal component analysis (PCA) decomposition of the diffusion data,
which is termed MPPCA \citep{Veraart2015a}.
\review{In all experiments, we used the suggested default local window size of $5 \times 5 \times 5$.}
Finally, we also compare to the Probabilistic Identification and Estimation of Noise (PIESNO) \citep{Koay2009b},
which originally proposed the change of variable to the Gamma distribution that is at the core of our proposed method.
PIESNO requires knowledge of $N$ (which is kept fixed by the algorithm) to iteratively estimate $\sigma_g$ until convergence
by removing voxels which do not belong to the distribution $\textit{Gamma}(N, 1)$ for a given slice.
We set $p = 0.05$ and $l = 50$ for the initial search of $\sigma_g$ in PIESNO and our proposed method, with
additional parameters set to $N_{min} = 1$ and $N_{max} = 12$ for all cases.
When estimating distributions from noise maps, we compute values in small local windows of size $3 \times 3 \times 3$.
\review{The list of the software implementations and their version used in this manuscript is available in the supplementary materials.}
To the best of our knowledge, ours is the first method which estimates both
$\sigma_g$ and $N$ jointly without requiring any prior information about the reconstruction process of the MRI scanner.
Because PIESNO and LANE both \textit{require} knowledge of the value of $N$, we set the correct value of $N$ for the spatially varying noise phantom experiments
and \review{assume a Rician distribution by setting} $N=1$ for the remaining experiments when $N$ is unknown.
We quantitatively assess the performance of each method on the synthetic datasets by measuring the
standard deviation of the noise and the \review{percentage} error inside the phantom against the known value of $\sigma_g$,
computed \review{for each voxel} as
\begin{equation}
    \text{\review{percentage} error} = 100 \times \left(\sigma_{g_{estimated}} - \sigma_{g_{true}}\right) / \sigma_{g_{true}}.
\end{equation}
As PIESNO and our proposed methods estimate a single value per slice whereas MPPCA and LANE provide estimates from small spatial neighborhood,
we report the mean value and the standard deviation \review{of the error} estimated inside the synthetic phantoms on each slice.
For the acquired phantom datasets, we report the estimated noise distributions using both the DWIs and the measured noise maps
for all 9 combinations of parallel imaging parameters that were acquired.
For the in vivo datasets, we report once again the noise distributions estimated by each method.
The reproducibility of the estimated distributions is assessed on the four GE datasets while the Connectom dataset is used
to evaluate the performance of each compared algorithm on a bias correction and denoising task.
In addition, we report $N$ as estimated by our proposed methods for all cases.

\paragraph{Bias correction and denoising of the Connectom dataset}

In a practical setting, small misestimation in the noise distribution (\eg spatially varying distribution vs nature of the distribution) might not impact much
the application of choice.
We evaluate this effect of misestimation on the Connectom dataset with a bias correction and a denoising task.
Specifically, we apply noncentral chi bias correction \citep{Koay2009a} on the in vivo dataset from the CDMRI challenge using \cref{eq:beta}.
The algorithm is initialized with a spherical harmonics decomposition of order 6 \citep{Descoteaux2007b} as done in \citep{St-Jean2016a}.
The data is then denoised using the non local spatial and angular matching (NLSAM) algorithm with 5 angular neighbors where each b-value is treated separately \citep{St-Jean2016a}.
Default parameters of a spatial patch size of $3\times 3\times 3$ were used and the estimation of $\sigma_g$ as computed by each method was given to the NLSAM algorithm.
For MPPCA, LANE and PIESNO, a default value of $N=1$ was used and the value of $N$ as computed by the moments and maximum likelihood equations for the proposed methods.
The bias correction algorithm was also generalized for non integer values of $N$ as detailed in \cref{sec:appendix_general_bias}.


\section{Results}
\label{sec:results}

We show here results obtained on the phantoms and in vivo datasets.
The first set of simulations uses a sum of square reconstruction with stationary and spatially varying noise profiles.
The second set of simulations includes SENSE and GRAPPA reconstructions, resulting in both spatially varying signal distribution profiles.
Finally, the distributions estimated by each algorithm for the in vivo dataset are used for a bias correction and denoising task.

\subsection{Synthetic phantom datasets}
\label{sec:synthetic_simulation}

\paragraph{Simulations with a sum of squares reconstruction}

\begin{figure}
    \raggedright{\textbf{A)}}~\includegraphics[width=\linewidth,valign=t]{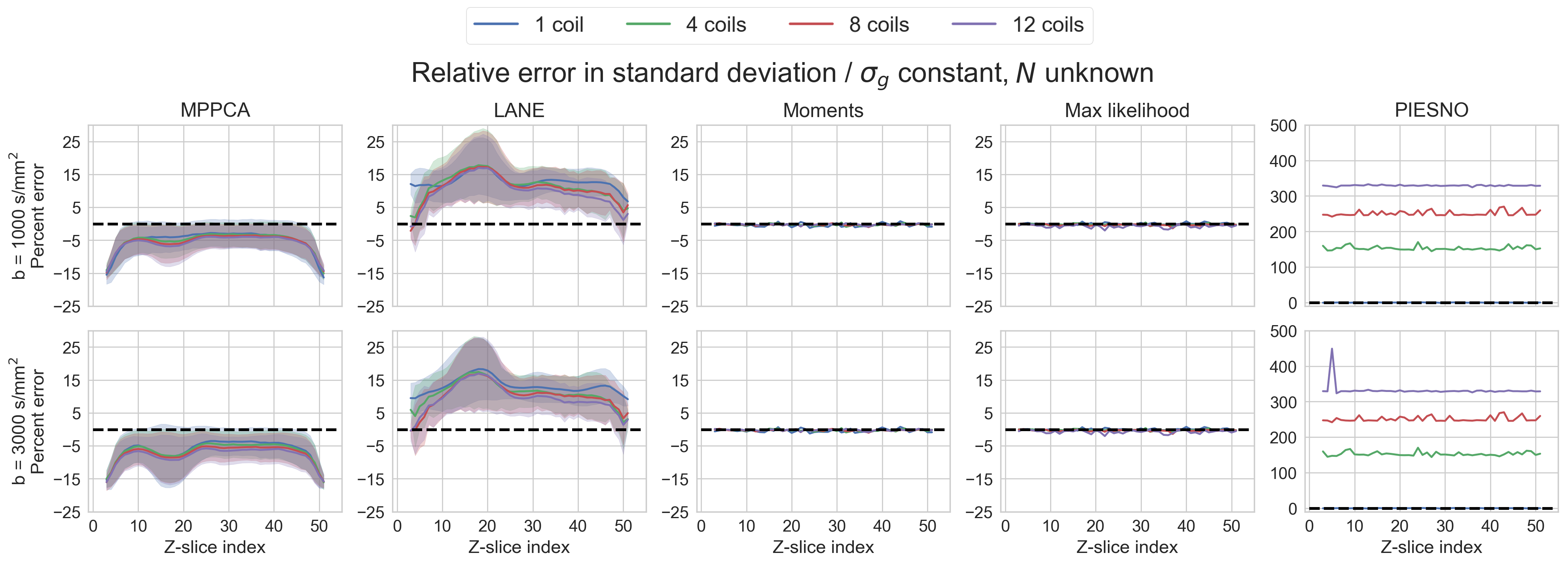}
    \raggedright{\textbf{B)}}~\includegraphics[width=\linewidth,valign=t]{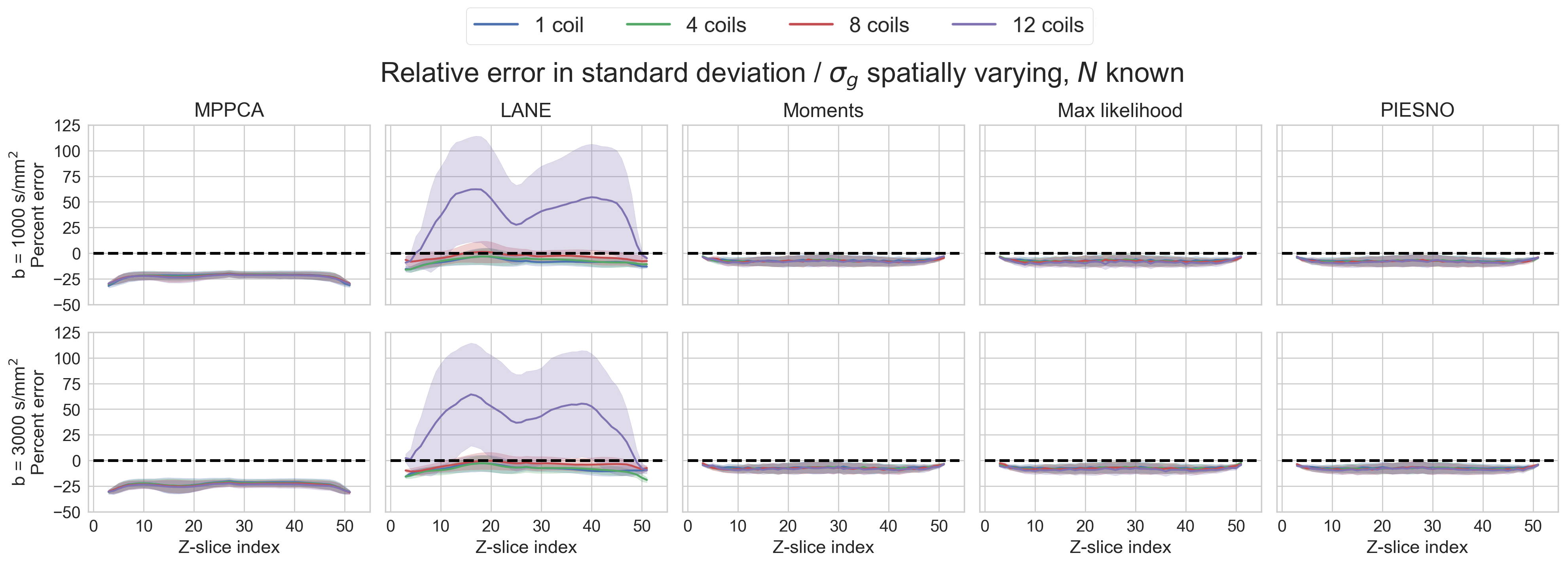}
    \caption{Percentage of error when the real value of $N$ is \review{unknown} and $\sigma_g$ is constant (in \textbf{A)})
    and $N$ is known with $\sigma_g$ spatially varying (in \textbf{B)}) with the mean (solid line) and standard deviation (shaded area).
    All methods underestimate spatially varying $\sigma_g$, except for LANE with $N=12$ which overestimates it instead.
    On average, all methods are tied at around 5\% of error with MPPCA reaching approximately 25\% of error.
    Of interesting note, the proposed methods are tied with PIESNO when the correct value of $N$ is given to the latter,
    but do not require an estimate of $N$, which is now an output instead of a prerequisite.}
    \label{fig:phantomas_sigma}
\end{figure}

\begin{figure}
    \includegraphics[width=\linewidth]{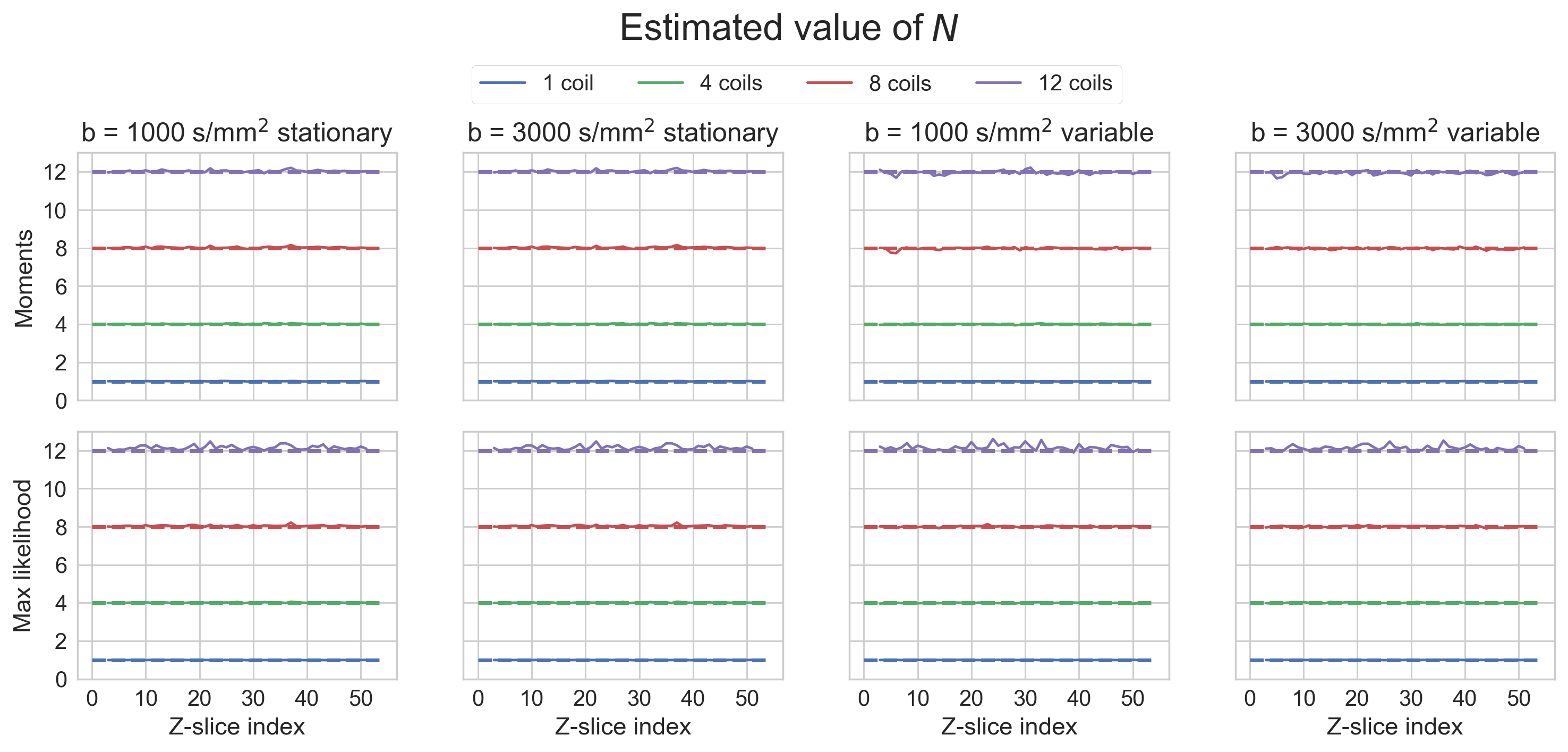}
    \caption{Estimated value of $N$ using equations from the moments (top) and with maximum likelihood (bottom) for the proposed methods.
    Even for the spatially variable case where $\sigma_g$ is slightly underestimated,
    the estimated values of $N$ are stable and correspond to the real values used in the synthetic simulations in every case.}
    \label{fig:phantomas_N}
\end{figure}

\cref{fig:phantomas_sigma} shows results from simulations with stationary and spatially varying noise profiles for all datasets
as estimated inside the \review{ISBI 2013 challenge} phantom.
For stationary noise profiles with $N$ unknown, estimation of $\sigma_g$ is the most accurate for the proposed methods with an error of about 1\%,
followed by MPPCA making an error of approximately 5\% and LANE of 15\%.
The error of PIESNO increases with the value of $N$, presumably due to misspecification in the signal distribution,
whereas MPPCA and LANE are both stable in their estimation with increasing values of $N$.
The proposed methods using equations based on the moments and maximum likelihood recovers the correct value of $\sigma_g$ in all cases with the lowest variance across slices,
indicating that the estimated value of $\sigma_g$ is similar in all slices as expected.
The same behavior is observed for PIESNO when $N=1$, but the estimated $\sigma_g$ is larger than the correct value by two to three times when $N$ is misspecified.
In the spatially varying noise case where $N$ is known, the moments, maximum likelihood equations and PIESNO all perform similarly with approximately 2\% of error.
LANE generally outperforms MPPCA except for the $N=12$ case, but still misestimates $\sigma_g$ by approximately 15\% and 25\% respectively.
Only the proposed methods and MPPCA are independent of correctly specifying $N$.
Finally, \cref{fig:phantomas_N} shows the estimated values of $N$ by the proposed methods.
Estimation generally follows the correct value, regardless of misestimation of $\sigma_g$.

\paragraph{Simulations with parallel imaging}

\begin{figure}
    \begin{subfigure}{0.49\linewidth}
        \textbf{A)}
        \includegraphics[width=\linewidth,valign=t]{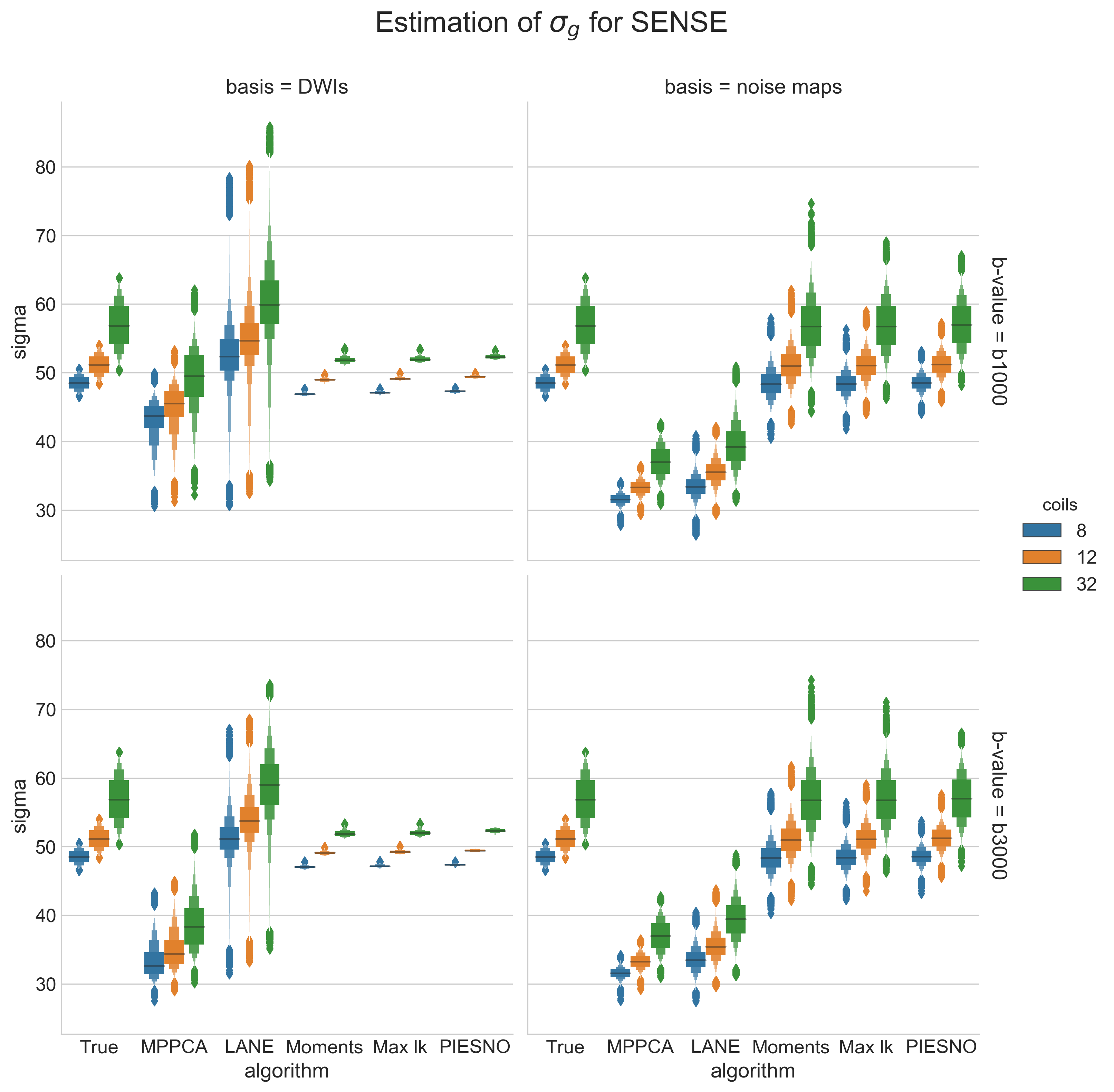}
    \end{subfigure}
    \begin{subfigure}{0.49\linewidth}
        \textbf{B)}
        \includegraphics[width=\linewidth,valign=t]{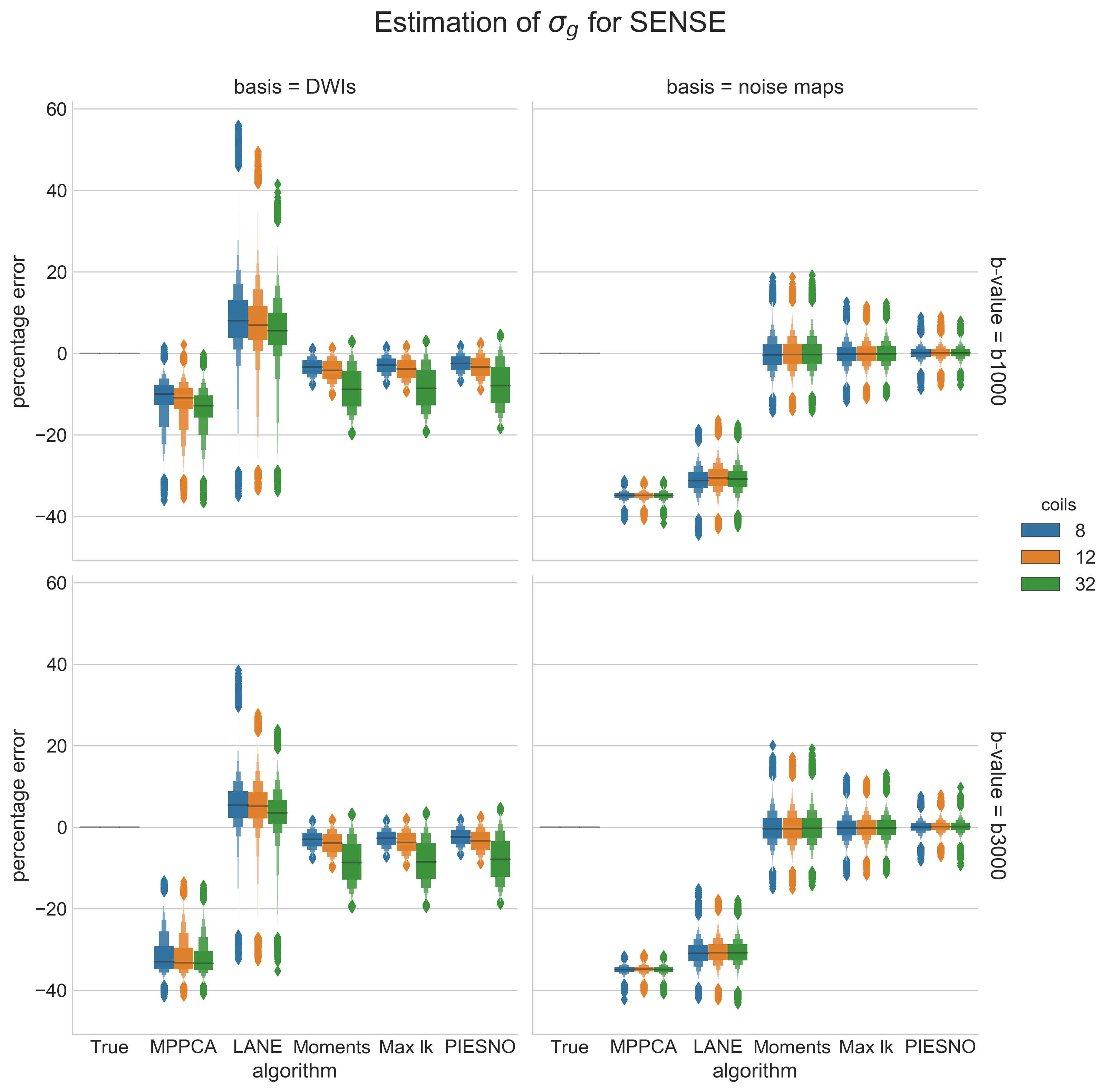}
    \end{subfigure}
    \centering
    \begin{subfigure}{0.75\linewidth}
        \textbf{C)}
        \includegraphics[width=\linewidth,valign=t]{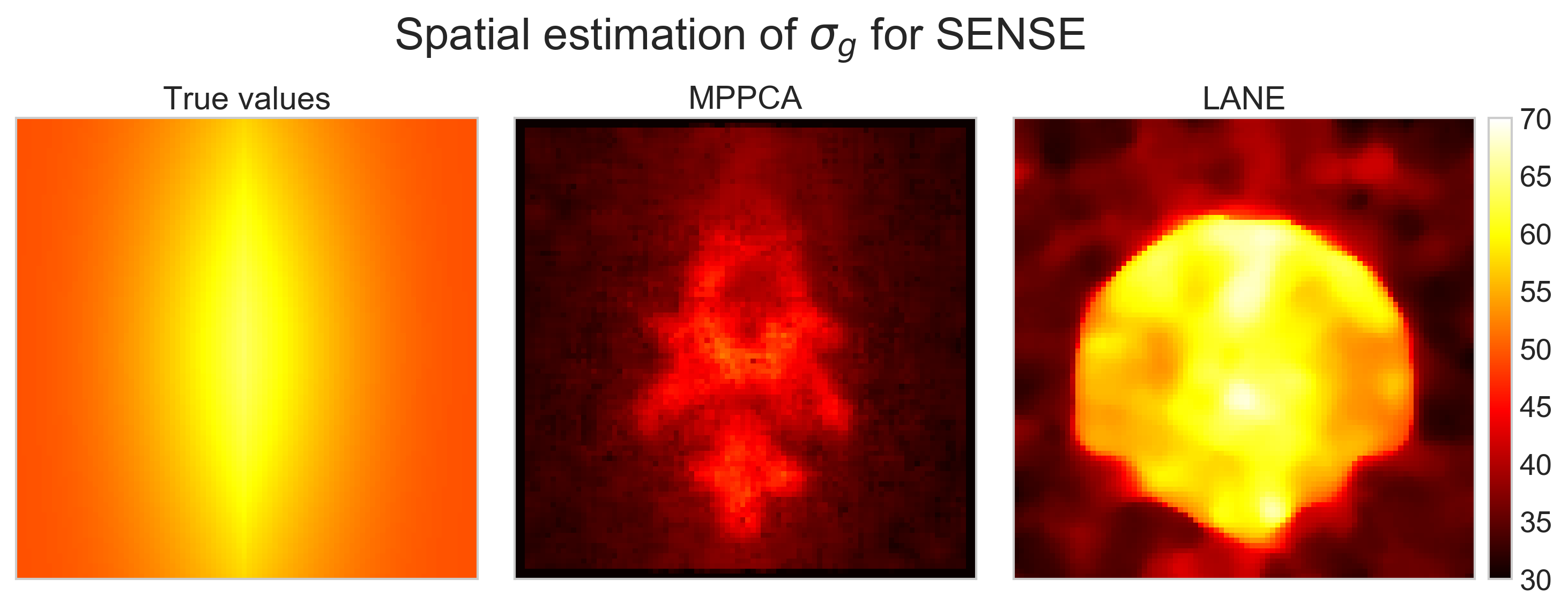}
        \textbf{D)}
        \includegraphics[width=\linewidth,valign=t]{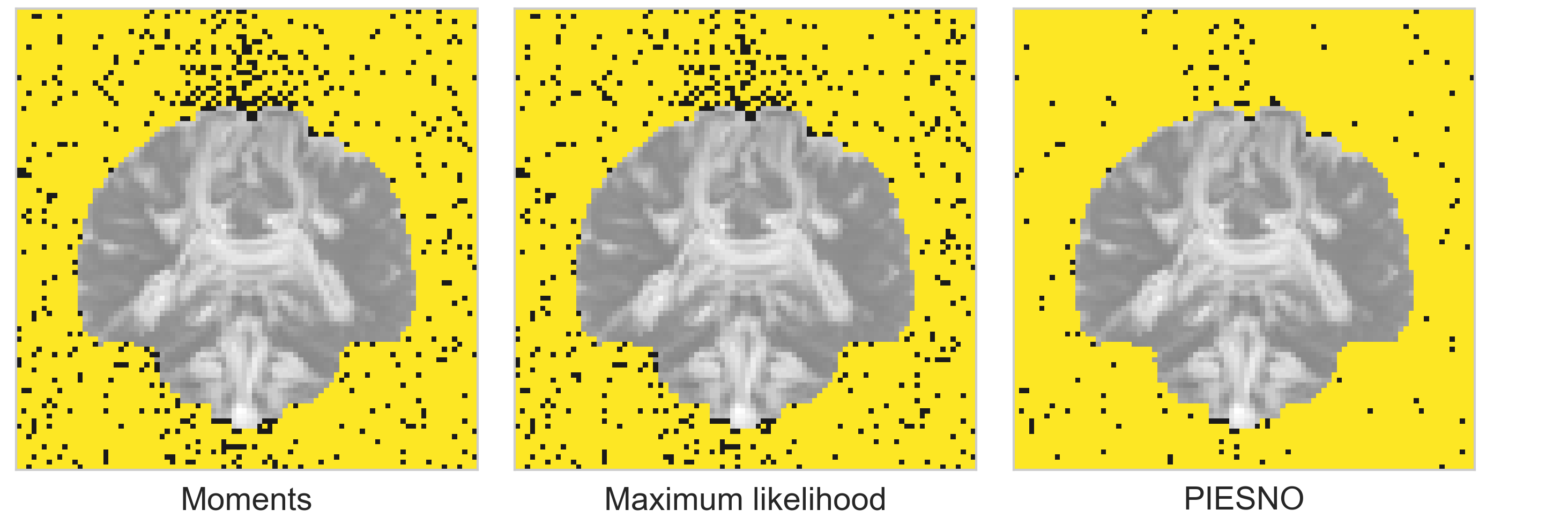}
    \end{subfigure}
    \caption{Estimation of the noise standard deviation $\sigma_g$ (in \textbf{A}) and the percentage error (in \textbf{B})
    inside the phantom only for each method using a SENSE reconstruction with 8, 12 or 32 coils.
    The left columns (basis = DWIs) shows estimation using all of the DWIs, while the right column (basis = noise maps) shows the estimated values from synthetic noise maps
    \review{in small windows of size $3\times 3\times 3$}.
    Results for \bval{1000} are on the top row, while the bottom row shows results for the \bval{3000} datasets.
    Figure \textbf{C)} shows the spatially estimated values of $\sigma_g$ using the \bval{3000} dataset with 32 coils for a single slice
    from the true distribution and local estimation as done by MPPCA and LANE.
    The general trend shows that even though MPPCA and LANE misestimate $\sigma_g$, they still follow the spatially varying pattern
    (lower at edges with the highest intensity near the middle) from the correct values.
    In \review{\textbf{D)}}, voxels identified as belonging to the same distribution Gamma$(N, 1)$ are overlaid in yellow over the sum of all DWIs.
    Note how voxels containing signal from the DWIs are excluded by all three methods.
    }
    \label{fig:fiberfox_sense}
\end{figure}

\begin{figure}
    \begin{subfigure}{0.49\linewidth}
        \textbf{A)}
        \includegraphics[width=\linewidth,valign=t]{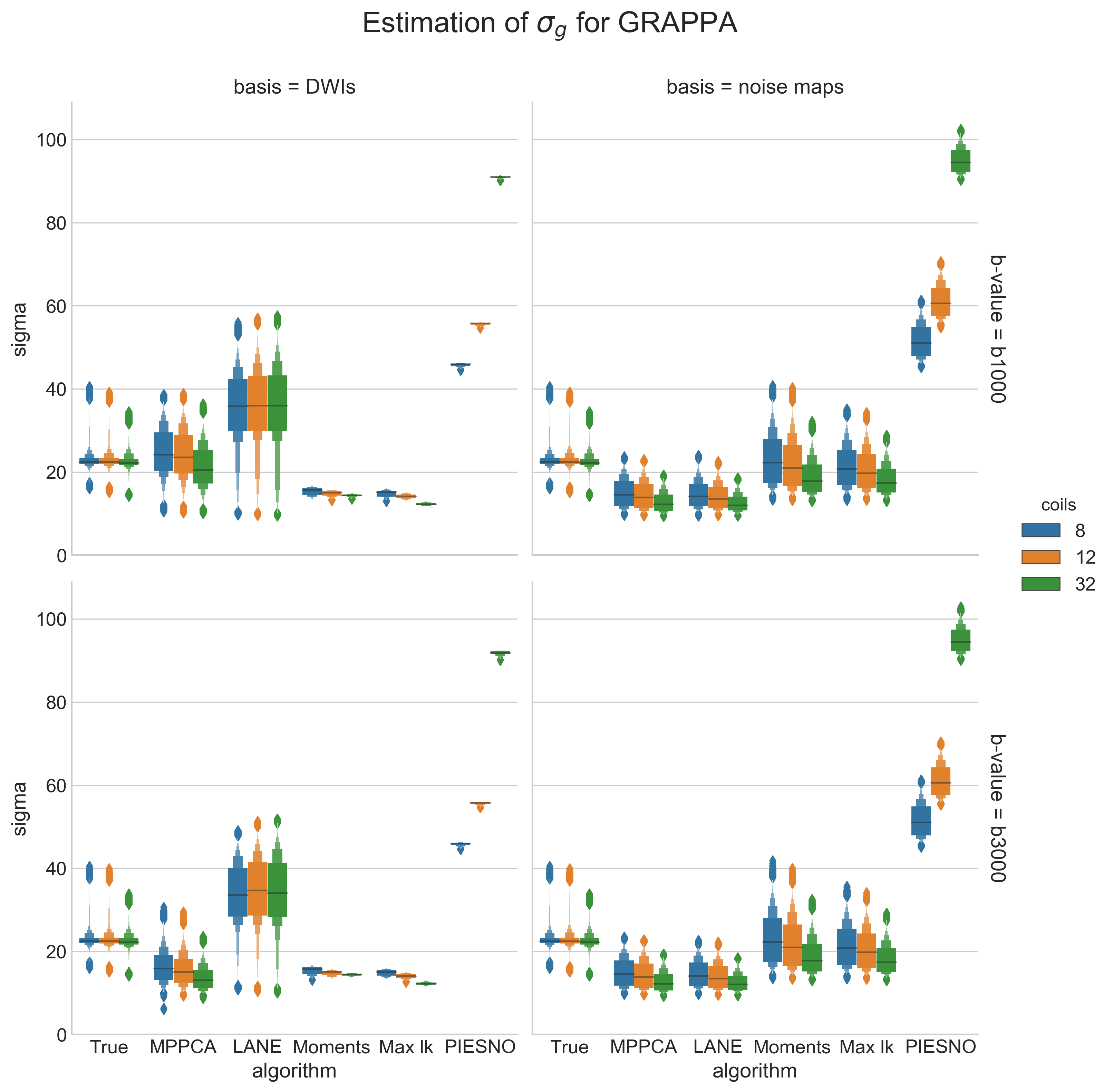}
    \end{subfigure}
    \begin{subfigure}{0.49\linewidth}
        \textbf{B)}
        \includegraphics[width=\linewidth,valign=t]{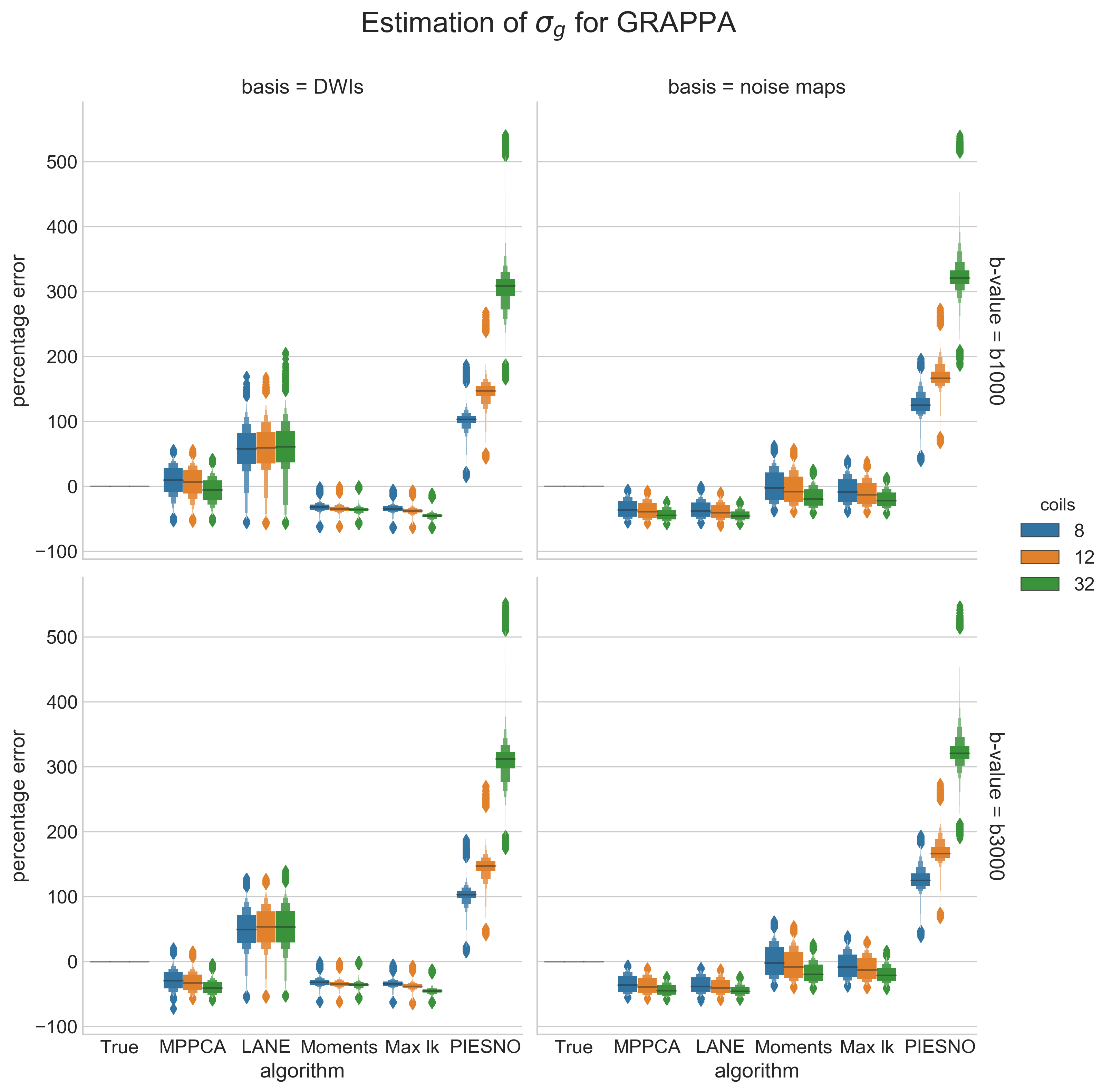}
    \end{subfigure}
    \centering
    \begin{subfigure}{0.75\linewidth}
        \textbf{C)}
        \includegraphics[width=\linewidth,valign=t]{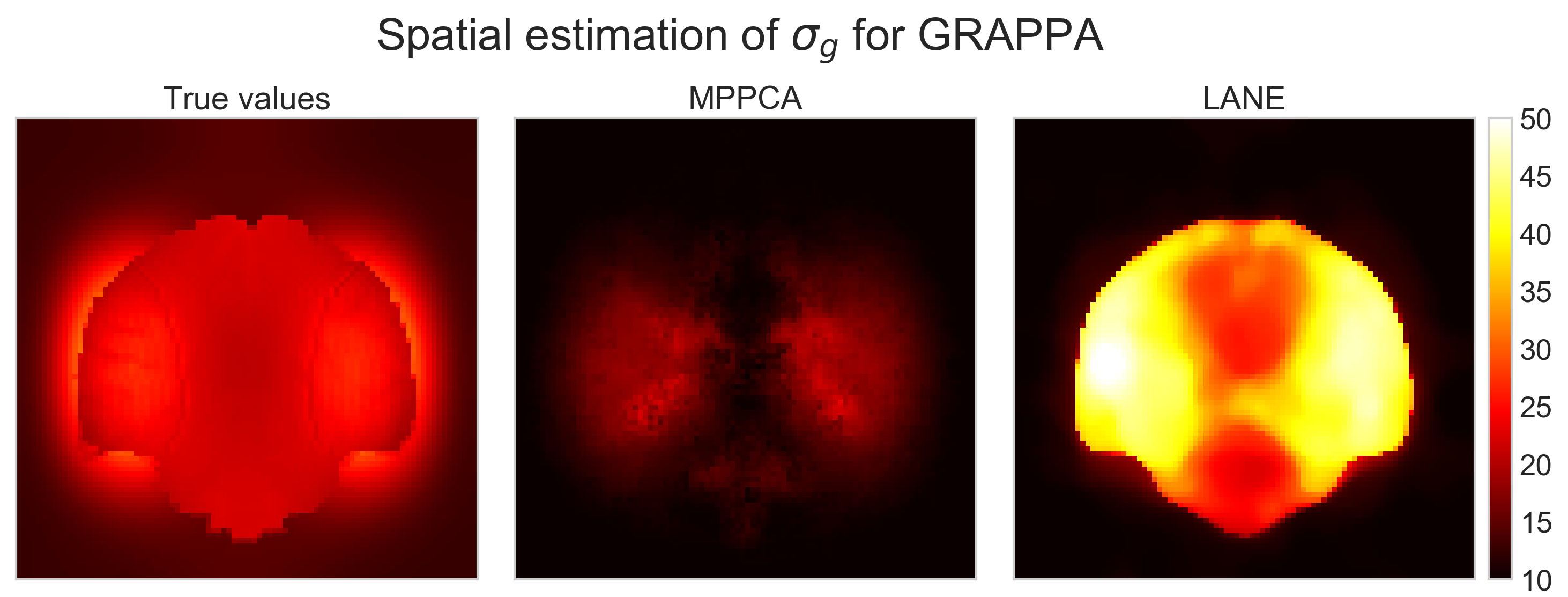}
        \textbf{D)}
        \includegraphics[width=\linewidth,valign=t]{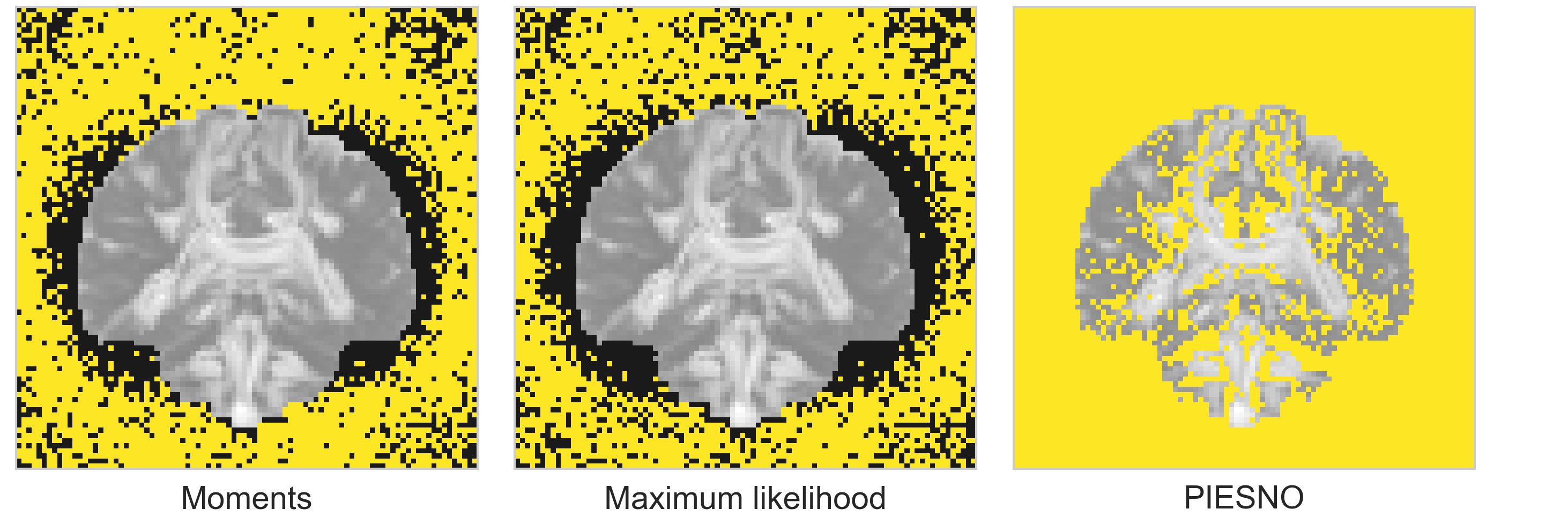}
    \end{subfigure}
    \caption{Estimation of the noise standard deviation $\sigma_g$ (in \textbf{A}) and the percentage error (in \textbf{B})
    inside the phantom only for each method using a GRAPPA reconstruction with 8, 12 or 32 coils, using the same conventions as \cref{fig:fiberfox_sense}.
    Figure \textbf{C)} shows the \review{the true value of $\sigma_g$ and the spatially estimated $\sigma_g$ from MPPCA and LANE}
    using the \bval{3000} dataset with 32 coils for a single slice.
    There is once again a misestimation for both methods while following the correct spatially varying pattern.
    \review{In \textbf{D)},} voxels identified as belonging to the same distribution Gamma$(N, 1)$ are overlaid in yellow over the sum of all DWIs.
    Note how PIESNO mistakenly selects some low intensity voxels belonging to the gray matter, in addition to all of the voxels in the background,
    which causes an overestimation of $\sigma_g$ with a fixed value of $N=1$.
    Both proposed methods instead select voxels with small variations in intensity as belonging to the same distribution
    without mistakenly selecting gray matter voxels.
    }
    \label{fig:fiberfox_grappa}
\end{figure}

\begin{figure}
    \begin{subfigure}{0.49\linewidth}
        \textbf{A)}
        \includegraphics[width=\linewidth,valign=t]{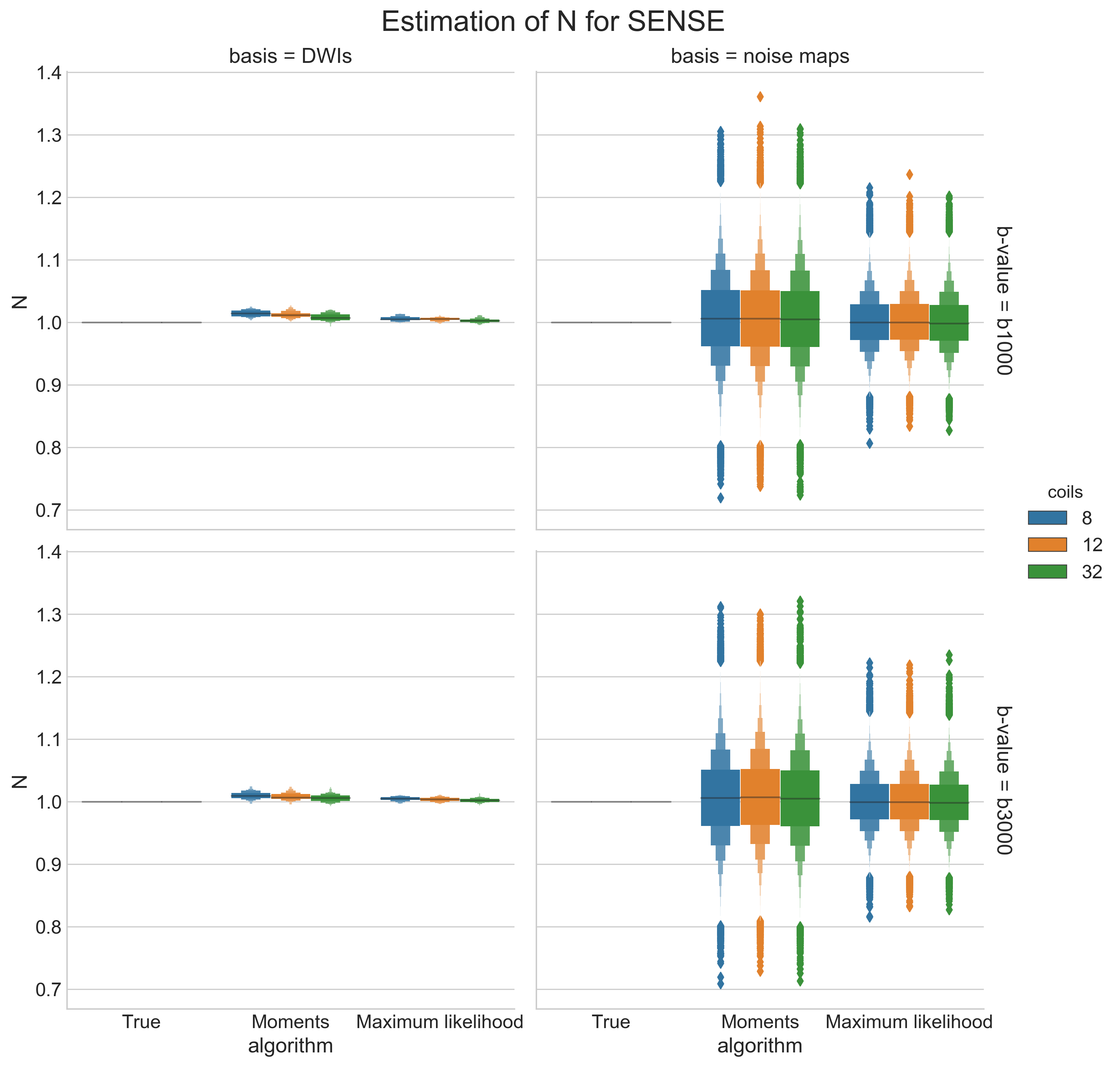}
    \end{subfigure}
    \begin{subfigure}{0.49\linewidth}
        \textbf{B)}
        \includegraphics[width=\linewidth,valign=t]{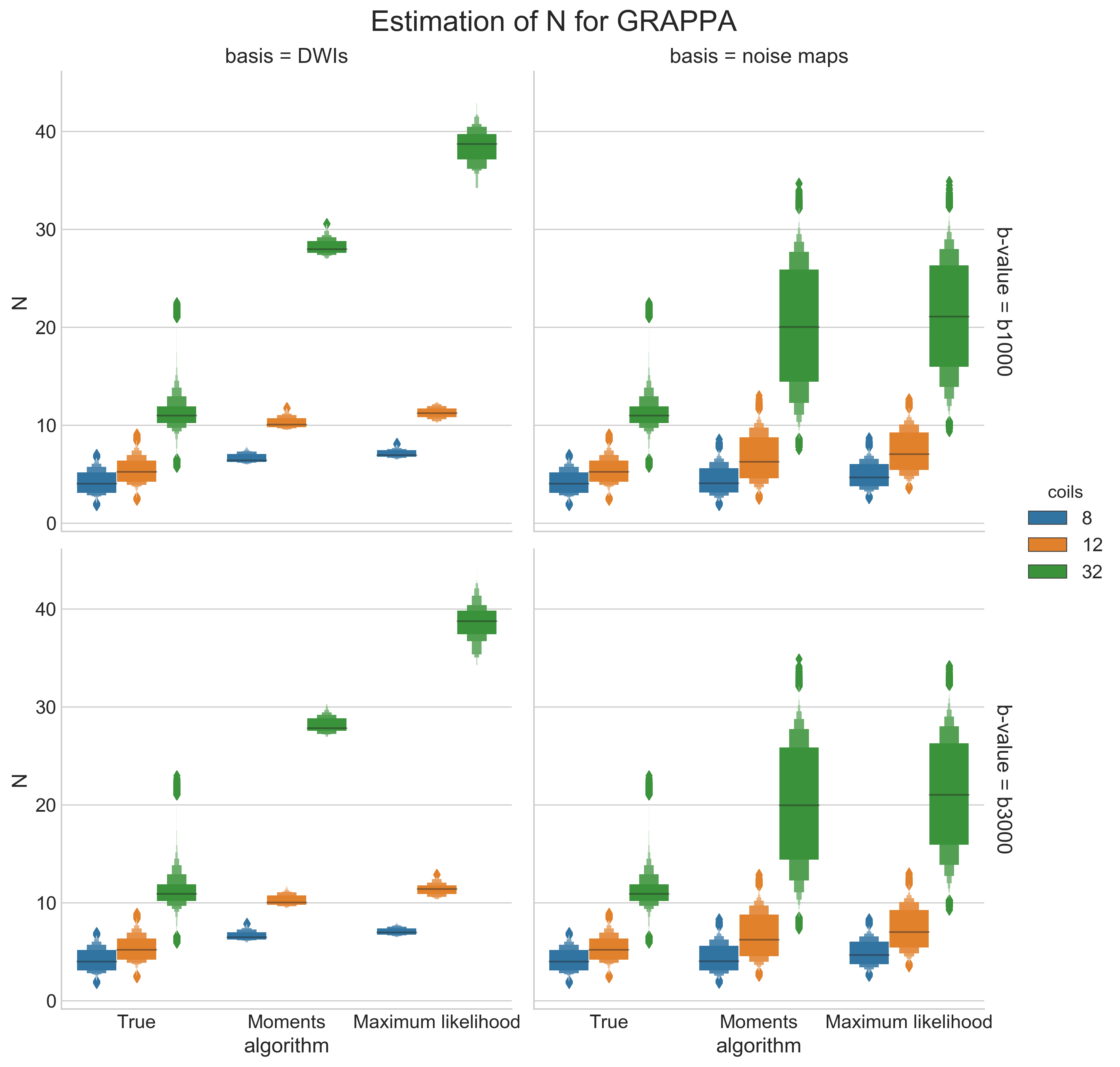}
    \end{subfigure}
    \begin{subfigure}{0.49\linewidth}
        \textbf{C)}
        \includegraphics[width=\linewidth,valign=t]{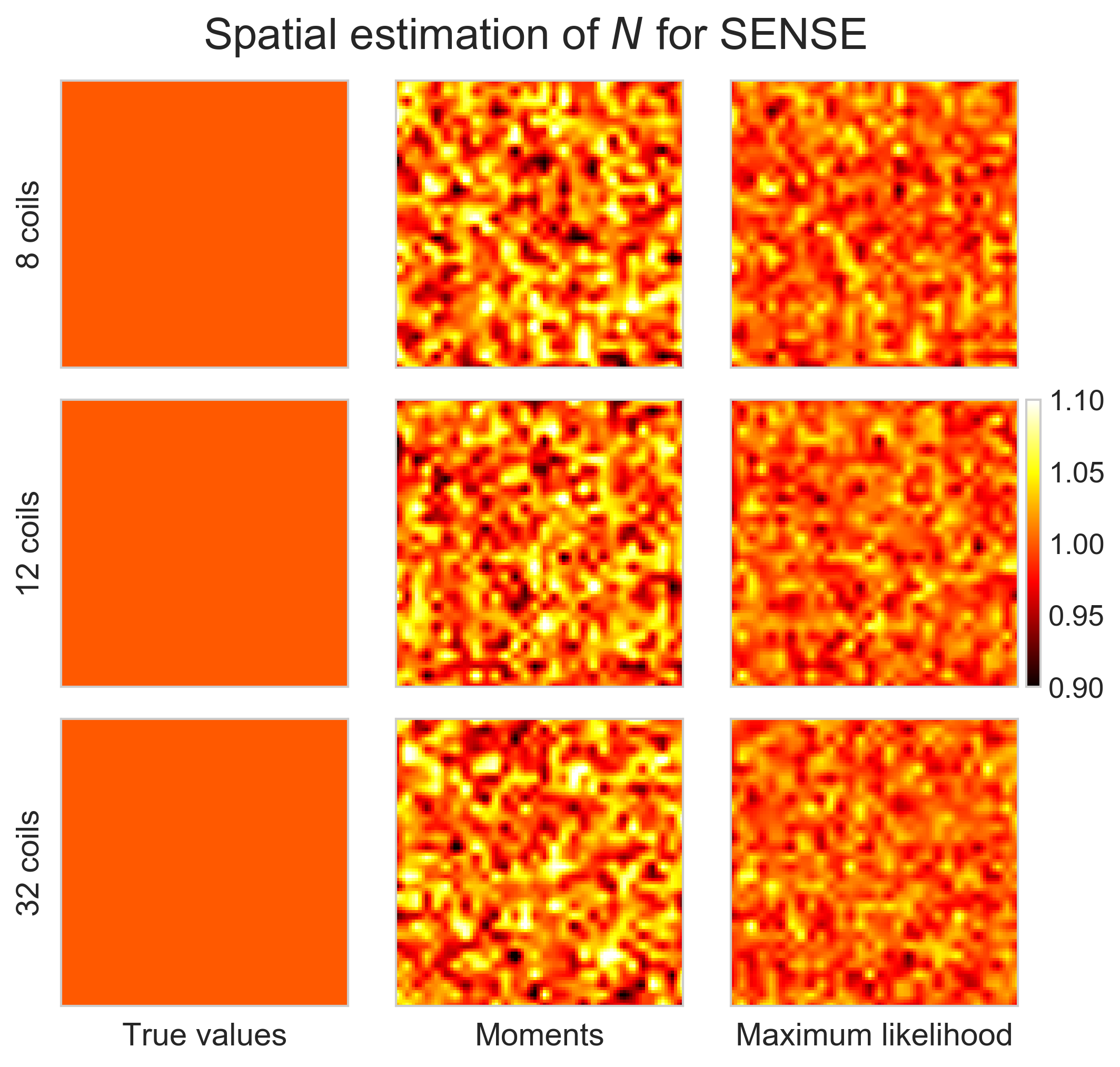}
    \end{subfigure}
    \begin{subfigure}{0.49\linewidth}
        \textbf{D)}
        \includegraphics[width=\linewidth,valign=t]{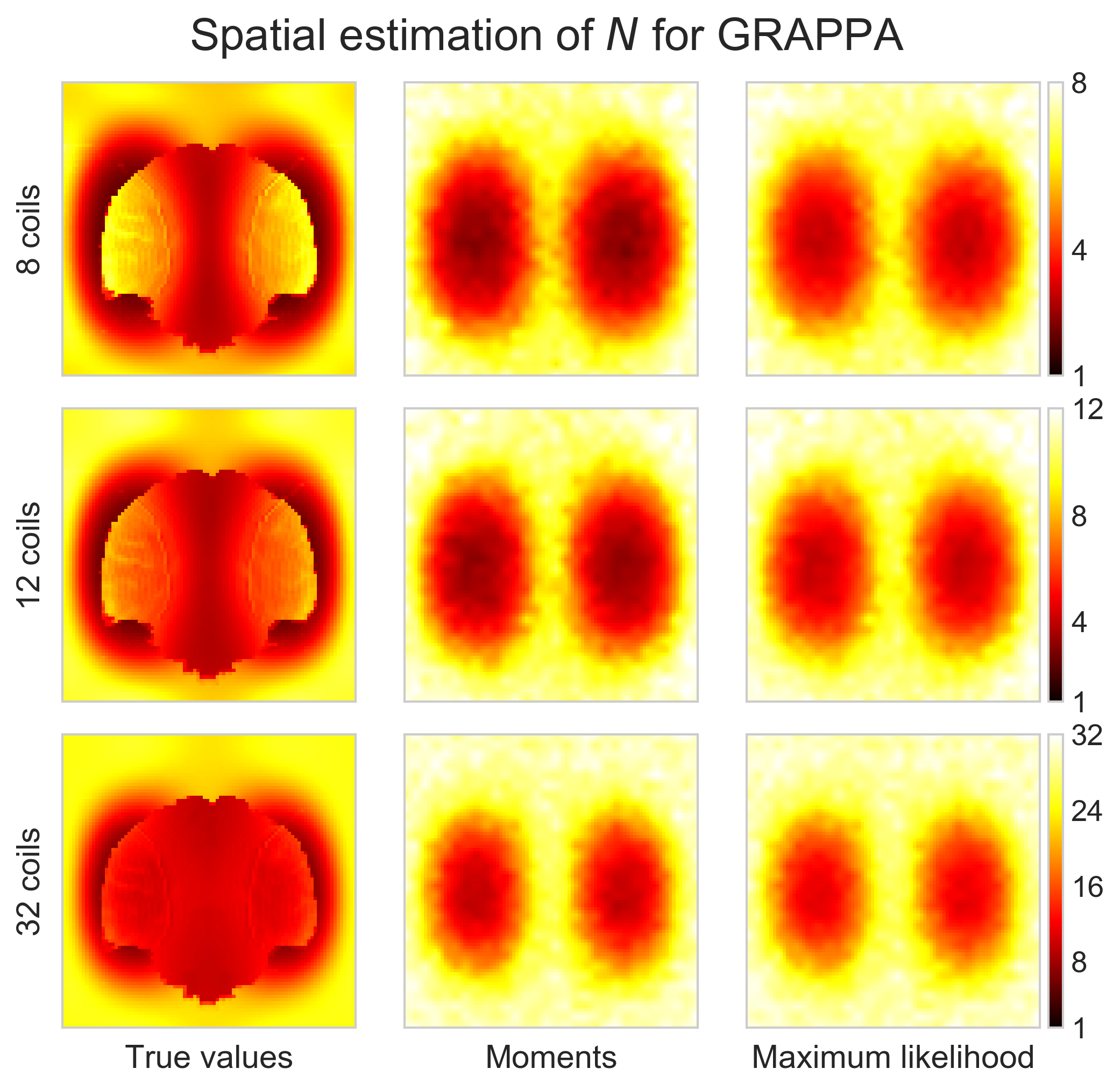}
    \end{subfigure}
    \caption{The estimated values of $N$ for SENSE (left column) and GRAPPA (right column) for the \bval{1000} (first row of boxplots) and \bval{3000} (second row of boxplots) datasets.
    The left column shows results computed from the automatically selected background voxels (basis = DWIs), while the right column shows local estimation using noise maps (basis = noise maps).
    In \textbf{A)} and \textbf{B)}, the boxplot of $N$ inside the phantom for the SENSE/GRAPPA algorithm with a spatial map of $N$
    shown in \textbf{C)} and \textbf{D)} computed using the noise maps from the \bval{3000} datasets.
    \review{Note how the colorbar is the same in \textbf{C)}, while each row of \textbf{D)} shares the same colorbar.}}
    \label{fig:fiberfox_N}
\end{figure}

\cref{fig:fiberfox_sense} shows the estimated values of $\sigma_g$ from a SENSE reconstruction
and \cref{fig:fiberfox_grappa} shows the results for the GRAPPA reconstructed datasets.
For SENSE, estimation using noise maps is the most precise for both proposed methods and PIESNO where the average error is around 0,
followed by LANE when using DWIs as the input which results in 10\% of overestimation.
MPPCA generally underestimates $\sigma_g$ by around 15\% for data at \bval{1000} and 30\% for data at \bval{3000}.
LANE instead overestimates when using DWIs and underestimates $\sigma_g$ when using noise maps and knowing the correct value of $N=1$.
The proposed methods (the moments and maximum likelihood equations) and PIESNO are performing similarly,
but PIESNO requires knowledge of $N=1$.
Estimation is also more precise for the three methods using the Gamma distribution (moments, maximum likelihood and PIESNO)
than those using local estimations (MPPCA and LANE) and closest to the true values when using noise maps.
In the case of GRAPPA, results are similar to the SENSE experiments with the exception of MPPCA being more precise than the compared methods
for the \bval{1000} case and performs equally well at \bval{3000} as the proposed methods with an average error of about 20\%.
Results using LANE are similar with increasing number of coils when assuming $N=1$, while the estimated value
from PIESNO also increases with the number of coils as previously seen in \cref{fig:phantomas_sigma}.
In this case, LANE overestimates $\sigma_g$ by around 50\% when using DWIs, but performs similarly to MPPCA when estimating $\sigma_g$ from the noise maps.
Estimation from noise maps using the moments or maximum likelihood equations is the most precise in all cases.
The error of PIESNO increases with $N$ as seen in \cref{fig:fiberfox_grappa} panel \textbf{C)}.
This is caused by mistakenly including gray matter voxels of low intensity in the estimated distribution
while they are correctly excluded automatically by the proposed methods.
Finally, \cref{fig:fiberfox_N} shows the estimated values of $\textit{N}_{eff}$ using the datasets from \cref{fig:fiberfox_sense,fig:fiberfox_grappa} by the proposed methods.
For the SENSE case, the true value is a constant $N=1$ by construction and the estimated values by both algorithms are on average correct
with the maximum likelihood equations having the lowest variance.
In the case of GRAPPA, values of $N$ vary spatially inside the phantom and depend on the per voxel signal intensity, just as $\sigma_g$ does in \cref{fig:fiberfox_grappa}.
This leads to some overestimation when only background voxels are considered, with the best estimation obtained when using the noise maps.
For simulations using 8 and 12 coils, estimated values of $N$ are, in general, following the expected values.
However, the spatially varying pattern can not be fully recovered as the correct value of $N$ depends on the true signal intensity $\eta$ in each voxel,
which is not present when collecting noise only measurements.

\subsection{Acquired phantom datasets}
\label{sec:water_phantom}

\begin{figure}
    \begin{subfigure}{0.49\linewidth}
        \textbf{A)}
        \includegraphics[width=\linewidth,valign=t]{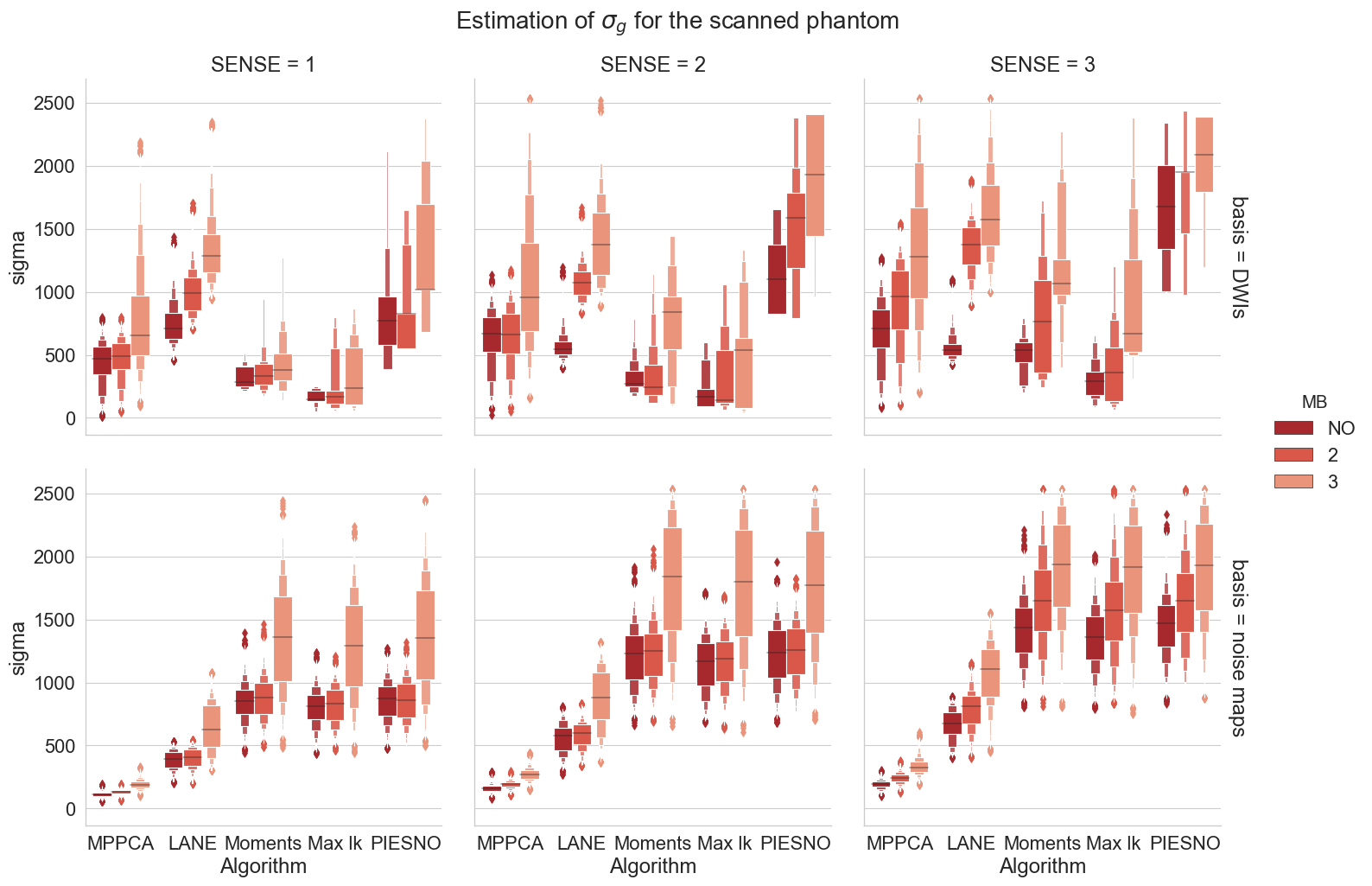}
    \end{subfigure}
    \hfill
    \begin{subfigure}{0.49\linewidth}
        \textbf{B)}
        \includegraphics[width=\linewidth,valign=t]{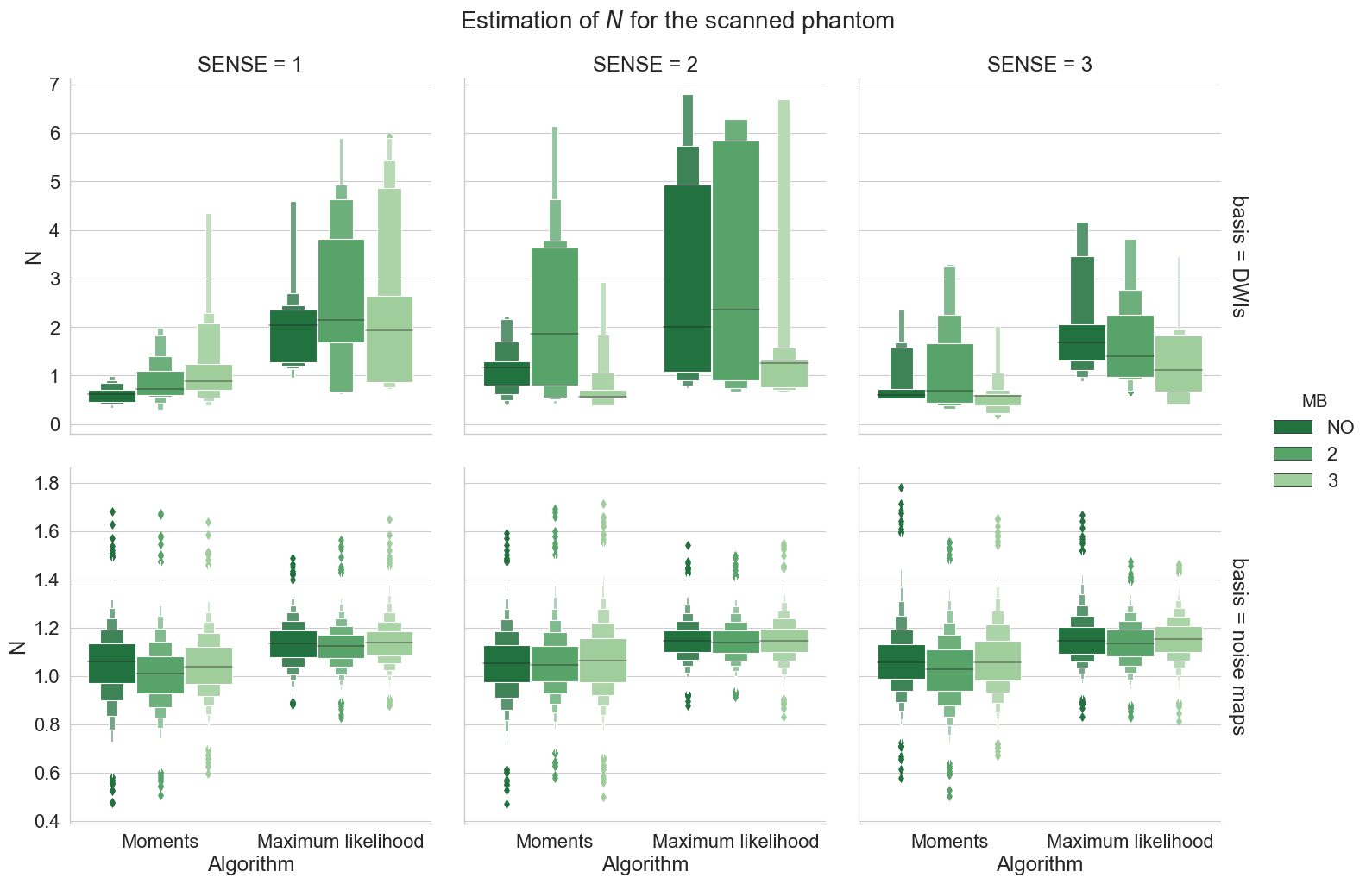}
    \end{subfigure}
    \begin{subfigure}{0.485\linewidth}
        \textbf{C)}
        \includegraphics[width=\linewidth,valign=t]{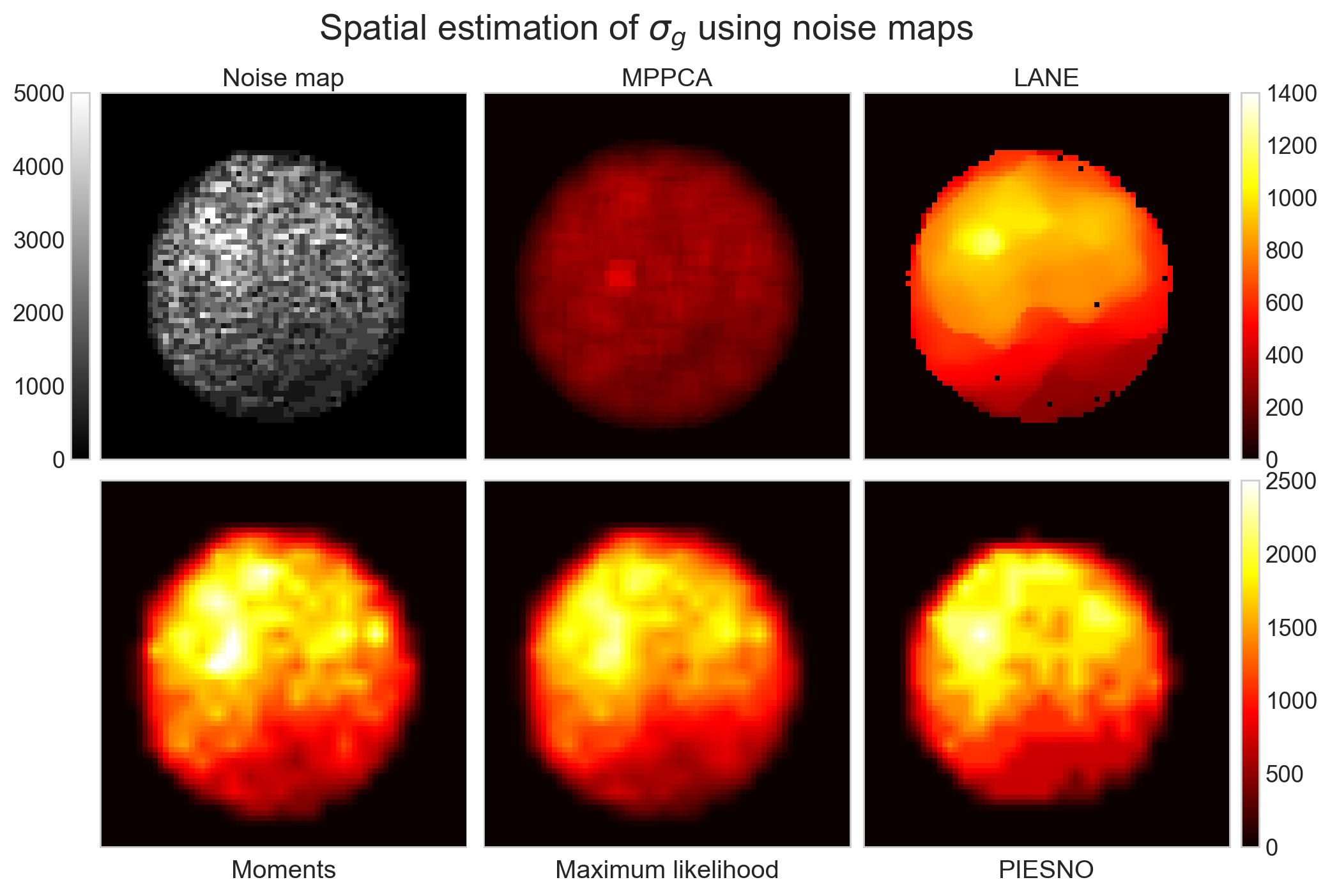}
    \end{subfigure}
    \hfill
    \begin{subfigure}{0.485\linewidth}
        \textbf{D)}
        \includegraphics[width=\linewidth,valign=t]{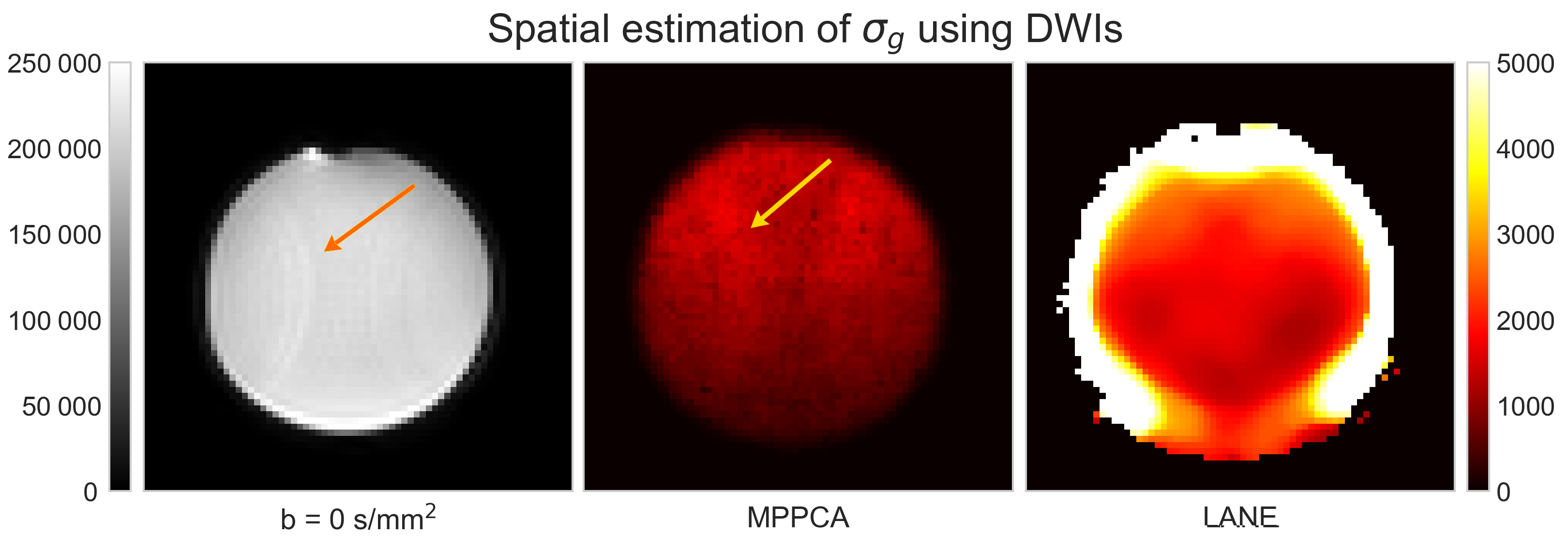}
        \textbf{E)}
        \includegraphics[width=\linewidth,valign=t]{water_phantom/figure7e}
    \end{subfigure}
    \caption{Estimation of noise distributions for the scanned phantom datasets inside a small ROI.
    Large outliers above the 95th percentile were removed to not skew the presented boxplots.
    In \textbf{A)}, the estimation of the noise standard deviation $\sigma_g$ for each method using DWIs (top row) and using noise maps (bottom row).
    Each column shows an increasing SENSE factor, where $\sigma_g$ increases (according to theory) with the square root of the SENSE factor.
    The different hues show an increasing multiband factor, which should not influence the estimation of $\sigma_g$.
    For the case \textit{MB} = 3, there may be signal leakage to adjacent slices, which would increase the measured values of $\sigma_g$
    even when the estimation uses only noise maps.
    In \textbf{B)}, boxplots for the values of $N$ estimated by both proposed methodologies for the experiments shown in \textbf{A)}.
    Estimated values using noise maps are always close to 1 on average while estimations using DWIs seems to be affected
    by the possible signal leakage inherent to the use of multiband imaging.
    In \textbf{C)}, an axial slice of a noise map and estimated values of $\sigma_g$ by all methods for the case $R = 3$ and \textit{MB} = 3,
    which is the highest rate of acceleration from all of the investigated cases.
    Note the different scaling between the top and bottom row as MPPCA and LANE estimates of $\sigma_g$ are two to three times lower than other methods.
    In \textbf{D)}, a \bval{0} image of the phantom and spatially estimated values of $\sigma_g$ for MPPCA and LANE.
    Note how some signal leakage (orange arrows) is affecting the \bval{0} volume due to using \textit{MB} = 3.
    In \textbf{E)}, location of the spherical ROI used for the boxplots overlaid on a noise map and spatially estimated values of $N$ for both proposed methods.
    As less voxels are available near the borders of the phantom, estimating the noise distributions parameters results in lower precision.
    }
    \label{fig:water_phantom}
\end{figure}

\cref{fig:water_phantom} shows the estimated values of $\sigma_g$ for all methods with a SENSE acceleration of rate $R = 1, 2 \text{ and } 3$
with multiband imaging at acceleration factors of \textit{MB} = 2, \textit{MB} = 3 or deactivated in panel \textbf{A)}.
Results show that $\sigma_g$ increases with $R$ and is higher when \textit{MB} = 3 for $R$ fixed,
even if in theory $\sigma_g$ should be similar for a given $R$ and increasing \textit{MB}.
Panel \textbf{C)} shows the estimated values of $\sigma_g$ when using noise maps as the input for $R=3$ and \textit{MB} = 3.
As in the synthetic experiments, MPPCA and LANE have the lowest estimates for $\sigma_g$ with PIESNO and the proposed methods estimating higher values.
Since the correct value is unknown, a reference sample slice of a noise map is also shown.
When compared to values from the measured noise map, estimated values of $\sigma_g$ are approximately fivefold lower for MPPCA,
four times lower for LANE and around half for the other methods.
Estimation on the noise maps yields a value of around $N=1$ for both proposed methods
as seen in panels \textbf{B)} and \textbf{E)}, irrespective of the acceleration used.
In the case of estimation using the DWIs, the range of estimated values is larger and increases at acceleration factors
of \textit{MB} = 2 and $R$ = 2 or 3.

\subsection{In vivo datasets}
\label{sec:invivo_dataset}

\paragraph{Multiple datasets from a single subject}

\begin{figure}
    \begin{subfigure}{\textwidth}
        \textbf{A)}
        \includegraphics[width=0.49\textwidth,valign=t]{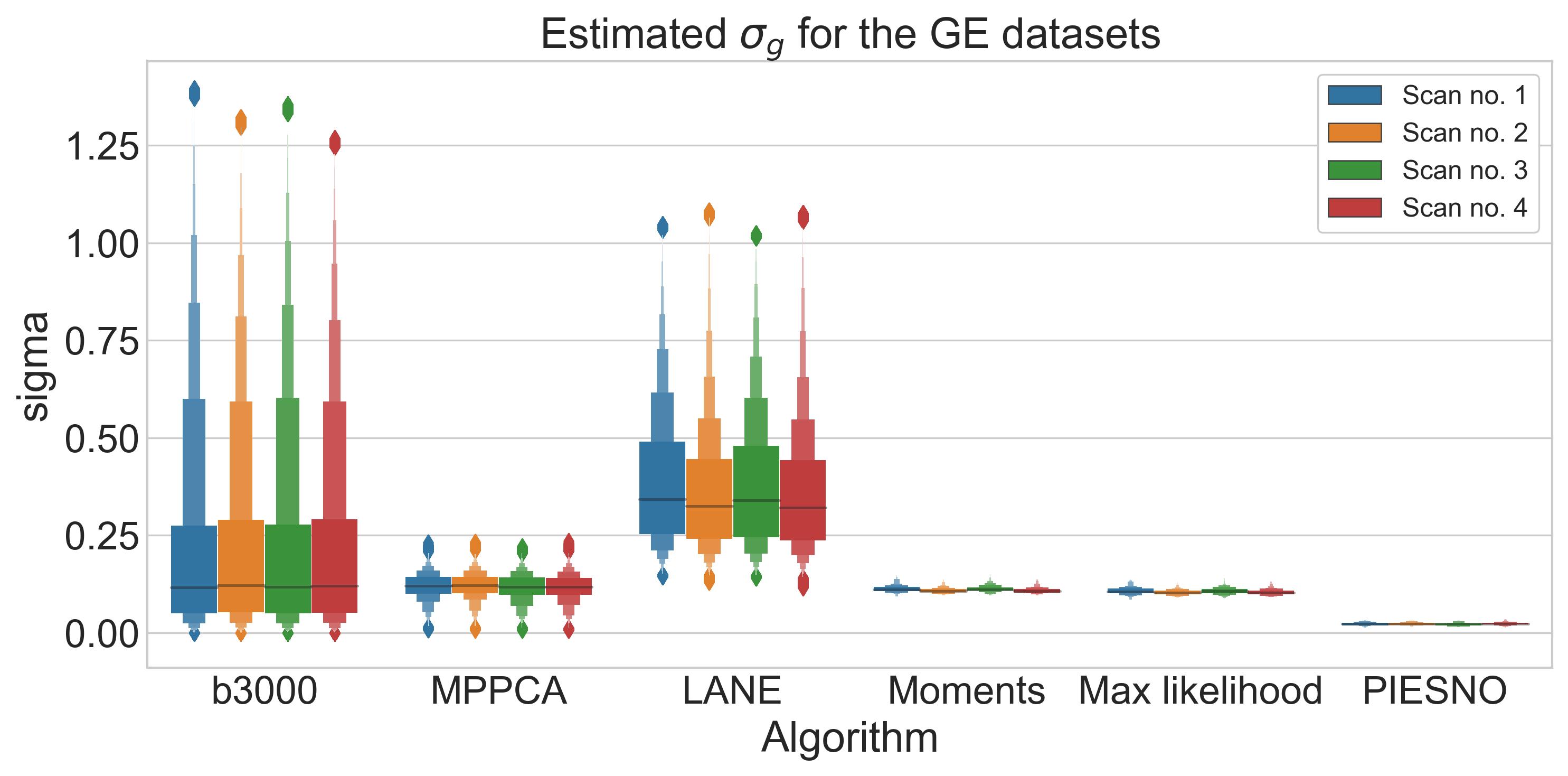}
        \textbf{B)}
        \includegraphics[width=0.49\textwidth,valign=t]{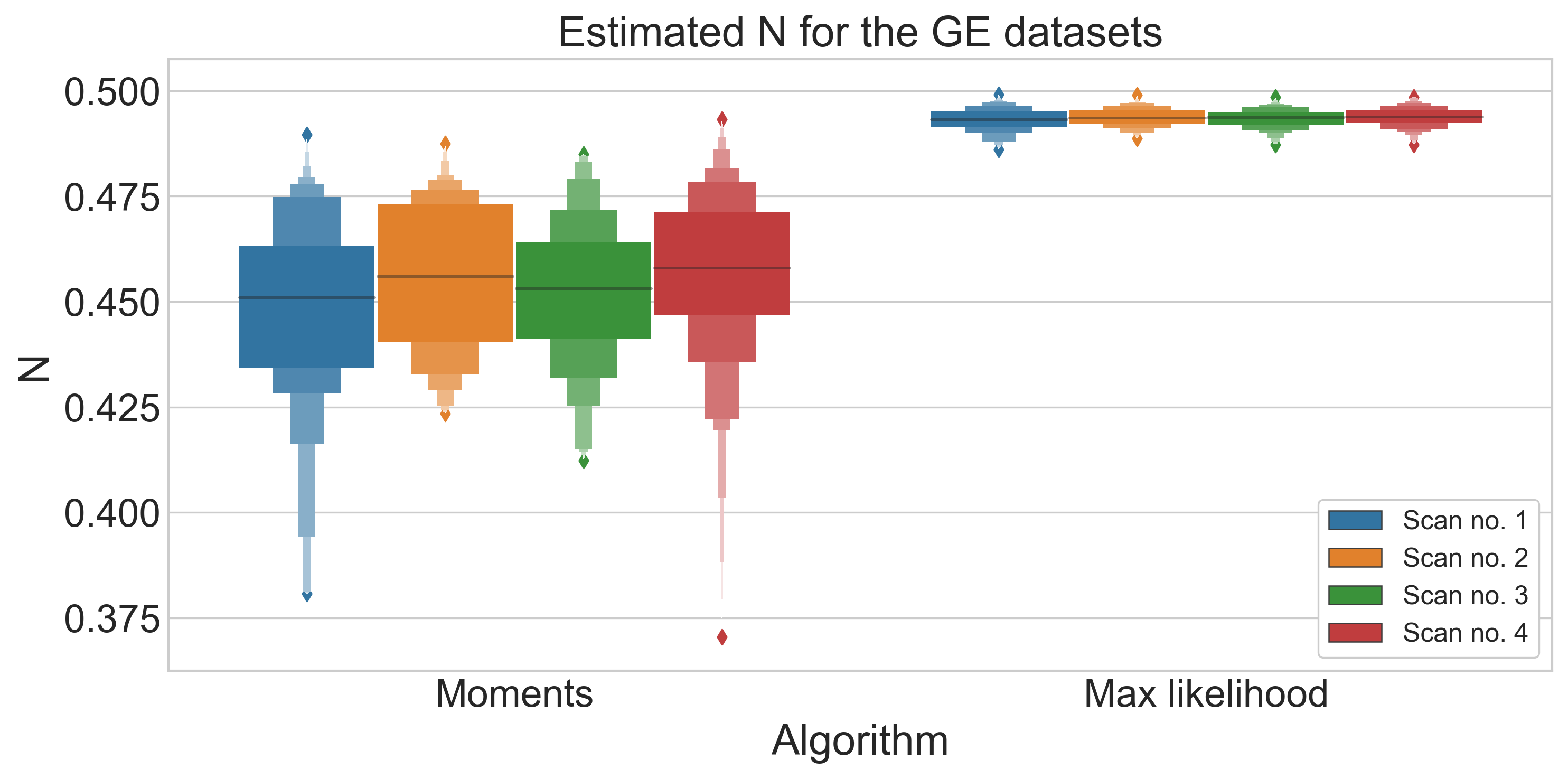}
    \end{subfigure}

    \begin{subfigure}[c]{\textwidth}
        \textbf{C)}
        \includegraphics[width=\textwidth,valign=t]{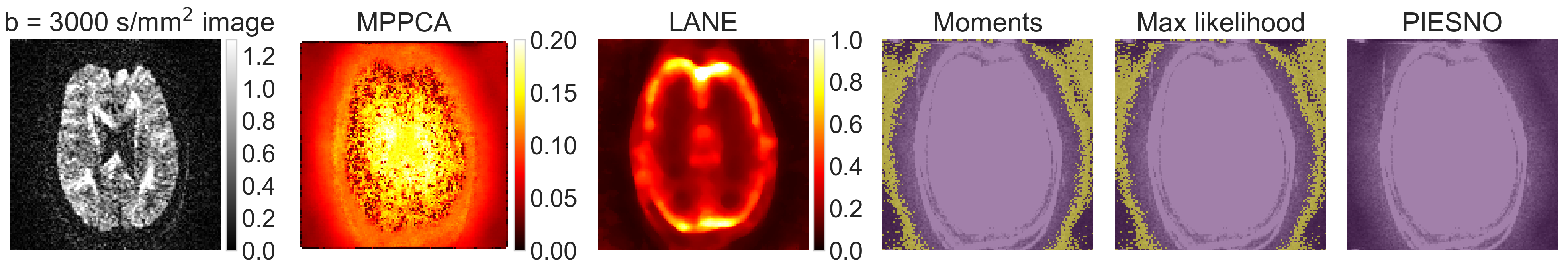}
    \end{subfigure}
    \caption{Estimation of the noise profiles on four repetitions of a single subject from a GE scanner.
    In \textbf{A)}, the baseline signal values of a \bval{3000} volume and estimated values of $\sigma_g$ for all methods inside a brain mask
    and \textbf{B)} estimated values of $N$ by the proposed methods are shown.
    Note that the values for LANE and the \bval{3000} volume were truncated at the 99 percentile to remove extreme outliers.
    In \textbf{C)}, an axial slice of a \bval{3000} image from one dataset and the estimated values of $\sigma_g$ for MPPCA and LANE.
    For the proposed methods and PIESNO, a mask of the identified background voxels (in yellow) overlaid on the data.}
    \label{fig:openfmri_std}
\end{figure}

\cref{fig:openfmri_std} shows the estimated value of $\sigma_g$
on four repetitions of the GE datasets for each method as computed inside a brain mask.
The values from a \bval{3000} volume (including background) is also shown as a reference for the values
present at the highest diffusion weighting in the dataset.
All methods show good reproducibility as their estimates are stable across the data.
The value of $N$ as computed by our proposed methods is also similar for all datasets with the median at
$N = 0.45$ for the moments and $N = 0.49$ for the maximum likelihood equations.
This corresponds to a half Gaussian distribution as would be obtained by a real part magnitude reconstruction \citep{Dietrich2008}.
However, LANE recovered the highest values of $\sigma_g$ amongst all methods with a large variance and a median higher than the \bval{3000} values,
which might indicate overestimation in some areas.
The median of MPPCA and the proposed methods are similar, while PIESNO estimates of $\sigma_g$ are approximately two times lower.
This could indicate that specifying $N=1$ was incorrect for these datasets, as PIESNO identified about 10 noise only voxels.

\begin{figure}
    \includegraphics[valign=c,width=0.7\textwidth]{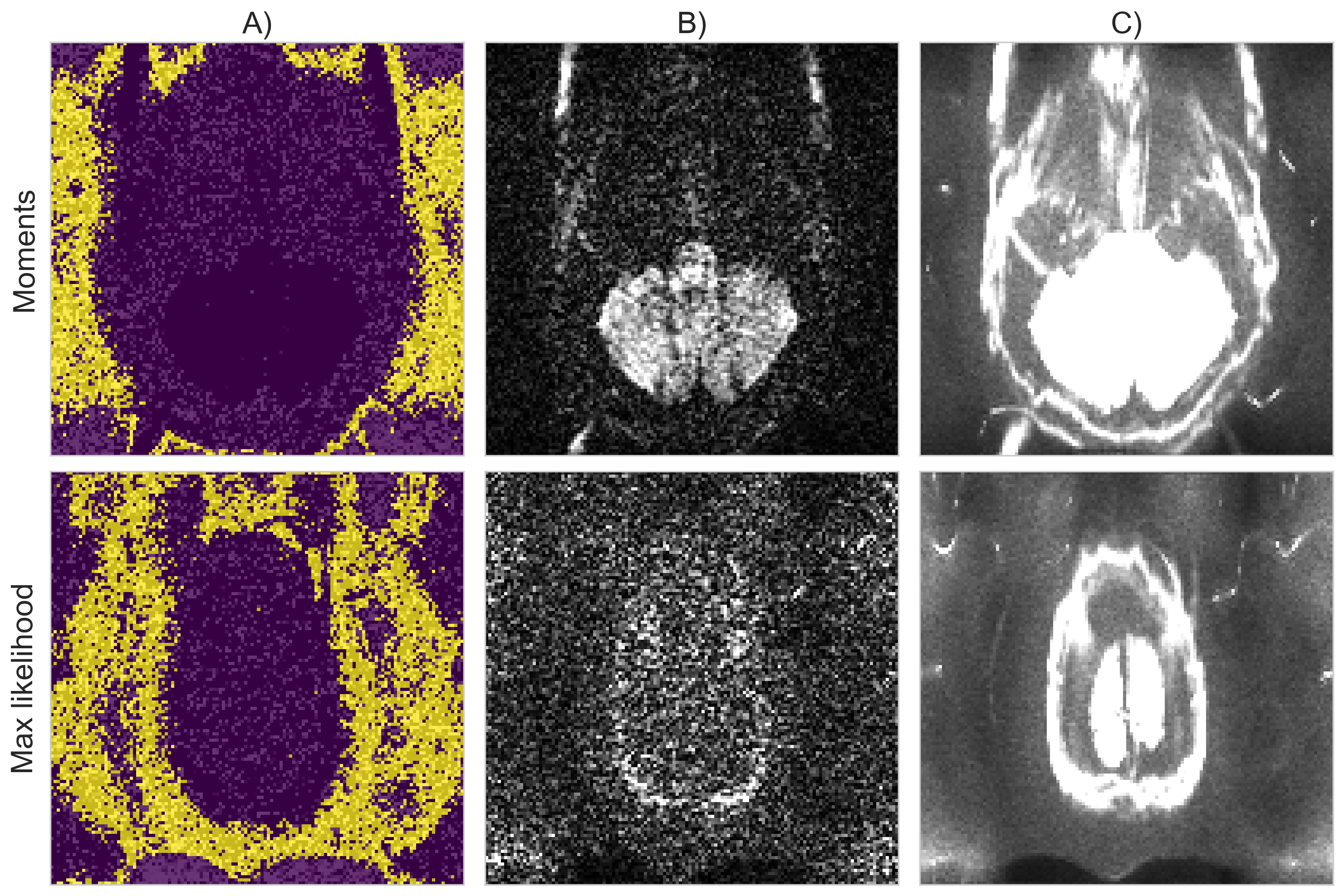}
    \hfill
    \includegraphics[valign=c,width=0.29\textwidth]{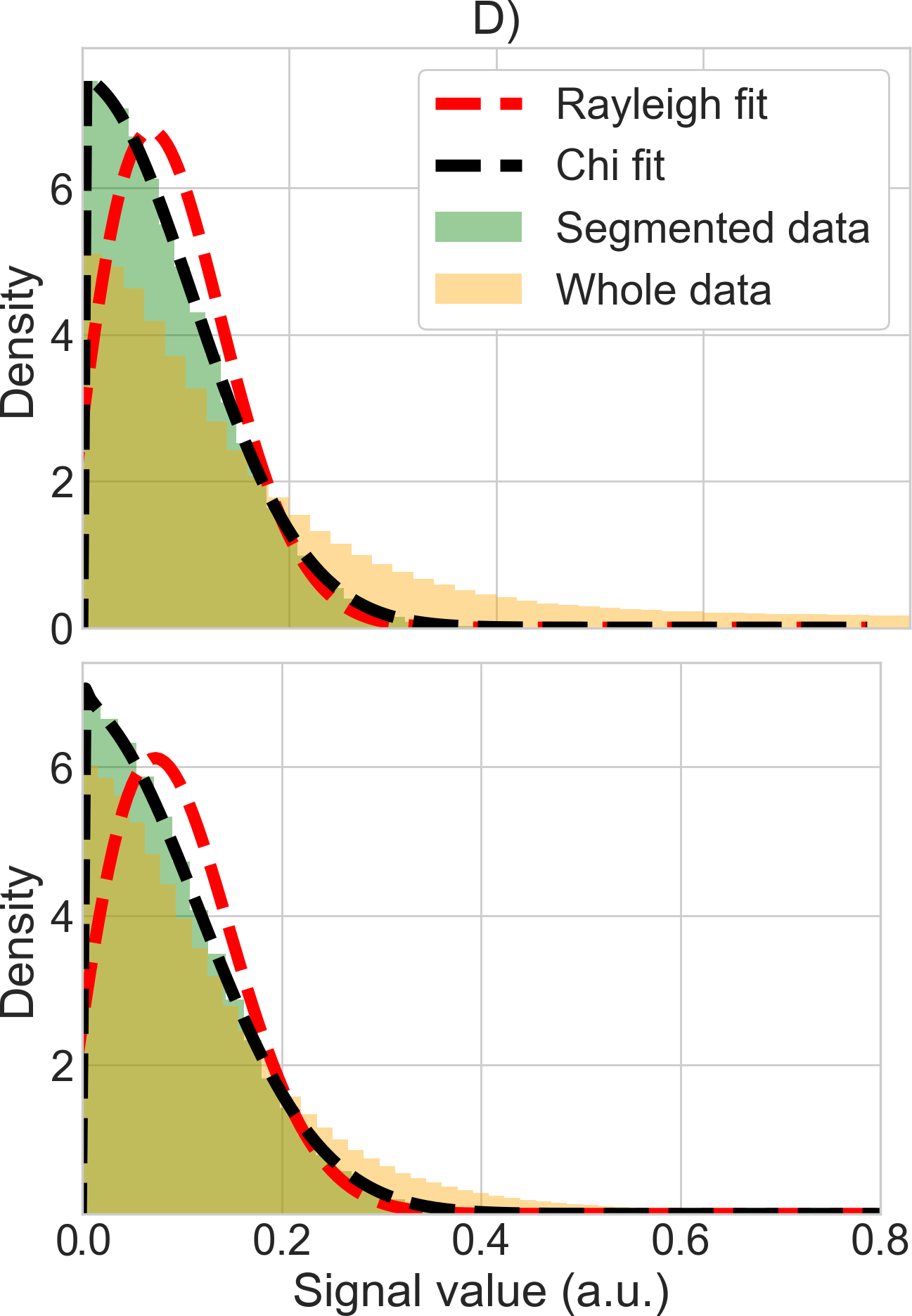}
    \caption{An axial slice in the cerebellum from one of the GE datasets.
    Voxels identified in \textbf{A)} as noise only (yellow) are free of artifacts in a single slice in \textbf{B)} or along the sum of all volumes in \textbf{C)}.
    In \textbf{D)}, the normalized density histogram using the selected voxels from \textbf{A)} (green)
    fit well a chi distribution (black dashed lines), while assuming a Rayleigh distribution (red dashed lines)
    or using all non brain voxels (orange) leads to a worse visual fit.}
    \label{fig:openfmri_mask}
\end{figure}

\cref{fig:openfmri_mask} shows an axial slice around the cerebellum and the top of the head which are corrupted by acquisition artifacts likely due to parallel imaging.
Voxels containing artifacts were automatically discarded by both methods, preventing misestimation of $\sigma_g$ and $N$.
The values computed from these voxels also offer
a better qualitative fit than assuming a Rayleigh distribution or selecting non-brain data.
We also timed each method to estimate $\sigma_g$ on one of the GE datasets using a standard desktop computer with a 3.5 GHz Intel Xeon processor.
The runtime to estimate $\sigma_g$ (and $N$) was around 5 seconds for the maximum likelihood equations, 9 seconds for the moments equations, 11 seconds for PIESNO,
3 minutes for MPPCA and 18 minutes for LANE.

\paragraph{Estimation with a Connectom dataset}

\begin{figure}
    \begin{subfigure}{0.48\linewidth}
        \textbf{A)}
        \includegraphics[width=\linewidth,valign=t]{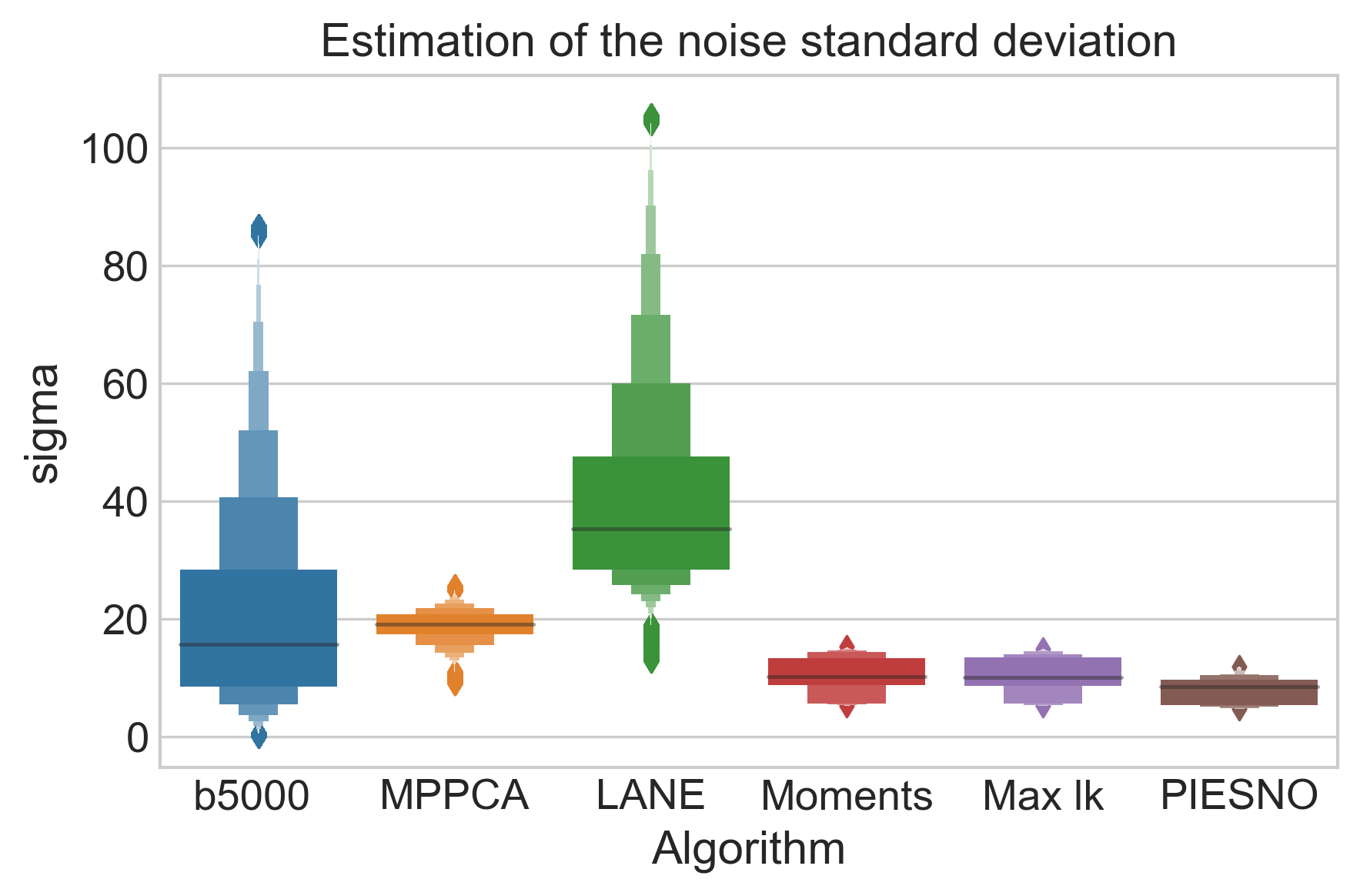}
    \end{subfigure}
    \hfill
    \begin{subfigure}{0.48\linewidth}
        \textbf{B)}
        \includegraphics[width=\linewidth,valign=t]{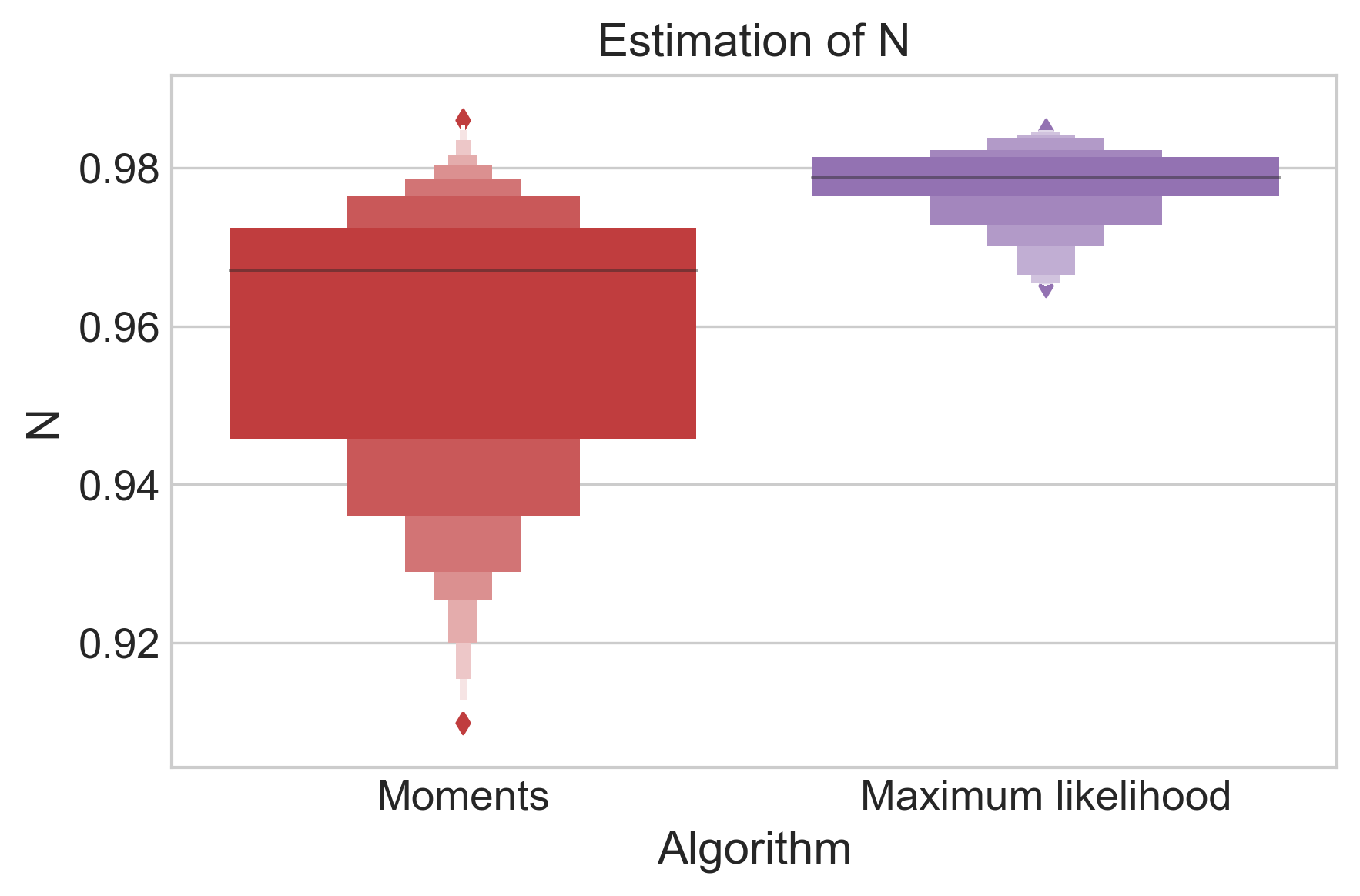}
    \end{subfigure}
    \begin{subfigure}{\linewidth}
        \textbf{C)}
        \includegraphics[width=\linewidth,valign=t]{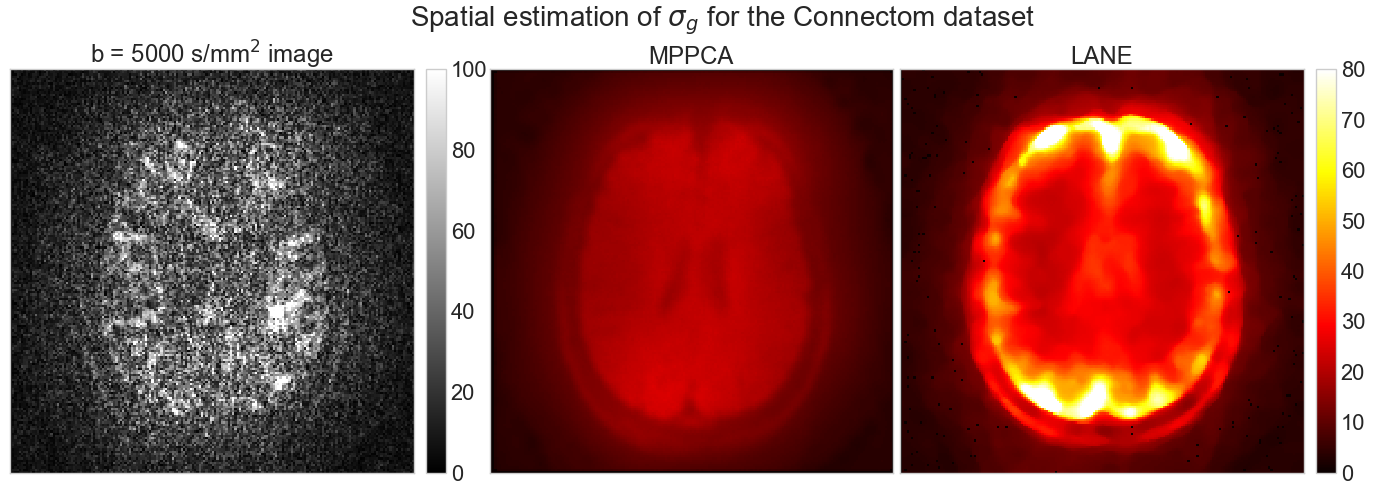}
        \textbf{D)}
        \includegraphics[width=\linewidth,valign=t]{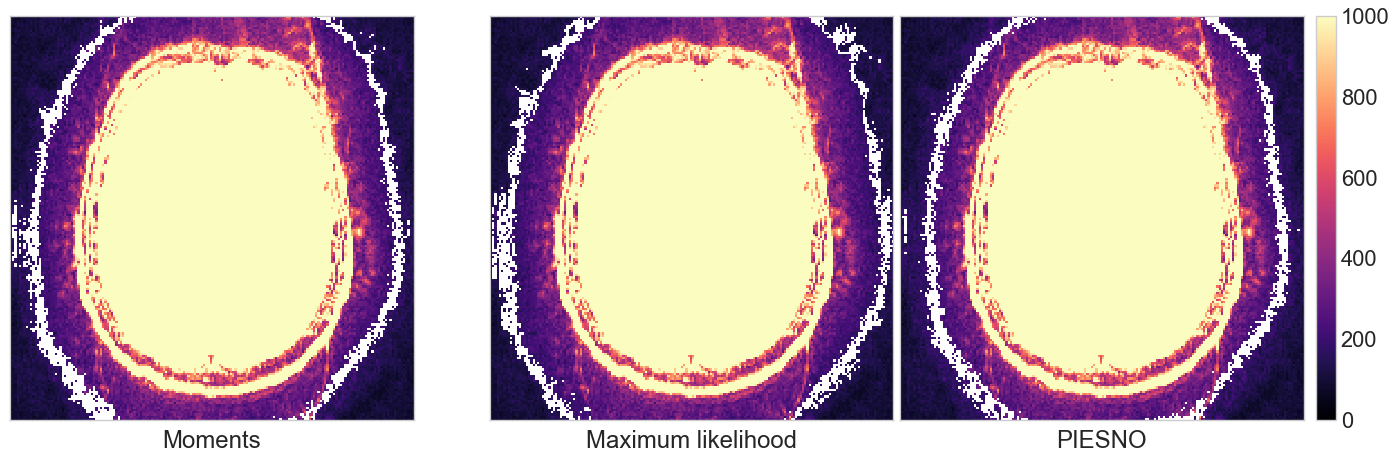}
    \end{subfigure}
    \caption{Estimation of noise distributions for the Connectom dataset.
    In \textbf{A)}, signal distribution of the original data and noise standard deviation $\sigma_g$ for all methods, where data above the 99th percentile for the \bval{5000} volume and LANE were discarded.
    In \textbf{B)}, values of $N$ as estimated using the moments (in red) and by maximum likelihood (in purple).
    In \textbf{C)} on the top row, a \bval{5000} volume and spatial estimation of $\sigma_g$ as measured by MPPCA and LANE.
    \review{In \textbf{D)}}, voxels identified as containing only noise (in white) by the moments, maximum likelihood and PIESNO overlaid on top of the sum of the \bval{0} volumes.
    Note how each algorithm identifies different voxels, while automatically ignoring voxels belonging to the data or contaminated with signal leakage from multiband imaging.
    }
    \label{fig:invivo_estimation}
\end{figure}

\begin{figure}
    \begin{subfigure}{0.8\linewidth}
        \textbf{A)}
        \includegraphics[width=\linewidth,valign=t]{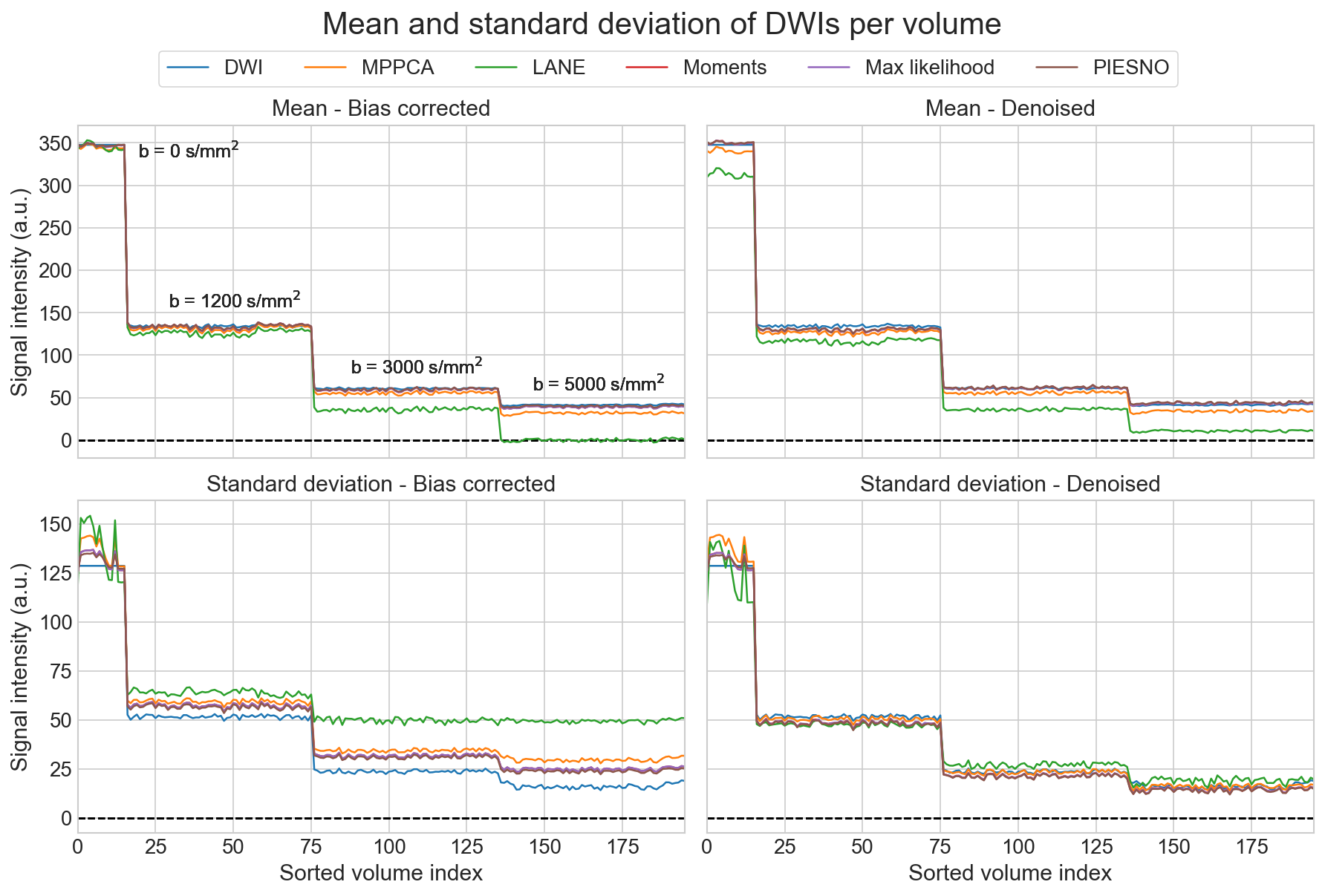}
    \end{subfigure}

    \begin{subfigure}{0.49\linewidth}
        \textbf{B)}
        \includegraphics[height=6.5cm,valign=t]{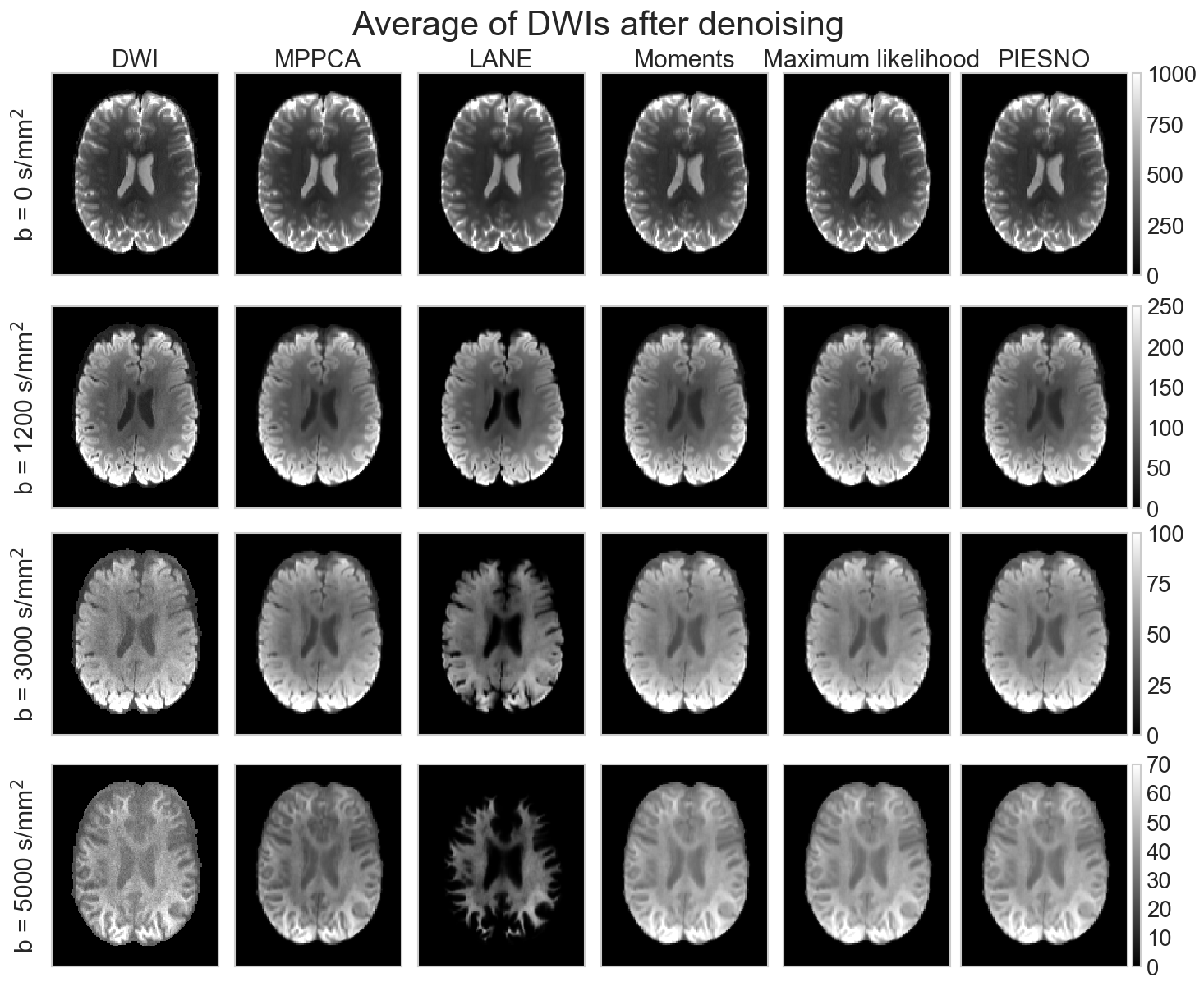}
    \end{subfigure}
    \hfill
    \begin{subfigure}{0.49\linewidth}
        \textbf{C)}
        \includegraphics[height=6.5cm,valign=t]{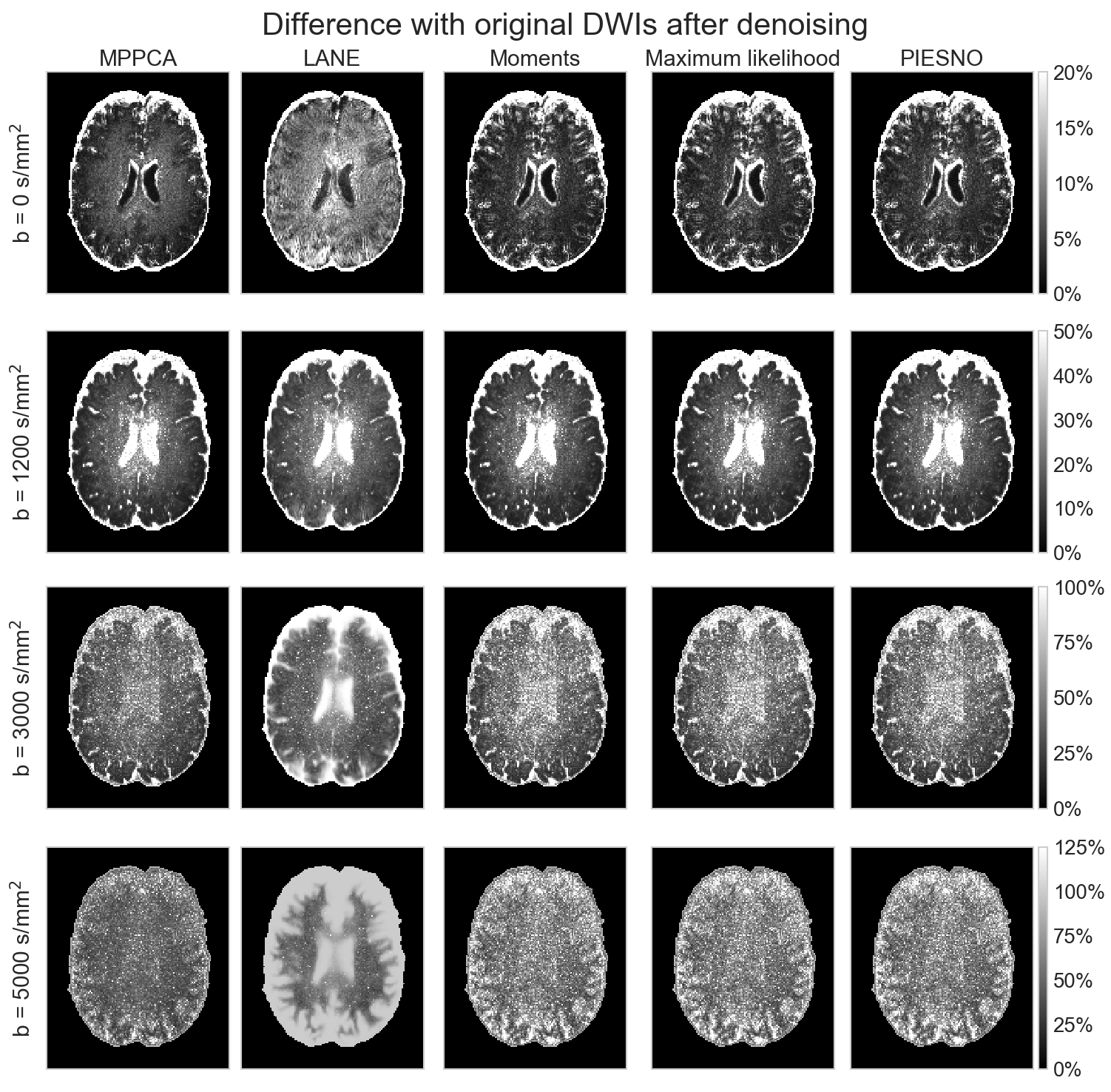}
    \end{subfigure}
    \caption{Bias correction and denoising \review{with the NLSAM algorithm} of the Connectom dataset from the noise distributions estimated by each method.
    In \textbf{A)}, the left column (resp. right column) shows the result of noncentral chi bias correction (resp. denoising) on the signal value.
    The top row (resp. bottom row) shows the mean (resp. standard deviation) of the signal inside a white and gray matter mask for each volume.
    Note how the bias corrected value of LANE goes below 0 (dashed line) due to its high estimation of $\sigma_g$.
    After denoising, the standard deviation of the signal decreases as the b-value increases, an effect which is less noticeable for the bias corrected signal only.
    However, this effect is less pronounced for the bias corrected signal only in the case of LANE and MPPCA.
    In \textbf{B)}, spatial maps of the original data and after denoising (in each column) \review{when averaging datasets at the same b-value} for \bval{0}, \bval{1200}, \bval{3000} and \bval{5000} (in each row) for each method.
    Note how each b-value uses a different scale to enhance visualization even though the signal intensity is lower for increasing b-values.
    Panel \textbf{C)} shows the difference in percentage between the original data and after denoising using parameters as estimated by each algorithm.
    }
    \label{fig:invivo_maps}
\end{figure}

\cref{fig:invivo_estimation} shows in \textbf{A)} the estimated values of $\sigma_g$ inside a brain mask and in
\textbf{B)} the values of $N$ computed by the proposed methods.
Estimated values of $\sigma_g$ vary by an order of magnitude between the different methods.
In the case of MPPCA and LANE, the median of the estimates is higher than the reference \bval{5000} data, while PIESNO and the proposed methods
estimate values lower than the reference and have lower variability in their estimated values.
For the estimation of $N$, recovered values are distributed close to 1 as is expected
from an adaptive combine reconstruction providing a Rician distribution.
Values estimated with the maximum likelihood equations have a lower variability than with the moments equations.
In \textbf{C)}, the top row shows the \bval{5000} volume and spatial maps of $\sigma_g$ as estimated by MPPCA and LANE.
The bottom row shows voxels identified as pure noise (in light purple) using the moments, the maximum likelihood equations and PIESNO.
Ghosting artifacts are excluded, but presumably affect estimation using the entire set of DWIs shown in the top row.
\cref{fig:invivo_maps} shows in \textbf{A)} the signal intensity after applying bias correction (left column)
and after denoising (right column) for each volume ordered by increasing b-value.
The top row (resp. bottom row) shows the mean (resp. standard deviation) as computed inside a white and gray matter mask.
The mean signal decays with increasing b-value as expected, but the standard deviation of the signal does not follow the same trend in the cases of LANE.
After denoising, the mean signal and its standard deviation decays once again as for the original data.
Panel \textbf{B)} shows the average DWI at a given b-value for the original dataset and after denoising using the noise distribution from each method.
Results are similar for all methods for the \bval{0} datasets,
but the overestimation of $\sigma_g$ by LANE produces missing values in the gray matter for \bval{3000} and \bval{5000}.
In general, averaging reduces the noise present at \bval{0} and \bval{1200} while only denoising is effective at \bval{3000} and \bval{5000}.
At \bval{5000}, the MPPCA denoised volume is of lower intensity than when obtained by the moments, maximum likelihood equations or PIESNO.
This is presumably due to LANE and MPPCA estimating higher values of $\sigma_g$ than the three other methods.
Finally, panel \textbf{C)} shows the absolute difference between the original and the denoised dataset obtained by each method.
At \bval{5000}, LANE removes most of the signal in the gray matter mistakenly due to overestimating $\sigma_g$.
Other methods perform comparably well on the end result, despite estimates of $\sigma_g$ of different magnitude.

\section{Discussion}
\label{sec:discussion}

We have shown how a change of variable to a Gamma distribution $\textit{Gamma}(N, 1)$
can be used to robustly and automatically identify voxels belonging only to the noise distribution.
At each iteration, the moments (\cref{eq:find_sigma,eq:find_N}) and maximum likelihood equations (\cref{eq:gamma_simplified_ml_sigma,eq:gamma_simplified_ml_N}) of the Gamma distribution can
be used to compute the number of degrees of freedom $N$ and the Gaussian noise standard deviation $\sigma_g$ relating to the original noise distribution.
Voxels not adhering to the distribution are discarded, therefore refining the estimated parameters until convergence.
One of the advantage of our proposed methods is that no \textit{a priori} knowledge is needed from the
acquisition or the reconstruction process itself, which is usually not stored or hard to obtain in a clinical setting.
Results from \cref{sec:synthetic_simulation} show that we can reliably estimate parameters from the magnitude data itself in the case of stationary distributions.
For spatially varying distributions without parallel acceleration, the proposed methods achieve an average \review{percentage} error of approximately 10\% when estimating $\sigma_g$,
which is equal or better than the other methods compared in this work.
Estimated values of $N$ are around the true values, even when $\sigma_g$ is misestimated.
While these experiments may still be considered to be simplistic when compared to modern scanning protocols where parallel acceleration is ubiquitous,
they highlight that even textbook cases can lead to misestimation if the correct signal distribution is not taken into account.
Practical tasks taking advantage of the signal distribution such as bias correction \citep{Pieciak2018}, noise floor removal \citep{Sakaie2017},
deep learning reconstruction with various signal distributions \citep{Lonning2019} or diffusion model estimation \citep{Collier2018,Zhang2012d,Landman2007a}
may be tolerant, but not perform optimally, to some misestimation of the noise distribution.
See \eg \citet{Hutchinson2017} for discussions on the impact of noise bias correction on diffusion metrics in an ex vivo rat brain dataset.
\review{Note that for some applications such as denoising, only the product of the parameters of the distribution might be needed (\ie $N \times \sigma_g^2$) \citep{Pieciak2016},
which is a case we did not cover in the present manuscript.}

\paragraph{Effects of misspecification of the noise distribution}

Experiments with SENSE from \cref{sec:synthetic_simulation} reveal that
using a local estimation with noise maps provides the best estimates for the proposed methods and PIESNO.
MPPCA and LANE perform better when using DWIs as the input rather than noise maps, but at the cost of a broader range of estimated values
for $\sigma_g$ and still underperform when compared to the three other methods.
This is presumably because the signal diverges from a Gaussian distribution at low SNR \citep{Gudbjartsson1995} and especially in noise maps,
leading to a misspecification of its parameters when the assumed noise distribution is incorrect.
Phantom experiments carried with GRAPPA show similar trends except for PIESNO,
which overestimates $\sigma_g$ as shown in \cref{fig:phantomas_sigma}.
When erroneously fixing $N=1$, low intensity voxels where $\eta>0$ (\eg gray matter) may be mistakenly included in the distribution after the change of variables, leading to overestimation of $\sigma_g$.

The presence of tissue in voxels used for noise estimation might compromise the accuracy of the estimated distributions as shown in \cref{sec:synthetic_simulation}.
This can be explained by the lower number of noise only voxels available to the proposed methods and PIESNO
and to difficulty in separating the signal from the noise for MPPCA and LANE at low SNR.
Using measured noise maps is not a foolproof solution as by definition they set $\eta = 0$, while the (unknown) noiseless signal from tissues is $\eta > 0$.
As the noise distribution may depend on $\eta$ \citep{Aja-Fernandez2014},
this means that its parameters (\eg from a GRAPPA reconstruction) will be inherently different than the one estimated from noise maps.
This effect can be seen in \cref{fig:fiberfox_N}, where the estimated values of $N$ from noise maps and DWIs are close to 1 for SENSE as expected in theory.
For GRAPPA, they are either overestimated and underestimated in regions of the phantom and overestimated in background regions as $N$ locally depends on $\eta$.
Accurate estimation of $\sigma_g$ and $N$ over signal regions still remains an open challenge.
Nevertheless, the median of the estimated distribution of $\sigma_g$ is closer to the true distribution when using noise maps than when using DWIs for the proposed methods.
Such noise map measurements could therefore provide improved signal distribution estimation for, \eg body or cardiac imaging, where no intrinsic background measurements are available.

\paragraph{Effects of parallel imaging and multiband in a phantom}

\cref{sec:water_phantom} presented results from a scanned phantom using SENSE coupled with multiband imaging.
While no ground truth is available, a SENSE acceleration should provide a Rician signal distribution $(N = 1)$
and $\sigma_g$ should increase with $\sqrt{R}$ \citep{Aja-Fernandez2014}.
\cref{fig:water_phantom} shows that for a common SENSE factor, all values of $\sigma_g$ estimated with \textit{MB} = 3 are larger than at lower factors.
The use of multiband imaging should not influence the estimation of $\sigma_g$ as it only reduces the measured signal, and not the noise component unlike SENSE does.
Indeed, estimated values of $\sigma_g$ are stable until \textit{MB} = 3 or $R = 2$ and \textit{MB} = 2 is used; this is possibly due
to signal leakage and aliasing signal from multiband folding over from adjacent slices with higher factors \citep{Todd2016a,Barth2016}.
Noise maps are less affected by this artifact, which is already present when $R = 2$ and \textit{MB} = 2,
as adjacent voxels have low values, similarly to unaffected voxels.
However, leaking signal in DWIs might impact parameters estimation as it can be interpreted
as an increase in SNR and therefore a lower noise contribution than expected.
Estimation of $\sigma_g$ is also increasing approximately with $\sqrt{R}$ for all methods as expected \citep{Aja-Fernandez2014}.
While we can not quantify these results, this follows the synthetic experiments with SENSE shown in \cref{fig:fiberfox_sense},
where PIESNO and the proposed methods were more precise in estimating $\sigma_g$ from noise maps.

In the case of estimation using DWIs as input, this expected increase in $\sigma_g$ for increasing SENSE factor
is less obvious \eg LANE estimates of $\sigma_g$ decrease from $R=2$ to $R=1$ for the no multiband case.
As MPPCA and LANE also estimate $\eta$, it could explain the larger variance of $\sigma_g$ as $\eta$
fundamentally depends on the microstructural content of each voxel, which is complex and subject to large spatial variations, \eg notably across DWIs.
This also means that estimation over DWIs is susceptible to signal leakage,
which would explain the increased estimated values of $\sigma_g$ for \textit{MB} = 2 and \textit{MB} = 3 for a given SENSE factor.
In the noise maps, we have observed that MPPCA and LANE estimated $\eta > \sigma_g$ in all cases (results not shown).
Overestimating the true value of $\eta = 0$, which is an implicit assumption in PIESNO and the proposed methods,
could explain underestimation of $\sigma_g$ when using noise maps.
This overestimation of $\eta$ in turn leads to lower estimates of $\sigma_g$.
The use of multiband and the inherent signal leakage at high factors could explain this overestimation of $\eta$ and underestimation of $\sigma_g$ for all tested cases.
In the case of SENSE, the proposed methods estimated approximately $N=1$ in all cases, suggesting robustness to multiband artifacts.

\paragraph{Estimation of noise distributions for in vivo datasets}

To complement earlier sections, two datasets acquired on different scanners
combining parallel and multiband imaging were analyzed in \cref{sec:invivo_dataset}.
\cref{fig:openfmri_std} shows that assuming a Rician distribution with $N=1$ can prove inadequate in some situations.
The four repetitions of a single subject acquired on a GE scanner point towards a half Gaussian
distribution instead as evidenced by the computed values of $N$ around 0.5.
This is further evidenced by the low number of voxels (less than 10) detected by PIESNO while assuming $N=1$.
In the preliminary results of our MICCAI submission \citep{St-jean2018a},
using $N=0.5$ for PIESNO gave similar results to the proposed methods,
suggesting the departure of the data from a pure Rician distribution.
Additionally, \cref{fig:openfmri_mask} shows that those voxels identified automatically as pure noise also adhere closer
to a chi distribution than a Rayleigh distribution (where $\eta = 0$ in both cases).
Considering the whole distribution of the data, which is contaminated by artifacts, would also lead to a different distribution.
Even if local methods can consider spatially varying noise profile,
the local estimation of $\sigma_g$ will invariably be affected whenever those same artifacts repeat over the data.
This introduce a compromise between avoiding artifacts at the cost of reduced spatial specificity
and local methods which may not be able to exclude artifacts, but provide local estimations of $\sigma_g$.
Measurements from noise maps, if available, could therefore offer a middle ground if $N$ is low or
does not depend locally on the coil geometries (\eg SENSE or homodyne reconstruction) as shown in \cref{sec:synthetic_simulation}.

\cref{fig:invivo_estimation} shows a large range of estimates for $\sigma_g$ across methods.
In particular, the moments and maximum likelihood equations estimate smaller values of $\sigma_g$ than MPPCA and LANE, but larger than PIESNO,
while still recovering values of $N$ close to 1 and successfully discard voxels contaminated by multiband artifacts.
The correct value of $\sigma_g$ most likely sits between these two results as parallel MRI produces spatially varying noise profile, which is higher in the center
and not fully captured by the background signal, but the local estimation methods also overestimated $\sigma_g$ in our synthetic simulations.
In panel \textbf{A)}, MPPCA and LANE estimates of $\sigma_g$ with DWIs are likely affected by multiband artifacts as the median is larger than the signal level at \bval{5000}.
This indicates a possible overestimation as $\sigma_g$ should be lower than the measured signal at the highest b-value.
For PIESNO and the proposed methods, the median $\sigma_g$ is lower than the median of the reference \bval{5000} data.
An overestimation of $N$ could explain the low values of $\sigma_g$ estimated by the proposed methods
just as misestimation of $\eta$ by MPPCA and LANE could affect their respective estimate of $\sigma_g$ by balancing out the misestimated values.

\cref{fig:invivo_maps} shows the result of each method on a bias correction and denoising task on the Connectom dataset.
In panel \textbf{A)}, the standard deviation of the signal (bottom left panel) is increased after bias correction for LANE (green line)
and decreased (around the same level) for the other methods when compared to the uncorrected data (blue line).
The situation is similar after denoising, but to a lesser extent, while the moments, maximum likelihood equations and PIESNO follow the same signal level as the unprocessed data on average.
Regarding the mean of the signal itself, LANE is on average lower or close to 0 after bias correction, indicating potential degeneracies due to overestimation of $\sigma_g$.
From panels \textbf{B)} and \textbf{C)}, the results of all methods are visually similar except for LANE (especially at \bval{3000} and \bval{5000}),
indicating that the NLSAM denoising algorithm treated different values of $\sigma_g$ in the same way.
This is because the optimal regularized solution (which depends on $\sigma_g$) is piecewise constant \citep{St-Jean2016a,Tibshirani2011}
and can tolerate small deviations in $\sigma_g$.
Finally, MPPCA, the moments and maximum likelihood equations and PIESNO perform similarly, even if they estimated different values of $\sigma_g$ and $N$,
with MPPCA showing slightly lower signal intensity at \bval{5000}.
This could be due to the bias correction having a larger effect when $\sigma_g$ is larger, increasing the standard deviation of the resulting signal.
As shown in panel \textbf{C)}, the difference with the original dataset for MPPCA is lower than the proposed methods or PIESNO,
even though the estimated value of $\sigma_g$ was larger.


\section{Conclusions}
\label{sec:conclusion}

We presented a new, fully automated framework for characterizing the noise distribution from a diffusion MRI dataset
using the moments or maximum likelihood equations of the Gamma distribution.
The estimated parameters can be subsequently used for \eg bias correction and denoising as we have shown or diffusion models taking advantage of this information.
This requires only magnitude data, without the use of dedicated maps or parameters intrinsic to the reconstruction process, which may be challenging to obtain in practice.
The proposed framework is fast and robust to artifacts as voxels not adhering to the noise distribution can be automatically discarded using an outlier rejection step.
This makes the proposed methods also applicable on previously acquired datasets,
which may not carry the necessary information required by more advanced estimation methods.
Experiments using parallel MRI and multiband imaging on simulations, an acquired phantom and in vivo datasets have shown how modern acquisition techniques
complicate estimation of the signal distribution due to artifacts at high acceleration factor.
This issue can be alleviated with the use of noise only measurements or by limiting the acceleration factor to prevent signal leakage.
Moreover, different vendors implement different default reconstruction algorithms which leads to different signal distributions,
challenging the strategy of assuming a Rician distribution or approximations of $N$ based on the physical amount of channels in the receiver coil.
We also have shown how signal bias correction and denoising can tolerate some misestimation of the noise distribution using an in vivo dataset.
Noteworthy is that the theory we presented also applies to any other MRI weighting using large samples of magnitude data (\eg functional MRI, dynamic contrast enhanced MRI).
This could help multicenter studies or data sharing initiatives to include knowledge of the noise distribution in their analysis in a fully automated way
to better account for inter-scanner effects.


\appendix
\gdef\thesection{\Alph{section}} 
\makeatletter
\renewcommand\@seccntformat[1]{Appendix \csname the#1\endcsname.\hspace{0.5em}}
\makeatother

\section{Estimating parameters of the Gamma distribution}
\label{sec:appendix_gamma}

\paragraph{Estimation using the method of moments}

For any given distribution, we can estimate its parameters by relating the samples and the theoretical expression of its moments.
The Gamma distribution is parametrized as $\textit{Gamma}(\alpha, \beta)$ and has a probability distribution function of
\begin{equation}
    \textit{pdf}(t | \alpha, \beta) = \frac{t^{\alpha-1}}{\Gamma(\alpha)\beta^\alpha } \exp{(-t/\beta)}
    \label{eq:pdf_gamma}
\end{equation}
with $t, \alpha, \beta > 0$ and $\Gamma(x)$ the gamma function.
The first moments are analytically given by \citep[Chap.~5][]{papoulis1991,weisstein_gamma}.
\begin{gather}
    \mu_{gamma} = \alpha\beta,\,
    \sigma^2_{gamma} = \alpha\beta^2,\,
    \label{eq:moments_gamma}
\end{gather}

In this paper, the Gamma distribution parameters are $\textit{Gamma}(\alpha = N, \beta = 1)$ after the change of variable $t=m^2 / (2\sigma^2_g)$ for our particular case.
Since we have $\beta = 1$, this leads to a special case where the mean and variance are \emph{equal}
with a value of $\alpha = N$ and can be expressed only in terms of the magnitude signal $m$.
For simplicity, we will only use the mean $\mu_{gamma}$ and variance $\sigma^2_{gamma}$ to estimate the required parameters $N$ and $\sigma^2_g$, but higher order moments could also be used.
However, in practice, they might accumulate numerical errors due to the higher powers involved and are not used here since two equations are enough to estimate the two parameters.
Starting from the analytical expression given by \cref{eq:moments_gamma}, we have for the special case $\textit{Gamma}(N, 1)$
\begin{align}
    &\mu_{gamma} = \alpha,\, \sigma^2_{gamma} = \alpha \\
    \intertext{Which we can compute using the sample mean and sample variance formulas such that }
    &\alpha = \frac{1}{K}\sum_{k=1}^K t_k = \frac{1}{K}\sum_{k=1}^K t_k^2 - \left(\frac{1}{K}\sum_{k=1}^K t_k\right)^2\\
    \intertext{Substituting the equation for the moments in terms of $t=m^2/2\sigma^2_g$, we obtain}
    & \frac{1}{K}\sum_{k=1}^K \frac{m^2_k}{2\sigma_g^2} = \frac{1}{K}\sum_{k=1}^K \left(\frac{m^2_k}{2\sigma_g^2}\right)^2 - \left(\frac{1}{K}\sum_{k=1}^K \frac{m^2_k}{2\sigma_g^2}\right)^2 \\
    \Rightarrow \quad & \frac{1}{2K\sigma_g^2}\sum_{k=1}^K m^2_k = \frac{1}{4K\sigma_g^4}\sum_{k=1}^K m^4_k - \frac{1}{4K^2\sigma_g^4}\left(\sum_{k=1}^K m^2_k\right)^2 \\
    \Rightarrow \quad & \sum_{k=1}^K m^2_k = \frac{1}{2K\sigma_g^2}\left(K\sum_{k=1}^K m^4_k - \left(\sum_{k=1}^K m^2_k\right)^2\right) \\
    \Rightarrow \quad & 2K\sigma_g^2 = \frac{K \sum_{k=1}^K m^4_k - \left(\sum_{k=1}^K m^2_k\right)^2}{\sum_{k=1}^K m^2_k} \\
    \Rightarrow \quad & \sigma_g = \frac{1}{\sqrt{2K}} \sqrt{\frac{K \sum_{k=1}^K m^4_k - \left(\sum_{k=1}^K m^2_k\right)^2}{\sum_{k=1}^K m^2_k}}\\
    \Rightarrow \quad & \sigma_g = \frac{1}{\sqrt{2}} \sqrt{\frac{\sum_{k=1}^K m^4_k}{\sum_{k=1}^K m^2_k} - \frac{1}{K}\sum_{k=1}^K m^2_k}
    \label{eq:find_sigma_appendix}
\end{align}
Therefore, it is possible to estimate the Gaussian noise standard deviation using \cref{eq:find_sigma_appendix}
and the values of magnitude data $m_k$, assuming that the voxels considered here do not contain any object signal.
With the value of the noise variance $\sigma^2_g$ now known,
going back to the original Gamma distribution $\textit{Gamma}(\alpha = N, \beta = 1)$ yields the number of coils $N$ as previously shown by \cref{eq:find_N}
\begin{equation}
    N = \alpha = \mu_{gamma} = \frac{1}{2K\sigma^2_g}\sum_{k=1}^K m^2_k
    \label{eq:gamma_moment_N}
\end{equation}

\paragraph{Estimation using maximum likelihood equations}

An alternative to the method of moments to estimate parameters from a given distribution
is to solve the equations derived from its likelihood function for each unknown parameter.
Given a set of observed data, maximizing the likelihood function from a known distribution (or equivalently, the log of the likelihood function)
yields a set of equations to estimate its parameters.
For the $\textit{Gamma}(\alpha, \beta)$ distribution, maximizing the log likelihood by equating the partial derivative to 0 for each parameter yields \citep{Thom1958}
\begin{align}
    \frac{1}{K\beta}\sum_{k=1}^K t_k - \alpha &= 0
    \label{eq:gamma_ml1}\\
    \log(\beta) + \frac{d}{d\alpha} \log(\Gamma(\alpha)) - \frac{1}{K}\sum_{k=1}^K \log(t_k) &= 0
    \label{eq:gamma_ml2}
\end{align}
Since we have $\alpha = N$ and $\beta = 1$, in this special case \cref{eq:gamma_ml1} is the same as \cref{eq:gamma_moment_N}.

Combining \cref{eq:gamma_ml1,eq:gamma_ml2} yields an implicit equation to estimate $\sigma_g$, which can be written as
\begin{align}
    f(\sigma_g)  &= \psi\left(\frac{1}{2K\sigma^2_g}\sum_{k=1}^K m^2_k\right) - \frac{1}{K}\sum_{k=1}^K \log (m^2_k) + \log(2\sigma_g^2) = 0 \\
    \label{eq:sigma_ml}
    f'(\sigma_g) &= \review{\frac{-1}{K\sigma_g^3} \left[\psi'\left(\frac{1}{2K\sigma^2_g}\sum_{k=1}^K m^2_k\right) \sum_{k=1}^K m^2_k\right] + \frac{2}{\sigma_g}  = 0}
\end{align}
and \cref{eq:gamma_ml2} can be rewritten as an implicit equation of $N$
\begin{align}
    f(N)  &= \psi(N) - \frac{1}{K}\sum_{k=1}^K \log (m^2_k / 2\sigma_g^2) = 0 \\
    \label{eq:gamma_ml_N}
    f'(N) &= \psi'(N) = 0
\end{align}
where $\psi(x) = \frac{d}{dx}\log(\Gamma(x))$ is the digamma function and $\psi'$ is the derivative of $\psi$, called the polygamma function.
\cref{eq:sigma_ml,eq:gamma_ml_N} can be solved numerically using Newton's method provided we have a starting estimate $x_0$.
The update rule for Newton's method at iteration $n$ is therefore
\begin{equation}
    x_{n+1} = x_{n} - \frac{f(x_n)}{f'(x_n)}
    \label{eq:newton}
\end{equation}
For the first iteration, a starting estimate $x_0$ to approximate the solution is needed.
For \cref{eq:sigma_ml}, we use $x_0 = \sigma_m$, \review{where $\sigma_m$ is the sample standard deviation of the identified noise only voxels.}
A starting estimate for \cref{eq:gamma_ml_N} is given by \citep{Minka2003}
considering $y = \frac{1}{K}\sum_{k=1}^K \log (m^2_k / 2\sigma_g^2)$.
\begin{equation}
    x_0 = \psi^{-1}(y)\approx
    \begin{dcases}
        \exp(y)+ 1/2        & \text{if } y \geq -2.22\\
        -1 / (y + \psi(1))  & \text{if } y <    -2.22\\
    \end{dcases}
\end{equation}
In practice, we have observed that 5 iterations of \cref{eq:newton}
were sufficient to reach $\abs{x_{n} - x_{n-1}} < 10^{-13}$.

\section{Generalized bias correction}
\label{sec:appendix_general_bias}

As an application which requires knowledge of both $\sigma_g$ and $N$, we now present a general version for non integer values of $N$ of the signal bias correction from \citet{Koay2006,Koay2009a}.
The correction factor $\xi(\eta | \sigma_g, N)$ can be used to obtain $\eta$ from the magnitude measurement $m_N$ given the values of $\sigma_g$ and $N$ such that
\begin{gather}
    \xi(\eta | \sigma_g, N) = 2N + \frac{\eta^2}{\sigma_g^2} - \left(\beta_N \, {}_1F_1 \left(-1/2, N, \frac{-\eta^2}{2\sigma_g^2} \right)\right)^2
    \label{eq:factor_xi}
\end{gather}
where ${}_1F_1$ is Kummer's function of the first kind.
By defining

\begin{align}
    \beta_N &= \sqrt{\pi / 2} \, \binom{N - 1/2}{1/2}\\
            &= \sqrt{\pi / 2} \left(\frac{\Gamma(N + 1/2)}{\Gamma(3/2)\Gamma(N)}\right)\\
            &= \sqrt{2} \left(\frac{\Gamma(N + 1/2)}{\Gamma(N)}\right)
    \label{eq:beta}
\end{align}
where $\binom{n}{k}$ is a binomial coefficient, we obtain a generalized version of \cref{eq:factor_xi} which can now be applied for non integer values of $N$,
such as in the case of a half Gaussian signal distribution $(N = 0.5)$ which occurs when employing half-Fourier reconstruction techniques \citep{Dietrich2008}.
Estimation of $\eta$ is finally done with

\begin{equation}
    \eta = \sqrt{\hat{m}^2 + (\xi(\eta | \sigma_g, N) - 2N)\sigma_g^2}
    \label{eq:find_eta}
\end{equation}
where $\hat{m}$ is an estimate of the first moment of a noncentral chi variable and is estimated from a spherical harmonics fit of order 6
\review{on the DWI datasets for each shell} in the present work.
\cref{eq:find_eta} can be solved iteratively w.r.t. $\eta$ until convergence, see \citep{Koay2009a} for further implementation details.

\section{Automated identification of noise only voxels}
\label{sec:appendix_algo}

This appendix outlines the proposed algorithm and details for a practical implementation.
Our implementation is also freely available at \url{https://github.com/samuelstjean/autodmri} \citep{St-Jean2019d}
and will be a part of ExploreDTI \citep{Leemans2009a}.
The synthetic and acquired datasets used in this manuscript are also available \citep{St-Jean2018d}.

\begin{algorithm}
\SetAlgoLined
\KwData{4D DWIs data, probability level $p = 0.05$, length of the search \review{grid} $l = 50$, $N_{min} = 1$, $N_{max} = 12$}
\KwResult{$\sigma_g$, $N$, mask of background only voxels}
    Compute the $median$ of the whole dataset\;
    \ForEach{2D Slice of the 4D dataset}
    {
        Compute the upper bound $\sigma_{g_{max}} = \textit{median} / \sqrt{2\, \textit{icdf}(N_{max}, 1/2)}$\;
        Compute the search interval $\Phi = [1 \sigma_{g_{max}} / l, 2 \sigma_{g_{max}} / l, \dotsc , l \sigma_{g_{max}} / l]$\;
        \While{$\sigma_g, N$ not converged}{
            Compute $\lambda_{-} = \textit{icdf}(\review{N}, p/2)$ and $\lambda_{+} = \textit{icdf}(\review{N}, 1-p/2)$\;
            \ForEach{$\sigma_{candidate} \in \Phi$}{
                Apply change of variable $t = \frac{data^2}{2\sigma_{candidate}^2}$\;
                Find voxels from the gamma distribution\;
                mask\_current $\displaystyle = \left(\lambda_{-} \le \sum_{k=1}^K t_k\right) \bigcap \displaystyle\left(\sum_{k=1}^K t_k \le \lambda_{+}\right)$\;
                \If{number of voxels in mask\_current $>$ mask}{
                    mask = mask\_current\;
                }
            }
            Compute $\sigma_g$ with the voxels inside the mask using \cref{eq:find_sigma} or \cref{eq:gamma_simplified_ml_sigma}\;
            Compute $N$ with the voxels inside the mask using \cref{eq:find_N} or \cref{eq:gamma_simplified_ml_N}\;
            Set $N_{min} = N_{max} = N$\;
            Set $\Phi = [0.95 \sigma_{g}, 0.96 \sigma_{g}, \dotsc, 1.05 \sigma_{g}]$\;
        }
    }
\caption{Main algorithm to identify voxels belonging to the Gamma distribution}
\label{alg:main}
\end{algorithm}

\section*{Acknowledgements}

\review{We would like to thank Michael Paquette for useful comments and discussion.}
The authors have declared no conflict of interest.
The funding agencies were not involved in the design, data collection nor interpretation of this study.
This research is supported by the Netherlands Organization for Scientific Research (NWO), Grant/Award Number: VIDI 639.072.411.
Samuel St-Jean is supported by the Fonds de recherche du Québec - Nature et technologies (FRQNT) (Dossier 192865) and
Chantal M. W. Tax is supported by the Netherlands Organization for Scientific Research (NWO), Grant/Award Number: Rubicon 680-50-1527.

\bibliography{stjean_etal_2019}

\begin{thebibliography}{54}
\expandafter\ifx\csname natexlab\endcsname\relax\def\natexlab#1{#1}\fi
\providecommand{\url}[1]{\texttt{#1}}
\providecommand{\href}[2]{#2}
\providecommand{\path}[1]{#1}
\providecommand{\DOIprefix}{doi:}
\providecommand{\ArXivprefix}{arXiv:}
\providecommand{\URLprefix}{URL: }
\providecommand{\Pubmedprefix}{pmid:}
\providecommand{\doi}[1]{\href{http://dx.doi.org/#1}{\path{#1}}}
\providecommand{\Pubmed}[1]{\href{pmid:#1}{\path{#1}}}
\providecommand{\bibinfo}[2]{#2}
\ifx\xfnm\relax \def\xfnm[#1]{\unskip,\space#1}\fi
\bibitem[{Aja-Fern{\'{a}}ndez et~al.(2013)Aja-Fern{\'{a}}ndez, Brion and
  Trist{\'{a}}n-Vega}]{Aja-Fernandez2013}
\bibinfo{author}{Aja-Fern{\'{a}}ndez, S.}, \bibinfo{author}{Brion, V.},
  \bibinfo{author}{Trist{\'{a}}n-Vega, A.}, \bibinfo{year}{2013}.
\newblock \bibinfo{title}{{Effective noise estimation and filtering from
  correlated multiple-coil MR data.}}
\newblock \bibinfo{journal}{Magnetic resonance imaging} \bibinfo{volume}{31},
  \bibinfo{pages}{272--85}.
\newblock \DOIprefix\doi{10.1016/j.mri.2012.07.006}.
\bibitem[{Aja-Fern{\'{a}}ndez and
  Trist{\'{a}}n-Vega(2015)}]{Aja-Fernandez2015a}
\bibinfo{author}{Aja-Fern{\'{a}}ndez, S.}, \bibinfo{author}{Trist{\'{a}}n-Vega,
  A.}, \bibinfo{year}{2015}.
\newblock \bibinfo{title}{{A review on statistical noise models for Magnetic
  Resonance Imaging}}.
\newblock \bibinfo{journal}{LPI, ETSI Telecomunicacion, Universidad de
  Valladolid, Spain, Tech. Rep} , \bibinfo{pages}{1--30}.
\bibitem[{Aja-Fern{\'{a}}ndez et~al.(2009)Aja-Fern{\'{a}}ndez,
  Trist{\'{a}}n-Vega and Alberola-L{\'{o}}pez}]{Aja-Fernandez2009}
\bibinfo{author}{Aja-Fern{\'{a}}ndez, S.}, \bibinfo{author}{Trist{\'{a}}n-Vega,
  A.}, \bibinfo{author}{Alberola-L{\'{o}}pez, C.}, \bibinfo{year}{2009}.
\newblock \bibinfo{title}{{Noise estimation in single- and multiple-coil
  magnetic resonance data based on statistical models.}}
\newblock \bibinfo{journal}{Magnetic resonance imaging} \bibinfo{volume}{27},
  \bibinfo{pages}{1397--409}.
\newblock \URLprefix \url{http://www.ncbi.nlm.nih.gov/pubmed/19570640},
  \DOIprefix\doi{10.1016/j.mri.2009.05.025}.
\bibitem[{Aja-Fern{\'{a}}ndez et~al.(2014)Aja-Fern{\'{a}}ndez,
  Vegas-S{\'{a}}nchez-Ferrero and Trist{\'{a}}n-Vega}]{Aja-Fernandez2014}
\bibinfo{author}{Aja-Fern{\'{a}}ndez, S.},
  \bibinfo{author}{Vegas-S{\'{a}}nchez-Ferrero, G.},
  \bibinfo{author}{Trist{\'{a}}n-Vega, A.}, \bibinfo{year}{2014}.
\newblock \bibinfo{title}{{Noise estimation in parallel MRI: GRAPPA and
  SENSE.}}
\newblock \bibinfo{journal}{Magnetic resonance imaging} \bibinfo{volume}{32},
  \bibinfo{pages}{281--90}.
\newblock \DOIprefix\doi{10.1016/j.mri.2013.12.001}.
\bibitem[{Barth et~al.(2016)Barth, Breuer, Koopmans, Norris and
  Poser}]{Barth2016}
\bibinfo{author}{Barth, M.}, \bibinfo{author}{Breuer, F.},
  \bibinfo{author}{Koopmans, P.J.}, \bibinfo{author}{Norris, D.G.},
  \bibinfo{author}{Poser, B.A.}, \bibinfo{year}{2016}.
\newblock \bibinfo{title}{{Simultaneous multislice (SMS) imaging techniques.}}
\newblock \bibinfo{journal}{Magnetic resonance in medicine}
  \bibinfo{volume}{75}, \bibinfo{pages}{63--81}.
\newblock \DOIprefix\doi{10.1002/mrm.25897}.
\bibitem[{Bernstein et~al.(2004)Bernstein, King and Zhou}]{Bernstein2004}
\bibinfo{author}{Bernstein, M.A.}, \bibinfo{author}{King, K.F.},
  \bibinfo{author}{Zhou, X.J.}, \bibinfo{year}{2004}.
\newblock \bibinfo{title}{{Handbook of MRI Pulse Sequences}}.
\newblock \bibinfo{publisher}{Elsevier Science}.
\bibitem[{Caruyer et~al.(2014)Caruyer, Daducci, Descoteaux, Houde, Thiran and
  Verma}]{Caruyer2014}
\bibinfo{author}{Caruyer, E.}, \bibinfo{author}{Daducci, A.},
  \bibinfo{author}{Descoteaux, M.}, \bibinfo{author}{Houde, J.C.},
  \bibinfo{author}{Thiran, J.P.}, \bibinfo{author}{Verma, R.},
  \bibinfo{year}{2014}.
\newblock \bibinfo{title}{{Phantomas: a flexible software library to simulate
  diffusion MR phantoms}}, in: \bibinfo{booktitle}{International Symposium on
  Magnetic Resonance in Medicine (ISMRM'14)}, p. \bibinfo{pages}{6407}.
\bibitem[{Collier et~al.(2018)Collier, Veraart, Jeurissen, Vanhevel, Pullens,
  Parizel, den Dekker and Sijbers}]{Collier2018}
\bibinfo{author}{Collier, Q.}, \bibinfo{author}{Veraart, J.},
  \bibinfo{author}{Jeurissen, B.}, \bibinfo{author}{Vanhevel, F.},
  \bibinfo{author}{Pullens, P.}, \bibinfo{author}{Parizel, P.M.},
  \bibinfo{author}{den Dekker, A.J.}, \bibinfo{author}{Sijbers, J.},
  \bibinfo{year}{2018}.
\newblock \bibinfo{title}{{Diffusion kurtosis imaging with free water
  elimination: A bayesian estimation approach}}.
\newblock \bibinfo{journal}{Magnetic Resonance in Medicine}
  \bibinfo{volume}{80}, \bibinfo{pages}{802--813}.
\newblock \DOIprefix\doi{10.1002/mrm.27075}.
\bibitem[{Constantinides et~al.(1997)Constantinides, Atalar and
  McVeigh}]{Constantinides1997}
\bibinfo{author}{Constantinides, C.D.}, \bibinfo{author}{Atalar, E.},
  \bibinfo{author}{McVeigh, E.R.}, \bibinfo{year}{1997}.
\newblock \bibinfo{title}{{Signal-to-noise measurements in magnitude images
  from NMR phased arrays}}.
\newblock \bibinfo{journal}{Magnetic Resonance in Medicine}
  \bibinfo{volume}{38}, \bibinfo{pages}{852--857}.
\newblock \DOIprefix\doi{10.1002/mrm.1910380524}.
\bibitem[{Descoteaux et~al.(2007)Descoteaux, Angelino, Fitzgibbons and
  Deriche}]{Descoteaux2007b}
\bibinfo{author}{Descoteaux, M.}, \bibinfo{author}{Angelino, E.},
  \bibinfo{author}{Fitzgibbons, S.}, \bibinfo{author}{Deriche, R.},
  \bibinfo{year}{2007}.
\newblock \bibinfo{title}{{Regularized, fast, and robust analytical Q-ball
  imaging}}.
\newblock \bibinfo{journal}{Magnetic Resonance in Medicine}
  \bibinfo{volume}{58}, \bibinfo{pages}{497--510}.
\newblock \DOIprefix\doi{10.1002/mrm.21277}.
\bibitem[{Dietrich et~al.(2008)Dietrich, Raya, Reeder, Ingrisch, Reiser and
  Schoenberg}]{Dietrich2008}
\bibinfo{author}{Dietrich, O.}, \bibinfo{author}{Raya, J.G.},
  \bibinfo{author}{Reeder, S.B.}, \bibinfo{author}{Ingrisch, M.},
  \bibinfo{author}{Reiser, M.F.}, \bibinfo{author}{Schoenberg, S.O.},
  \bibinfo{year}{2008}.
\newblock \bibinfo{title}{{Influence of multichannel combination, parallel
  imaging and other reconstruction techniques on MRI noise characteristics.}}
\newblock \bibinfo{journal}{Magnetic resonance imaging} \bibinfo{volume}{26},
  \bibinfo{pages}{754--62}.
\newblock \DOIprefix\doi{10.1016/j.mri.2008.02.001}.
\bibitem[{Duchesne et~al.(2019)Duchesne, Chouinard, Potvin, Fonov, Khademi,
  Bartha, Bellec, Collins, Descoteaux, Hoge, McCreary, Ramirez, Scott, Smith,
  Strother and Black}]{Duchesne2019}
\bibinfo{author}{Duchesne, S.}, \bibinfo{author}{Chouinard, I.},
  \bibinfo{author}{Potvin, O.}, \bibinfo{author}{Fonov, V.S.},
  \bibinfo{author}{Khademi, A.}, \bibinfo{author}{Bartha, R.},
  \bibinfo{author}{Bellec, P.}, \bibinfo{author}{Collins, D.L.},
  \bibinfo{author}{Descoteaux, M.}, \bibinfo{author}{Hoge, R.},
  \bibinfo{author}{McCreary, C.R.}, \bibinfo{author}{Ramirez, J.},
  \bibinfo{author}{Scott, C.J.}, \bibinfo{author}{Smith, E.E.},
  \bibinfo{author}{Strother, S.C.}, \bibinfo{author}{Black, S.E.},
  \bibinfo{year}{2019}.
\newblock \bibinfo{title}{{The Canadian Dementia Imaging Protocol: Harmonizing
  National Cohorts}}.
\newblock \bibinfo{journal}{Journal of Magnetic Resonance Imaging}
  \bibinfo{volume}{49}, \bibinfo{pages}{456--465}.
\newblock \DOIprefix\doi{10.1002/jmri.26197}.
\bibitem[{Emaus et~al.(2015)Emaus, Bakker, Peeters, Loo, Mann, de~Jong,
  Bisschops, Veltman, Duvivier, Lobbes, Pijnappel, Karssemeijer, de~Koning,
  van~den Bosch, Monninkhof, Mali, Veldhuis and van Gils}]{Emaus2015}
\bibinfo{author}{Emaus, M.J.}, \bibinfo{author}{Bakker, M.F.},
  \bibinfo{author}{Peeters, P.H.M.}, \bibinfo{author}{Loo, C.E.},
  \bibinfo{author}{Mann, R.M.}, \bibinfo{author}{de~Jong, M.D.F.},
  \bibinfo{author}{Bisschops, R.H.C.}, \bibinfo{author}{Veltman, J.},
  \bibinfo{author}{Duvivier, K.M.}, \bibinfo{author}{Lobbes, M.B.I.},
  \bibinfo{author}{Pijnappel, R.M.}, \bibinfo{author}{Karssemeijer, N.},
  \bibinfo{author}{de~Koning, H.J.}, \bibinfo{author}{van~den Bosch, M.A.A.J.},
  \bibinfo{author}{Monninkhof, E.M.}, \bibinfo{author}{Mali, W.P.T.M.},
  \bibinfo{author}{Veldhuis, W.B.}, \bibinfo{author}{van Gils, C.H.},
  \bibinfo{year}{2015}.
\newblock \bibinfo{title}{{MR Imaging as an Additional Screening Modality for
  the Detection of Breast Cancer in Women Aged 50–75 Years with Extremely
  Dense Breasts: The DENSE Trial Study Design}}.
\newblock \bibinfo{journal}{Radiology} \bibinfo{volume}{277},
  \bibinfo{pages}{527--537}.
\newblock \DOIprefix\doi{10.1148/radiol.2015141827}.
\bibitem[{Griswold et~al.(2002)Griswold, Jakob, Heidemann, Nittka, Jellus,
  Wang, Kiefer and Haase}]{Griswold2002}
\bibinfo{author}{Griswold, M.A.}, \bibinfo{author}{Jakob, P.M.},
  \bibinfo{author}{Heidemann, R.M.}, \bibinfo{author}{Nittka, M.},
  \bibinfo{author}{Jellus, V.}, \bibinfo{author}{Wang, J.},
  \bibinfo{author}{Kiefer, B.}, \bibinfo{author}{Haase, A.},
  \bibinfo{year}{2002}.
\newblock \bibinfo{title}{{Generalized Autocalibrating Partially Parallel
  Acquisitions (GRAPPA)}}.
\newblock \bibinfo{journal}{Magnetic Resonance in Medicine}
  \bibinfo{volume}{47}, \bibinfo{pages}{1202--1210}.
\newblock \DOIprefix\doi{10.1002/mrm.10171}.
\bibitem[{Gudbjartsson and Patz(1995)}]{Gudbjartsson1995}
\bibinfo{author}{Gudbjartsson, H.}, \bibinfo{author}{Patz, S.},
  \bibinfo{year}{1995}.
\newblock \bibinfo{title}{{The Rician distribution of noisy mri data}}.
\newblock \bibinfo{journal}{Magnetic Resonance in Medicine}
  \bibinfo{volume}{34}, \bibinfo{pages}{910--914}.
\newblock \DOIprefix\doi{10.1002/mrm.1910340618}.
\bibitem[{Heidemann et~al.(2012)Heidemann, Anwander, Feiweier, Kn{\"{o}}sche
  and Turner}]{Heidemann2012}
\bibinfo{author}{Heidemann, R.M.}, \bibinfo{author}{Anwander, A.},
  \bibinfo{author}{Feiweier, T.}, \bibinfo{author}{Kn{\"{o}}sche, T.R.},
  \bibinfo{author}{Turner, R.}, \bibinfo{year}{2012}.
\newblock \bibinfo{title}{{k-space and q-space: Combining ultra-high spatial
  and angular resolution in diffusion imaging using ZOOPPA at 7T}}.
\newblock \bibinfo{journal}{NeuroImage} \bibinfo{volume}{60},
  \bibinfo{pages}{967--978}.
\newblock \DOIprefix\doi{10.1016/j.neuroimage.2011.12.081}.
\bibitem[{Holdsworth et~al.(2019)Holdsworth, O'Halloran and
  Setsompop}]{Holdsworth2019}
\bibinfo{author}{Holdsworth, S.J.}, \bibinfo{author}{O'Halloran, R.},
  \bibinfo{author}{Setsompop, K.}, \bibinfo{year}{2019}.
\newblock \bibinfo{title}{{The quest for high spatial resolution
  diffusion-weighted imaging of the human brain in vivo}}.
\newblock \bibinfo{journal}{NMR in Biomedicine} \bibinfo{volume}{32},
  \bibinfo{pages}{e4056}.
\newblock \DOIprefix\doi{10.1002/nbm.4056}.
\bibitem[{Hutchinson et~al.(2017)Hutchinson, Avram, Irfanoglu, Koay, Barnett,
  Komlosh, {\"{O}}zarslan, Schwerin, Juliano and Pierpaoli}]{Hutchinson2017}
\bibinfo{author}{Hutchinson, E.B.}, \bibinfo{author}{Avram, A.V.},
  \bibinfo{author}{Irfanoglu, M.O.}, \bibinfo{author}{Koay, C.G.},
  \bibinfo{author}{Barnett, A.S.}, \bibinfo{author}{Komlosh, M.E.},
  \bibinfo{author}{{\"{O}}zarslan, E.}, \bibinfo{author}{Schwerin, S.C.},
  \bibinfo{author}{Juliano, S.L.}, \bibinfo{author}{Pierpaoli, C.},
  \bibinfo{year}{2017}.
\newblock \bibinfo{title}{{Analysis of the effects of noise, DWI sampling, and
  value of assumed parameters in diffusion MRI models.}}
\newblock \bibinfo{journal}{Magnetic resonance in medicine}
  \bibinfo{volume}{78}, \bibinfo{pages}{1767--1780}.
\newblock \DOIprefix\doi{10.1002/mrm.26575}.
\bibitem[{Koay and Basser(2006)}]{Koay2006}
\bibinfo{author}{Koay, C.G.}, \bibinfo{author}{Basser, P.J.},
  \bibinfo{year}{2006}.
\newblock \bibinfo{title}{{Analytically exact correction scheme for signal
  extraction from noisy magnitude MR signals}}.
\newblock \bibinfo{journal}{Journal of Magnetic Resonance}
  \bibinfo{volume}{179}, \bibinfo{pages}{317--322}.
\newblock \DOIprefix\doi{10.1016/j.jmr.2006.01.016}.
\bibitem[{Koay et~al.(2009a)Koay, {\"{O}}zarslan and Basser}]{Koay2009a}
\bibinfo{author}{Koay, C.G.}, \bibinfo{author}{{\"{O}}zarslan, E.},
  \bibinfo{author}{Basser, P.J.}, \bibinfo{year}{2009}a.
\newblock \bibinfo{title}{{A signal transformational framework for breaking the
  noise floor and its applications in MRI}}.
\newblock \bibinfo{journal}{Journal of Magnetic Resonance}
  \bibinfo{volume}{197}, \bibinfo{pages}{108--119}.
\newblock \DOIprefix\doi{10.1016/j.jmr.2008.11.015}.
\bibitem[{Koay et~al.(2009b)Koay, {\"{O}}zarslan and Pierpaoli}]{Koay2009b}
\bibinfo{author}{Koay, C.G.}, \bibinfo{author}{{\"{O}}zarslan, E.},
  \bibinfo{author}{Pierpaoli, C.}, \bibinfo{year}{2009}b.
\newblock \bibinfo{title}{{Probabilistic Identification and Estimation of Noise
  (PIESNO): A self-consistent approach and its applications in MRI}}.
\newblock \bibinfo{journal}{Journal of Magnetic Resonance}
  \bibinfo{volume}{199}, \bibinfo{pages}{94--103}.
\newblock \DOIprefix\doi{10.1016/j.jmr.2009.03.005}.
\bibitem[{Landman et~al.(2007)Landman, Bazin and Prince}]{Landman2007a}
\bibinfo{author}{Landman, B.}, \bibinfo{author}{Bazin, P.L.},
  \bibinfo{author}{Prince, J.}, \bibinfo{year}{2007}.
\newblock \bibinfo{title}{{Diffusion tensor estimation by maximizing Rician
  likelihood}}.
\newblock \bibinfo{journal}{Proceedings of the IEEE International Conference on
  Computer Vision} ,
  \bibinfo{pages}{1--8}\DOIprefix\doi{10.1109/ICCV.2007.4409140}.
\bibitem[{Leemans et~al.(2009)Leemans, Jeurissen, Sijbers and
  Jones}]{Leemans2009a}
\bibinfo{author}{Leemans, A.}, \bibinfo{author}{Jeurissen, B.},
  \bibinfo{author}{Sijbers, J.}, \bibinfo{author}{Jones, D.},
  \bibinfo{year}{2009}.
\newblock \bibinfo{title}{{ExploreDTI: a graphical toolbox for processing,
  analyzing, and visualizing diffusion MR data}}.
\newblock \bibinfo{journal}{Proceedings 17th Scientific Meeting, International
  Society for Magnetic Resonance in Medicine} \bibinfo{volume}{17},
  \bibinfo{pages}{3537}.
\bibitem[{L{\o}nning et~al.(2019)L{\o}nning, Putzky, Sonke, Reneman, Caan and
  Welling}]{Lonning2019}
\bibinfo{author}{L{\o}nning, K.}, \bibinfo{author}{Putzky, P.},
  \bibinfo{author}{Sonke, J.J.}, \bibinfo{author}{Reneman, L.},
  \bibinfo{author}{Caan, M.W.}, \bibinfo{author}{Welling, M.},
  \bibinfo{year}{2019}.
\newblock \bibinfo{title}{{Recurrent inference machines for reconstructing
  heterogeneous MRI data}}.
\newblock \bibinfo{journal}{Medical Image Analysis} \bibinfo{volume}{53},
  \bibinfo{pages}{64--78}.
\newblock \DOIprefix\doi{10.1016/j.media.2019.01.005}.
\bibitem[{Lustig et~al.(2007)Lustig, Donoho and Pauly}]{Lustig2007}
\bibinfo{author}{Lustig, M.}, \bibinfo{author}{Donoho, D.},
  \bibinfo{author}{Pauly, J.M.}, \bibinfo{year}{2007}.
\newblock \bibinfo{title}{{Sparse MRI: The application of compressed sensing
  for rapid MR imaging}}.
\newblock \bibinfo{journal}{Magnetic Resonance in Medicine}
  \bibinfo{volume}{58}, \bibinfo{pages}{1182--1195}.
\newblock \DOIprefix\doi{10.1002/mrm.21391}.
\bibitem[{Maier-Hein et~al.(2017)Maier-Hein, Neher, Houde, C{\^{o}}t{\'{e}},
  Garyfallidis, Zhong, Chamberland, Yeh, Lin, Ji, Reddick, Glass, Chen, Feng,
  Gao, Wu, Ma, Renjie, Li, Westin, Deslauriers-Gauthier, Gonz{\'{a}}lez,
  Paquette, St-Jean, Girard, Rheault, Sidhu, Tax, Guo, Mesri, D{\'{a}}vid,
  Froeling, Heemskerk, Leemans, Bor{\'{e}}, Pinsard, Bedetti, Desrosiers,
  Brambati, Doyon, Sarica, Vasta, Cerasa, Quattrone, Yeatman, Khan, Hodges,
  Alexander, Romascano, Barakovic, Aur{\'{i}}a, Esteban, Lemkaddem, Thiran,
  Cetingul, Odry, Mailhe, Nadar, Pizzagalli, Prasad, Villalon-Reina, Galvis,
  Thompson, Requejo, Laguna, Lacerda, Barrett, Dell'Acqua, Catani, Petit,
  Caruyer, Daducci, Dyrby, Holland-Letz, Hilgetag, Stieltjes and
  Descoteaux}]{Maier-Hein2017}
\bibinfo{author}{Maier-Hein, K.H.}, \bibinfo{author}{Neher, P.F.},
  \bibinfo{author}{Houde, J.C.}, \bibinfo{author}{C{\^{o}}t{\'{e}}, M.A.},
  \bibinfo{author}{Garyfallidis, E.}, \bibinfo{author}{Zhong, J.},
  \bibinfo{author}{Chamberland, M.}, \bibinfo{author}{Yeh, F.C.},
  \bibinfo{author}{Lin, Y.C.}, \bibinfo{author}{Ji, Q.},
  \bibinfo{author}{Reddick, W.E.}, \bibinfo{author}{Glass, J.O.},
  \bibinfo{author}{Chen, D.Q.}, \bibinfo{author}{Feng, Y.},
  \bibinfo{author}{Gao, C.}, \bibinfo{author}{Wu, Y.}, \bibinfo{author}{Ma,
  J.}, \bibinfo{author}{Renjie, H.}, \bibinfo{author}{Li, Q.},
  \bibinfo{author}{Westin, C.F.}, \bibinfo{author}{Deslauriers-Gauthier, S.},
  \bibinfo{author}{Gonz{\'{a}}lez, J.O.O.}, \bibinfo{author}{Paquette, M.},
  \bibinfo{author}{St-Jean, S.}, \bibinfo{author}{Girard, G.},
  \bibinfo{author}{Rheault, F.}, \bibinfo{author}{Sidhu, J.},
  \bibinfo{author}{Tax, C.M.W.}, \bibinfo{author}{Guo, F.},
  \bibinfo{author}{Mesri, H.Y.}, \bibinfo{author}{D{\'{a}}vid, S.},
  \bibinfo{author}{Froeling, M.}, \bibinfo{author}{Heemskerk, A.M.},
  \bibinfo{author}{Leemans, A.}, \bibinfo{author}{Bor{\'{e}}, A.},
  \bibinfo{author}{Pinsard, B.}, \bibinfo{author}{Bedetti, C.},
  \bibinfo{author}{Desrosiers, M.}, \bibinfo{author}{Brambati, S.},
  \bibinfo{author}{Doyon, J.}, \bibinfo{author}{Sarica, A.},
  \bibinfo{author}{Vasta, R.}, \bibinfo{author}{Cerasa, A.},
  \bibinfo{author}{Quattrone, A.}, \bibinfo{author}{Yeatman, J.},
  \bibinfo{author}{Khan, A.R.}, \bibinfo{author}{Hodges, W.},
  \bibinfo{author}{Alexander, S.}, \bibinfo{author}{Romascano, D.},
  \bibinfo{author}{Barakovic, M.}, \bibinfo{author}{Aur{\'{i}}a, A.},
  \bibinfo{author}{Esteban, O.}, \bibinfo{author}{Lemkaddem, A.},
  \bibinfo{author}{Thiran, J.P.}, \bibinfo{author}{Cetingul, H.E.},
  \bibinfo{author}{Odry, B.L.}, \bibinfo{author}{Mailhe, B.},
  \bibinfo{author}{Nadar, M.S.}, \bibinfo{author}{Pizzagalli, F.},
  \bibinfo{author}{Prasad, G.}, \bibinfo{author}{Villalon-Reina, J.E.},
  \bibinfo{author}{Galvis, J.}, \bibinfo{author}{Thompson, P.M.},
  \bibinfo{author}{Requejo, F.D.S.}, \bibinfo{author}{Laguna, P.L.},
  \bibinfo{author}{Lacerda, L.M.}, \bibinfo{author}{Barrett, R.},
  \bibinfo{author}{Dell'Acqua, F.}, \bibinfo{author}{Catani, M.},
  \bibinfo{author}{Petit, L.}, \bibinfo{author}{Caruyer, E.},
  \bibinfo{author}{Daducci, A.}, \bibinfo{author}{Dyrby, T.B.},
  \bibinfo{author}{Holland-Letz, T.}, \bibinfo{author}{Hilgetag, C.C.},
  \bibinfo{author}{Stieltjes, B.}, \bibinfo{author}{Descoteaux, M.},
  \bibinfo{year}{2017}.
\newblock \bibinfo{title}{{The challenge of mapping the human connectome based
  on diffusion tractography}}.
\newblock \bibinfo{journal}{Nature Communications} \bibinfo{volume}{8},
  \bibinfo{pages}{1349}.
\newblock \DOIprefix\doi{10.1038/s41467-017-01285-x}.
\bibitem[{Minka(2012)}]{Minka2003}
\bibinfo{author}{Minka, T.P.}, \bibinfo{year}{2012}.
\newblock \bibinfo{title}{{Estimating a Dirichlet distribution}}.
\newblock \bibinfo{journal}{Microsoft research} .
\bibitem[{Mirzaalian et~al.(2018)Mirzaalian, Ning, Savadjiev, Pasternak, Bouix,
  Michailovich, Karmacharya, Grant, Marx, Morey, Flashman, George, McAllister,
  Andaluz, Shutter, Coimbra, Zafonte, Coleman, Kubicki, Westin, Stein, Shenton
  and Rathi}]{Mirzaalian2018}
\bibinfo{author}{Mirzaalian, H.}, \bibinfo{author}{Ning, L.},
  \bibinfo{author}{Savadjiev, P.}, \bibinfo{author}{Pasternak, O.},
  \bibinfo{author}{Bouix, S.}, \bibinfo{author}{Michailovich, O.},
  \bibinfo{author}{Karmacharya, S.}, \bibinfo{author}{Grant, G.},
  \bibinfo{author}{Marx, C.E.}, \bibinfo{author}{Morey, R.A.},
  \bibinfo{author}{Flashman, L.A.}, \bibinfo{author}{George, M.S.},
  \bibinfo{author}{McAllister, T.W.}, \bibinfo{author}{Andaluz, N.},
  \bibinfo{author}{Shutter, L.}, \bibinfo{author}{Coimbra, R.},
  \bibinfo{author}{Zafonte, R.D.}, \bibinfo{author}{Coleman, M.J.},
  \bibinfo{author}{Kubicki, M.}, \bibinfo{author}{Westin, C.F.},
  \bibinfo{author}{Stein, M.B.}, \bibinfo{author}{Shenton, M.E.},
  \bibinfo{author}{Rathi, Y.}, \bibinfo{year}{2018}.
\newblock \bibinfo{title}{{Multi-site harmonization of diffusion MRI data in a
  registration framework}}.
\newblock \bibinfo{journal}{Brain Imaging and Behavior} \bibinfo{volume}{12},
  \bibinfo{pages}{284--295}.
\newblock \DOIprefix\doi{10.1007/s11682-016-9670-y}.
\bibitem[{Moeller et~al.(2010)Moeller, Yacoub, Olman, Auerbach, Strupp, Harel
  and Uǧurbil}]{Moeller2010}
\bibinfo{author}{Moeller, S.}, \bibinfo{author}{Yacoub, E.},
  \bibinfo{author}{Olman, C.A.}, \bibinfo{author}{Auerbach, E.},
  \bibinfo{author}{Strupp, J.}, \bibinfo{author}{Harel, N.},
  \bibinfo{author}{Uǧurbil, K.}, \bibinfo{year}{2010}.
\newblock \bibinfo{title}{{Multiband multislice GE-EPI at 7 tesla, with 16-fold
  acceleration using partial parallel imaging with application to high spatial
  and temporal whole-brain FMRI}}.
\newblock \bibinfo{journal}{Magnetic Resonance in Medicine}
  \bibinfo{volume}{63}, \bibinfo{pages}{1144--1153}.
\newblock \DOIprefix\doi{10.1002/mrm.22361}.
\bibitem[{Neher et~al.(2014)Neher, Laun, Stieltjes and Maier-Hein}]{Neher2013e}
\bibinfo{author}{Neher, P.F.}, \bibinfo{author}{Laun, F.B.},
  \bibinfo{author}{Stieltjes, B.}, \bibinfo{author}{Maier-Hein, K.H.},
  \bibinfo{year}{2014}.
\newblock \bibinfo{title}{{Fiberfox: Facilitating the creation of realistic
  white matter software phantoms}}.
\newblock \bibinfo{journal}{Magnetic Resonance in Medicine}
  \bibinfo{volume}{72}, \bibinfo{pages}{1460--1470}.
\newblock \DOIprefix\doi{10.1002/mrm.25045}.
\bibitem[{Noll et~al.(1991)Noll, Nishimura and Macovski}]{Noll1991}
\bibinfo{author}{Noll, D.C.}, \bibinfo{author}{Nishimura, D.G.},
  \bibinfo{author}{Macovski, A.}, \bibinfo{year}{1991}.
\newblock \bibinfo{title}{{Homodyne Detection in Magnetic Resonance Imaging}}.
\newblock \bibinfo{journal}{IEEE Transactions on Medical Imaging}
  \bibinfo{volume}{10}, \bibinfo{pages}{154--163}.
\newblock \DOIprefix\doi{10.1109/42.79473}.
\bibitem[{Nunes et~al.(2006)Nunes, Hajnal, Golay and Larkman}]{Nunes2006}
\bibinfo{author}{Nunes, R.G.}, \bibinfo{author}{Hajnal, J.V.},
  \bibinfo{author}{Golay, X.}, \bibinfo{author}{Larkman, D.J.},
  \bibinfo{year}{2006}.
\newblock \bibinfo{title}{{Simultaneous slice excitation and reconstruction for
  single shot EPI}}.
\newblock \bibinfo{journal}{Proc. Intl. Soc. Mag. Reson. Med}
  \bibinfo{volume}{13}, \bibinfo{pages}{293}.
\bibitem[{Papoulis(1991)}]{papoulis1991}
\bibinfo{author}{Papoulis, A.}, \bibinfo{year}{1991}.
\newblock \bibinfo{title}{{Probability, random variables, and stochastic
  processes}}.
\newblock \bibinfo{publisher}{Tata McGraw-Hill Education}.
\bibitem[{Paquette et~al.(2015)Paquette, Merlet, Gilbert, Deriche and
  Descoteaux}]{Paquette2014a}
\bibinfo{author}{Paquette, M.}, \bibinfo{author}{Merlet, S.},
  \bibinfo{author}{Gilbert, G.}, \bibinfo{author}{Deriche, R.},
  \bibinfo{author}{Descoteaux, M.}, \bibinfo{year}{2015}.
\newblock \bibinfo{title}{{Comparison of sampling strategies and sparsifying
  transforms to improve compressed sensing diffusion spectrum imaging}}.
\newblock \bibinfo{journal}{Magnetic Resonance in Medicine}
  \bibinfo{volume}{73}, \bibinfo{pages}{401--416}.
\newblock \DOIprefix\doi{10.1002/mrm.25093}.
\bibitem[{Pieciak et~al.(2018)Pieciak, Rabanillo-Viloria and
  Aja-Fernandez}]{Pieciak2018}
\bibinfo{author}{Pieciak, T.}, \bibinfo{author}{Rabanillo-Viloria, I.},
  \bibinfo{author}{Aja-Fernandez, S.}, \bibinfo{year}{2018}.
\newblock \bibinfo{title}{{Bias correction for non-stationary noise filtering
  in MRI}}, in: \bibinfo{booktitle}{2018 IEEE 15th International Symposium on
  Biomedical Imaging (ISBI 2018)}, \bibinfo{publisher}{IEEE}. pp.
  \bibinfo{pages}{307--310}.
\newblock \DOIprefix\doi{10.1109/ISBI.2018.8363580}.
\bibitem[{Pieciak et~al.(2016)Pieciak, Vegas-Sanchez-Ferrero and
  Aja-Fernandez}]{Pieciak2016}
\bibinfo{author}{Pieciak, T.}, \bibinfo{author}{Vegas-Sanchez-Ferrero, G.},
  \bibinfo{author}{Aja-Fernandez, S.}, \bibinfo{year}{2016}.
\newblock \bibinfo{title}{{Variance stabilization of noncentral-chi data:
  Application to noise estimation in MRI}}, in: \bibinfo{booktitle}{2016 IEEE
  13th International Symposium on Biomedical Imaging (ISBI)},
  \bibinfo{publisher}{IEEE}. pp. \bibinfo{pages}{1376--1379}.
\newblock \URLprefix \url{http://ieeexplore.ieee.org/document/7493523/},
  \DOIprefix\doi{10.1109/ISBI.2016.7493523}.
\bibitem[{Poldrack et~al.(2015)Poldrack, Laumann, Koyejo, Gregory, Hover, Chen,
  Gorgolewski, Luci, Joo, Boyd, Hunicke-Smith, Simpson, Caven, Sochat, Shine,
  Gordon, Snyder, Adeyemo, Petersen, Glahn, {Reese Mckay}, Curran,
  G{\"{o}}ring, Carless, Blangero, Dougherty, Leemans, Handwerker, Frick,
  Marcotte and Mumford}]{Poldrack2015a}
\bibinfo{author}{Poldrack, R.A.}, \bibinfo{author}{Laumann, T.O.},
  \bibinfo{author}{Koyejo, O.}, \bibinfo{author}{Gregory, B.},
  \bibinfo{author}{Hover, A.}, \bibinfo{author}{Chen, M.Y.},
  \bibinfo{author}{Gorgolewski, K.J.}, \bibinfo{author}{Luci, J.},
  \bibinfo{author}{Joo, S.J.}, \bibinfo{author}{Boyd, R.L.},
  \bibinfo{author}{Hunicke-Smith, S.}, \bibinfo{author}{Simpson, Z.B.},
  \bibinfo{author}{Caven, T.}, \bibinfo{author}{Sochat, V.},
  \bibinfo{author}{Shine, J.M.}, \bibinfo{author}{Gordon, E.},
  \bibinfo{author}{Snyder, A.Z.}, \bibinfo{author}{Adeyemo, B.},
  \bibinfo{author}{Petersen, S.E.}, \bibinfo{author}{Glahn, D.C.},
  \bibinfo{author}{{Reese Mckay}, D.}, \bibinfo{author}{Curran, J.E.},
  \bibinfo{author}{G{\"{o}}ring, H.H.H.}, \bibinfo{author}{Carless, M.A.},
  \bibinfo{author}{Blangero, J.}, \bibinfo{author}{Dougherty, R.},
  \bibinfo{author}{Leemans, A.}, \bibinfo{author}{Handwerker, D.A.},
  \bibinfo{author}{Frick, L.}, \bibinfo{author}{Marcotte, E.M.},
  \bibinfo{author}{Mumford, J.A.}, \bibinfo{year}{2015}.
\newblock \bibinfo{title}{{Long-term neural and physiological phenotyping of a
  single human}}.
\newblock \bibinfo{journal}{Nature Communications} \bibinfo{volume}{6},
  \bibinfo{pages}{8885}.
\newblock \DOIprefix\doi{10.1038/ncomms9885}.
\bibitem[{Pruessmann et~al.(1999)Pruessmann, Weiger, Scheidegger and
  Boesiger}]{Pruessmann1999}
\bibinfo{author}{Pruessmann, K.P.}, \bibinfo{author}{Weiger, M.},
  \bibinfo{author}{Scheidegger, M.B.}, \bibinfo{author}{Boesiger, P.},
  \bibinfo{year}{1999}.
\newblock \bibinfo{title}{{SENSE: Sensitivity encoding for fast MRI}}.
\newblock \bibinfo{journal}{Magnetic Resonance in Medicine}
  \bibinfo{volume}{42}, \bibinfo{pages}{952--962}.
\newblock
  \DOIprefix\doi{10.1002/(SICI)1522-2594(199911)42:5<952::AID-MRM16>3.0.CO;2-S}.
\bibitem[{Sakaie and Lowe(2017)}]{Sakaie2017}
\bibinfo{author}{Sakaie, K.}, \bibinfo{author}{Lowe, M.}, \bibinfo{year}{2017}.
\newblock \bibinfo{title}{{Retrospective correction of bias in diffusion tensor
  imaging arising from coil combination mode}}.
\newblock \bibinfo{journal}{Magnetic Resonance Imaging} \bibinfo{volume}{37},
  \bibinfo{pages}{203--208}.
\newblock \DOIprefix\doi{10.1016/j.mri.2016.12.004}.
\bibitem[{Sakaie et~al.(2018)Sakaie, Zhou, Lin, Debbins, Lowe and
  Fox}]{Sakaie2018}
\bibinfo{author}{Sakaie, K.}, \bibinfo{author}{Zhou, X.}, \bibinfo{author}{Lin,
  J.}, \bibinfo{author}{Debbins, J.}, \bibinfo{author}{Lowe, M.},
  \bibinfo{author}{Fox, R.J.}, \bibinfo{year}{2018}.
\newblock \bibinfo{title}{{Technical Note: Retrospective reduction in
  systematic differences across scanner changes by accounting for noise floor
  effects in diffusion tensor imaging}}.
\newblock \bibinfo{journal}{Medical Physics} \bibinfo{volume}{45},
  \bibinfo{pages}{4171--4178}.
\newblock \DOIprefix\doi{10.1002/mp.13088}.
\bibitem[{Sotiropoulos et~al.(2013)Sotiropoulos, Moeller, Jbabdi, Xu,
  Andersson, Auerbach, Yacoub, Feinberg, Setsompop, Wald, Behrens, Ugurbil and
  Lenglet}]{Sotiropoulos2013b}
\bibinfo{author}{Sotiropoulos, S.N.}, \bibinfo{author}{Moeller, S.},
  \bibinfo{author}{Jbabdi, S.}, \bibinfo{author}{Xu, J.},
  \bibinfo{author}{Andersson, J.L.}, \bibinfo{author}{Auerbach, E.J.},
  \bibinfo{author}{Yacoub, E.}, \bibinfo{author}{Feinberg, D.},
  \bibinfo{author}{Setsompop, K.}, \bibinfo{author}{Wald, L.L.},
  \bibinfo{author}{Behrens, T.E.J.}, \bibinfo{author}{Ugurbil, K.},
  \bibinfo{author}{Lenglet, C.}, \bibinfo{year}{2013}.
\newblock \bibinfo{title}{{Effects of image reconstruction on fiber orientation
  mapping from multichannel diffusion MRI: Reducing the noise floor using
  SENSE}}.
\newblock \bibinfo{journal}{Magnetic Resonance in Medicine}
  \bibinfo{volume}{70}, \bibinfo{pages}{1682--1689}.
\newblock \DOIprefix\doi{10.1002/mrm.24623}.
\bibitem[{St-Jean et~al.(2016)St-Jean, Coup{\'{e}} and
  Descoteaux}]{St-Jean2016a}
\bibinfo{author}{St-Jean, S.}, \bibinfo{author}{Coup{\'{e}}, P.},
  \bibinfo{author}{Descoteaux, M.}, \bibinfo{year}{2016}.
\newblock \bibinfo{title}{{Non Local Spatial and Angular Matching: Enabling
  higher spatial resolution diffusion MRI datasets through adaptive
  denoising}}.
\newblock \bibinfo{journal}{Medical Image Analysis} \bibinfo{volume}{32},
  \bibinfo{pages}{115--130}.
\newblock \DOIprefix\doi{10.1016/j.media.2016.02.010}.
\bibitem[{St-Jean et~al.(2018a)St-Jean, {De Luca}, Tax, Viergever and
  Leemans}]{St-Jean2018d}
\bibinfo{author}{St-Jean, S.}, \bibinfo{author}{{De Luca}, A.},
  \bibinfo{author}{Tax, C.M.W.}, \bibinfo{author}{Viergever, M.A.},
  \bibinfo{author}{Leemans, A.}, \bibinfo{year}{2018}a.
\newblock \bibinfo{title}{{Datasets for 'Automated characterization of noise
  distributions in diffusion MRI data'}}.
\newblock \bibinfo{journal}{Zenodo} \DOIprefix\doi{10.5281/zenodo.2483105}.
\bibitem[{St-Jean et~al.(2019)St-Jean, {De Luca}, Tax, Viergever and
  Leemans}]{St-Jean2019d}
\bibinfo{author}{St-Jean, S.}, \bibinfo{author}{{De Luca}, A.},
  \bibinfo{author}{Tax, C.M.W.}, \bibinfo{author}{Viergever, M.A.},
  \bibinfo{author}{Leemans, A.}, \bibinfo{year}{2019}.
\newblock \bibinfo{title}{{samuelstjean/autodmri: First release - 2019-07-17}}.
\newblock \DOIprefix\doi{10.5281/zenodo.3339158}.
\bibitem[{St-Jean et~al.(2018b)St-Jean, {De Luca}, Viergever and
  Leemans}]{St-jean2018a}
\bibinfo{author}{St-Jean, S.}, \bibinfo{author}{{De Luca}, A.},
  \bibinfo{author}{Viergever, M.A.}, \bibinfo{author}{Leemans, A.},
  \bibinfo{year}{2018}b.
\newblock \bibinfo{title}{{Automatic, Fast and Robust Characterization of Noise
  Distributions for Diffusion MRI}}, in: \bibinfo{editor}{Frangi, A.F.},
  \bibinfo{editor}{Schnabel, J.A.}, \bibinfo{editor}{Davatzikos, C.},
  \bibinfo{editor}{Alberola-L{\'{o}}pez, C.}, \bibinfo{editor}{Fichtinger, G.}
  (Eds.), \bibinfo{booktitle}{Medical Image Computing and Computer Assisted
  Intervention -- MICCAI 2018}. \bibinfo{publisher}{Springer International
  Publishing}, pp. \bibinfo{pages}{304--312}.
\newblock \DOIprefix\doi{10.1007/978-3-030-00928-1_35}.
\bibitem[{Storey et~al.(2007)Storey, Frigo, Hinks, Mock, Collick, Baker,
  Marmurek and Graham}]{Storey2007a}
\bibinfo{author}{Storey, P.}, \bibinfo{author}{Frigo, F.J.},
  \bibinfo{author}{Hinks, R.S.}, \bibinfo{author}{Mock, B.J.},
  \bibinfo{author}{Collick, B.D.}, \bibinfo{author}{Baker, N.},
  \bibinfo{author}{Marmurek, J.}, \bibinfo{author}{Graham, S.J.},
  \bibinfo{year}{2007}.
\newblock \bibinfo{title}{{Partial k-space reconstruction in single-shot
  diffusion-weighted echo-planar imaging}}.
\newblock \bibinfo{journal}{Magnetic Resonance in Medicine}
  \bibinfo{volume}{57}, \bibinfo{pages}{614--619}.
\newblock \DOIprefix\doi{10.1002/mrm.21132}.
\bibitem[{Tabelow et~al.(2015)Tabelow, Voss and Polzehl}]{Tabelow2014}
\bibinfo{author}{Tabelow, K.}, \bibinfo{author}{Voss, H.U.},
  \bibinfo{author}{Polzehl, J.}, \bibinfo{year}{2015}.
\newblock \bibinfo{title}{{Local estimation of the noise level in MRI using
  structural adaptation}}.
\newblock \bibinfo{journal}{Medical Image Analysis} \bibinfo{volume}{20},
  \bibinfo{pages}{76--86}.
\newblock \DOIprefix\doi{10.1016/j.media.2014.10.008}.
\bibitem[{Tax et~al.(2019)Tax, Grussu, Kaden, Ning, Rudrapatna, {John Evans},
  St-Jean, Leemans, Koppers, Merhof, Ghosh, Tanno, Alexander, Zappal{\`{a}},
  Charron, Kusmia, Linden, Jones and Veraart}]{Tax2019}
\bibinfo{author}{Tax, C.M.}, \bibinfo{author}{Grussu, F.},
  \bibinfo{author}{Kaden, E.}, \bibinfo{author}{Ning, L.},
  \bibinfo{author}{Rudrapatna, U.}, \bibinfo{author}{{John Evans}, C.},
  \bibinfo{author}{St-Jean, S.}, \bibinfo{author}{Leemans, A.},
  \bibinfo{author}{Koppers, S.}, \bibinfo{author}{Merhof, D.},
  \bibinfo{author}{Ghosh, A.}, \bibinfo{author}{Tanno, R.},
  \bibinfo{author}{Alexander, D.C.}, \bibinfo{author}{Zappal{\`{a}}, S.},
  \bibinfo{author}{Charron, C.}, \bibinfo{author}{Kusmia, S.},
  \bibinfo{author}{Linden, D.E.}, \bibinfo{author}{Jones, D.K.},
  \bibinfo{author}{Veraart, J.}, \bibinfo{year}{2019}.
\newblock \bibinfo{title}{{Cross-scanner and cross-protocol diffusion MRI data
  harmonisation: A benchmark database and evaluation of algorithms}}.
\newblock \bibinfo{journal}{NeuroImage} \bibinfo{volume}{195},
  \bibinfo{pages}{285--299}.
\newblock \DOIprefix\doi{10.1016/j.neuroimage.2019.01.077}.
\bibitem[{Thom(1958)}]{Thom1958}
\bibinfo{author}{Thom, H.C.S.}, \bibinfo{year}{1958}.
\newblock \bibinfo{title}{{A Note on the Gamma Distribution}}.
\newblock \bibinfo{journal}{Monthly Weather Review} \bibinfo{volume}{86},
  \bibinfo{pages}{117--122}.
\bibitem[{Tibshirani and Taylor(2011)}]{Tibshirani2011}
\bibinfo{author}{Tibshirani, R.J.}, \bibinfo{author}{Taylor, J.},
  \bibinfo{year}{2011}.
\newblock \bibinfo{title}{{The solution path of the generalized lasso}}.
\newblock \bibinfo{journal}{The Annals of Statistics} \bibinfo{volume}{39},
  \bibinfo{pages}{1335--1371}.
\newblock \DOIprefix\doi{10.1214/11-AOS878}.
\bibitem[{Todd et~al.(2016)Todd, Moeller, Auerbach, Yacoub, Flandin and
  Weiskopf}]{Todd2016a}
\bibinfo{author}{Todd, N.}, \bibinfo{author}{Moeller, S.},
  \bibinfo{author}{Auerbach, E.J.}, \bibinfo{author}{Yacoub, E.},
  \bibinfo{author}{Flandin, G.}, \bibinfo{author}{Weiskopf, N.},
  \bibinfo{year}{2016}.
\newblock \bibinfo{title}{{Evaluation of 2D multiband EPI imaging for
  high-resolution, whole-brain, task-based fMRI studies at 3T: Sensitivity and
  slice leakage artifacts}}.
\newblock \bibinfo{journal}{NeuroImage} \bibinfo{volume}{124},
  \bibinfo{pages}{32--42}.
\newblock \DOIprefix\doi{10.1016/j.neuroimage.2015.08.056}.
\bibitem[{Veraart et~al.(2016)Veraart, Fieremans and Novikov}]{Veraart2015a}
\bibinfo{author}{Veraart, J.}, \bibinfo{author}{Fieremans, E.},
  \bibinfo{author}{Novikov, D.S.}, \bibinfo{year}{2016}.
\newblock \bibinfo{title}{{Diffusion MRI noise mapping using random matrix
  theory}}.
\newblock \bibinfo{journal}{Magnetic Resonance in Medicine}
  \bibinfo{volume}{76}, \bibinfo{pages}{1582--1593}.
\newblock \DOIprefix\doi{10.1002/mrm.26059}.
\bibitem[{Weisstein(2017)}]{weisstein_gamma}
\bibinfo{author}{Weisstein, E.W.}, \bibinfo{year}{2017}.
\newblock \bibinfo{title}{{Gamma Distribution. From MathWorld---A Wolfram Web
  Resource http://mathworld.wolfram.com/GammaDistribution.html Last accessed
  2017-10-09}}.
\bibitem[{Zhang et~al.(2012)Zhang, Schneider, Wheeler-Kingshott and
  Alexander}]{Zhang2012d}
\bibinfo{author}{Zhang, H.}, \bibinfo{author}{Schneider, T.},
  \bibinfo{author}{Wheeler-Kingshott, C.A.}, \bibinfo{author}{Alexander, D.C.},
  \bibinfo{year}{2012}.
\newblock \bibinfo{title}{{NODDI: practical in vivo neurite orientation
  dispersion and density imaging of the human brain.}}
\newblock \bibinfo{journal}{NeuroImage} \bibinfo{volume}{61},
  \bibinfo{pages}{1000--16}.
\newblock \DOIprefix\doi{10.1016/j.neuroimage.2012.03.072}.

\end{thebibliography}

\end{document}